\documentclass[english]{article}
\usepackage[T1]{fontenc}
\usepackage[utf8]{inputenc}
\usepackage{color}
\usepackage{enumitem}
\usepackage{amsmath}
\usepackage{amssymb}

\makeatletter
\numberwithin{equation}{section}

\usepackage{jheppub}

\author[a,b]{Tamas Gombor}

\affiliation[a]{MTA-ELTE “Momentum” Integrable Quantum Dynamics Research Group, Department of Theoretical Physics, Eötvös
  Loránd University}
\affiliation[b]{Holographic QFT Group, Wigner Research Centre for Physics, Budapest, Hungary}

\emailAdd{gombort@caesar.elte.hu}

\abstract{We study the integrable crosscap states of the integrable quantum spin chains 
and we classify them for the $\mathfrak{gl}(N)$ symmetric models. 
We also give a derivation for the exact overlaps between the integrable crosscap states and the Bethe states. 
The first part of the derivation is to calculate sum formula for the off-shell overlap. 
Using this formula we prove that the normalized overlaps of the multi-particle states are ratios of the Gaudin-like determinants. 
Furthermore we collect the integrable crosscap states which can be relevant in the AdS/CFT correspondence.}

\makeatother

\usepackage{babel}
\begin{document}
\title{Integrable crosscap states in $\mathfrak{gl}(N)$ spin chains}
\maketitle

\section{Introduction}

In recent years there has been renewed interest for the integrable
boundary states of 1+1 dimensional field theories \cite{Ghoshal:1993tm}
and their spin chain versions \cite{Piroli:2017sei,Pozsgay_2019}.
These states appear in quite distinct parts of theoretical physics
including statistical physics and the gauge/string duality. In statistical
physics these quantities appear in the context of non-equilibrium
dynamics of the integrable models. In a quantum quench (a parameter
of the Hamiltonian is suddenly changed) the goal is to study the non-equilibrium
dynamics, the emergence of steady states, and their properties \cite{Essler:2016ufo}.
The so-called Quench Action method \cite{Caux:2013ra} is one of the
main methods for the investigation of the steady states where the
knowledge of the exact overlaps between the boundary states and energy
eigenstates is an important input \cite{De_Nardis_2014,Rylands:2022gev}.
The exact overlap formulas also played a central role in the early
studies of the Generalized Gibbs Ensemble (GGE) in interacting integrable
models, see \cite{Wouters_2014,Pozsgay_2014}.

The boundary states also play crucial role in the AdS/CFT correspondence.
It turned out that the one-point functions in defect CFT \cite{deLeeuw:2015hxa,Buhl-Mortensen:2015gfd,deLeeuw:2016umh,DeLeeuw:2018cal,deLeeuw:2019ebw,Kristjansen:2020mhn,Kristjansen:2021abc,Gombor:2022aqj}
and three-point functions involving determinant operators \cite{Jiang:2019xdz,Jiang:2019zig,Yang:2021hrl},
can be mapped to the overlaps between boundary states and energy eigenstates
of integrable spin chains at weak coupling. Some of these results
are also extended to any coupling at the asymptotic limit \cite{Jiang:2019xdz,Gombor:2020kgu,Gombor:2020auk,Komatsu:2020sup}.

In recent years several exact overlaps were determined. Based on the
early results on the overlaps of integrable boundary states, it was
conjectured that the normalized overlap always can be written in the
following form\footnote{The boundary states have two types: two-site states and matrix product
states (MPS). The form (\ref{eq:gen}) corresponds to the two-site
states. For MPS there is also a summation for the boundary dependent
terms.}
\begin{equation}
{\color{blue}\frac{\left|{\color{red}\langle\mathcal{B}|}{\color{blue}\mathbb{B}(\bar{u})}\right|^{2}}{||\mathbb{B}(\bar{u})||^{2}}}={\color{red}\underbrace{\prod_{j,\nu}\mathcal{F}_{\nu}(u_{j}^{\nu})}_{\text{boundary dependent}}}\times{\color{blue}\underbrace{\frac{\det G^{+}}{\det G^{-}}}_{\text{universal}}},\label{eq:gen}
\end{equation}
where $\langle\mathcal{B}|$ is the integrable boundary state and
$\mathbb{B}(\bar{u})$ is a Bethe state\footnote{In this paper we use the notations of \cite{Hutsalyuk:2016srn,Hutsalyuk:2017tcx,Hutsalyuk:2017way,Hutsalyuk:2020dlw}
where the symbol $\mathbb{B}(\bar{u})$ without ket denotes the Bethe
state.}. This conjecture was proved in several ways \cite{Brockmann_2014,Brockmann_2014odd,Foda:2015nfk,Jiang:2020sdw,Gombor:2021uxz}
but these derivations could be used only for XXZ type spin chains
(rank 1 symmetry). The first derivation for nested spin chains was
presented in \cite{Gombor:2021hmj}. The derivation is based on the
algebraic relation of the boundary state, the so-called the $KT$-relation.
This $KT$-relation can be used to derive a recurrence equation for
the off-shell overlap and applying the original Korepin's idea this
recursion is enough the prove the form (\ref{eq:gen}). In \cite{Gombor:2021hmj}
the exact overlaps of $\mathfrak{gl}(N)$ spin chain were proved for
all of the $\mathfrak{so}(N)$ and $\mathfrak{gl}(\left\lfloor \frac{N}{2}\right\rfloor )\oplus\mathfrak{gl}(\left\lceil \frac{N}{2}\right\rceil )$
symmetric boundary states.

Recently, in \cite{Caetano:2021dbh} the authors introduced an other
type of initial states of spin chains which are the analogous versions
of the crosscap states of 2d CFT \cite{Ishibashi:1988kg}. Geometrically,
these states correspond to non-orientable surfaces such as $\mathbb{RP}^{2}$
and the Klein bottle. In \cite{Caetano:2021dbh} the spin chain version
of crosscap states was introduced for XXX spin chain. It was shown
that this crosscap state satisfies the same integrability condition
which was proposed for the boundary states in \cite{Piroli:2017sei}.
The authors of \cite{Caetano:2021dbh} also proposed a formula for
the overlap:
\begin{equation}
{\color{blue}\frac{\left|{\color{red}\langle\mathcal{C}|}{\color{blue}\mathbb{B}(\bar{u})}\right|^{2}}{||\mathbb{B}(\bar{u})||^{2}}}={\color{blue}\underbrace{\frac{\det G^{+}}{\det G^{-}}}_{\text{universal}}},\label{eq:gen-1}
\end{equation}
where $\langle\mathcal{C}|$ is the integrable crosscap state. This
overlap formula has not been proved, it was guessed and validated
by numerical computations. We can see that the overlap of the crosscap
state is similar to the boundary state but now the boundary dependent
term is missing.

Recently a holographic description of $\mathcal{N}=4$ super Yang-Mills
on the four-dimensional real projective space $\mathbb{RP}^{4}$ was
proposed in \cite{Caetano:2022mus}. The $\mathbb{RP}^{4}$ is the
simplest unorientable four-manifold which is obtained by modding out
the four dimensional sphere $S^{4}$ by the involution that identifies
antipodal points. To formulate $\mathcal{N}=4$ SYM on $\mathbb{RP}^{4}$
one has to specify how this involution acts on the elementary fields.
In \cite{Caetano:2022mus} the authors found two possibility which
preserves the half of the supersymmetry: the involution acts as identity
or charge conjugation on the fields. They investigated the first version
and showed that the one-point functions of scalar operators do not
have selection rules what we expect for an integrability preserving
configuration. However the authors stated that for the second configuration
there is a compelling guess for the holographic dual, as an orientifold
projection of $AdS_{5}\times S^{5}$, which in particular adds a crosscap
on the worldsheet. In \cite{Caetano:2021dbh} it was shown that integrability
survives in their presence. This suggests that the orientifold setup
is in fact integrable. Based on these observations, the classification
of the integrable crosscap states of $SO(6)$ spin chain can help
to find the integrability preserving orientifold setups of the $\mathcal{N}=4$
SYM.

The goals of this paper are the following: classifying the boundary
crosscap states for every $\mathfrak{gl}(N)$ symmetric spin chain
and deriving the proposed overlap formula (\ref{eq:gen-1}) for as
many states as possible. As mentioned earlier, for boundary states
these results are consequences of the $KT$-relation therefore at
first we generalize the algebraic framework of \cite{Gombor:2021hmj}
for crosscap states, i.e., we introduce the $KT$-relation for crosscap
states of the $\mathfrak{gl}(N)$ spin chains. Using this $KT$-relation
we can classify the integrable crosscap states. It turns out that
the possible residual symmetries are $\mathfrak{so}(N)$, $\mathfrak{sp}(N)$
and $\mathfrak{gl}(M)\oplus\mathfrak{gl}(N-M)$. We also discuss which
of these states may be relevant in AdS/CFT. An other benefit of the
$KT$-relation is that we can derive a sum formula for the off-shell
overlap which can be used to derive the on-shell formula (\ref{eq:gen-1})
for the $\mathfrak{so}(N)$ and $\mathfrak{gl}(\left\lfloor \frac{N}{2}\right\rfloor )\oplus\mathfrak{gl}(\left\lceil \frac{N}{2}\right\rceil )$
symmetric crosscap states.

The paper is organized as follows. In section \ref{sec:Definitions}
we review the basic definitions of the $\mathfrak{gl}(N)$ spin chains.
In section \ref{sec:Integrable-crosscap-states} we introduce the
$KT$-relations and we also classify the solutions which give us the
integrable crosscap states. In section \ref{sec:Application-to-AdS/CFT}
we discuss the potentially relevant states for the AdS/CFT correspondence.
In section \ref{sec:Sum-formulas-for} we derive a sum formula for
the off-shell overlaps. In section \ref{sec:On-shell-limit} we take
the on-shell limit of the sum formula and show that the normalized
overlap is equal to the ratio of Gaudin-like determinants.

\section{Definitions\label{sec:Definitions}}

In this section we review the definitions of the $\mathfrak{gl}(N)$
spin chains. We use the notations of \cite{Hutsalyuk:2017tcx,Hutsalyuk:2017way,Liashyk:2018egk,Hutsalyuk:2020dlw,Gombor:2021hmj}.
At first let us define the monodromy matrix $T(u)=\sum_{i,j=1}^{N}E_{i,j}\otimes T_{i,j}(u)\in\mathrm{End}(\mathbb{C}^{N})\otimes\mathrm{End}(\mathcal{H})$
($\mathcal{H}$ is the quantum space and \textbf{$E_{i,j}$}-s are
the generators of $\mathfrak{gl}(N)$) by the usual $RTT$-relation
\begin{equation}
R_{12}(u-v)T_{1}(u)T_{2}(v)=T_{2}(v)T_{1}(u)R_{12}(u-v),
\end{equation}
where we used the $\mathfrak{gl}(N)$ R-matrix
\begin{equation}
R(u)=u\mathbf{1}+c\mathbf{P},
\end{equation}
where $c$ is a constant and $\mathbf{I}$ and $\mathbf{P}$ are the
identity and the permutation operators in the vector space $\mathbb{C}^{N}\otimes\mathbb{C}^{N}$.
The monodromy matrix entries $T_{i,j}(u)$ generate the famous Yangian
algebra $Y(N)$ \cite{Molev:1994rs}. We can define representations
of the Yangian $Y(N)$ on the quantum space $\mathcal{H}$. A representation
is highest weight if there exists a unique pseudo-vacuum $\left|0\right\rangle \in\mathcal{H}$
such that
\begin{equation}
\begin{split}T_{i,i}\left|0\right\rangle  & =\lambda_{i}(u)\left|0\right\rangle ,\qquad\text{for }i=1,\dots,N,\\
T_{j,i}\left|0\right\rangle  & =0,\qquad\qquad\text{for }1\leq i<j\leq N.
\end{split}
\end{equation}
The $\lambda_{i}(u)$-s are the vacuum eigenvalues. The irreducible
representations of $\mathfrak{gl}(N)$ can be generalized for representations
of $Y(N)$. For the $N$-tuples $\Lambda=(\Lambda_{1},\dots,\Lambda_{N})$
we can define a representation $\mathcal{V}^{\Lambda}$. Let $E_{i,j}^{\Lambda}\in\mathrm{End}(\mathcal{V}^{\Lambda}),$
$|0^{\Lambda}\rangle\in\mathcal{V}^{\Lambda}$ be the corresponding
generators and highest weight state for which
\begin{equation}
\begin{split}E_{i,j}^{\Lambda}|0^{\Lambda}\rangle & =\Lambda_{i}|0^{\Lambda}\rangle,\qquad\text{for }i=1,\dots,N,\\
E_{i,j}^{\Lambda}|0^{\Lambda}\rangle & =0,\qquad\qquad\text{for }1\leq i<j\leq N.
\end{split}
\end{equation}
Using this representation we can define the matrices (Lax-operators)
\begin{equation}
L^{\Lambda}(u)=\mathbf{1}+\frac{c}{u}\sum_{i,j=1}^{N}E_{i,j}\otimes E_{j,i}^{\Lambda}\in\mathrm{End}(\mathbb{C}^{N})\otimes\mathrm{End}(\mathcal{V}^{\Lambda}),\label{eq:Lax}
\end{equation}
which are solutions of the $RTT$-relation. We use the following convention
for the representation
\begin{equation}
\left(E_{i,j}^{\Lambda}\right)^{t}=E_{j,i}^{\Lambda},
\end{equation}
where $t$ denotes the transposition. We can define the twisted Lax
operator
\begin{align}
\bar{L}_{1,2}^{\Lambda}(u)=V_{1}\left(L_{1,2}^{\Lambda}(-u)\right)^{t_{1}}V_{1} & =\mathbf{1}-\frac{c}{u}E_{N+1-j,N+1-i}\otimes E_{j,i}^{\Lambda}=\nonumber \\
 & =\mathbf{1}-\frac{c}{u}E_{i,j}\otimes E_{N+1-i,N+1-j}^{\Lambda},\label{eq:twLax}
\end{align}
where $\left[V\right]_{i,j}=\delta_{i,N+1-j}$. Defining the operator
$V^{\Lambda}\in GL(\mathcal{V}_{\Lambda})$ as
\begin{equation}
E_{N+1-i,N+1-j}^{\Lambda}=V^{\Lambda}E_{i,j}^{\Lambda}V^{\Lambda},
\end{equation}
we obtain that
\begin{equation}
\bar{L}_{1,2}^{\Lambda}(u)=V_{2}^{\Lambda}\left(L_{1,2}^{\Lambda}(-u)\right)^{t_{2}}V_{2}^{\Lambda}.
\end{equation}
Let us also define the contra-gradient reps $\bar{\Lambda}=(-\Lambda_{N},-\Lambda_{N-1},\dots,-\Lambda_{1})$.
Since the generators
\begin{equation}
\bar{E}_{i,j}^{\Lambda}:=-V^{\Lambda}E_{j,i}^{\Lambda}V^{\Lambda}=-E_{N+1-j,N+1-i}^{\Lambda}
\end{equation}
have highest weights $(-\Lambda_{N},-\Lambda_{N-1},\dots,-\Lambda_{1})$,
we can define the generators of rep $\mathcal{V}_{\bar{\Lambda}}$
as
\begin{equation}
E_{i,j}^{\bar{\Lambda}}=\bar{E}_{i,j}^{\Lambda}.
\end{equation}
We can also obtain the following identity:
\begin{equation}
\left(L_{1,2}^{\Lambda}(u)\right)^{t_{1}t_{2}}=L_{1,2}^{\Lambda}(u).
\end{equation}

Let us consider the following tensor product quantum space $\mathcal{H}=\mathcal{H}^{(1)}\otimes\mathcal{H}^{(2)}$
and define monodromy matrices on each sub-spaces $T^{(i)}(u)\in\mathrm{End}(\mathbb{C}^{N})\otimes\mathcal{H}^{(i)}$
for $i=1,2$. We can define a monodromy matrix (which satisfy the
$RTT$-relation) on the tensor product space as
\begin{equation}
T(u)=T^{(2)}(u)T^{(1)}(u).
\end{equation}
The consequence of this co-product property is that we can build more
general monodromy matrices using the elementary ones (\ref{eq:Lax}):

\begin{equation}
T_{0}(u)=L_{0,J}^{\Lambda^{(J)}}(u-\xi_{J})\dots L_{0,1}^{\Lambda^{(1)}}(u-\xi_{1}).\label{eq:genmod}
\end{equation}

We can also define the transfer matrix
\begin{equation}
\mathcal{T}(u)=\mathrm{tr}T(u),
\end{equation}
which gives commuting quantities
\begin{equation}
[\mathcal{T}(u),\mathcal{T}(v)]=0.
\end{equation}
For a given sets of complex numbers $\bar{t}^{\mu}=\{t_{k}^{\mu}\}_{k=1}^{r_{k}}$,
$\mu=1,\dots,N-1$, following \cite{Hutsalyuk:2017tcx}, one can define
off-shell Bethe vectors
\[
\mathbb{B}(\bar{t})\equiv\mathbb{B}(\bar{t}^{1},\dots,\bar{t}^{N-1}).
\]
The recursion for the definition of the off-shell Bethe vectors can
be found in the appendix \ref{sec:Off-shell-Bethe-vectors}. We call
the Bethe vector on-shell if the Bethe roots $\bar{t}^{\mu}$ satisfies
the Bethe Ansatz equations
\begin{equation}
\alpha_{\mu}(t_{k}^{\mu}):=\frac{\lambda_{\mu}(t_{k}^{\mu})}{\lambda_{\mu+1}(t_{k}^{\mu})}=\frac{f(t_{k}^{\mu},\bar{t}_{k}^{\mu})}{f(\bar{t}_{k}^{\mu},t_{k}^{\mu})}\frac{f(\bar{t}^{\mu+1},t_{k}^{\mu})}{f(t_{k}^{\mu},\bar{t}^{\mu-1})},\label{eq:BE}
\end{equation}
where we used the following notations
\begin{equation}
\begin{split}f(u,v) & =1+g(u,v)=\frac{u-v+c}{u-v},\qquad\bar{t}_{k}^{\mu}=\bar{t}^{\mu}\backslash t_{k}^{\mu},\\
f(u,\bar{t}^{i}) & =\prod_{k=1}^{r_{i}}f(u,t_{k}^{i}),\quad f(\bar{t}^{i},u)=\prod_{k=1}^{r_{i}}f(t_{k}^{i},u),\quad f(\bar{t}^{i},\bar{t}^{j})=\prod_{k=1}^{r_{i}}f(t_{k}^{i},\bar{t}^{j}).
\end{split}
\end{equation}
The on-shell Bethe vectors are eigenvectors of the transfer matrix
\begin{equation}
\mathcal{T}(u)\mathbb{B}(\bar{t})=\tau(u|\bar{t})\mathbb{B}(\bar{t}),
\end{equation}
with the eigenvalue
\begin{equation}
\tau(u|\bar{t})=\sum_{i=1}^{N}\lambda_{i}(u)f(\bar{t}^{i},u)f(u,\bar{t}^{i-1}),\label{eq:eig}
\end{equation}
where $r_{0}=r_{N}=0$.

One can also define the left eigenvectors of the transfer matrix
\begin{equation}
\mathbb{C}(\bar{t})\mathcal{T}(u)=\tau(u|\bar{t})\mathbb{C}(\bar{t}),
\end{equation}
and the square of the norm of the on-shell Bethe states satisfies
the Gaudin hypothesis \cite{Hutsalyuk:2017way}
\begin{equation}
\mathbb{C}(\bar{t})\mathbb{B}(\bar{t})=\frac{\prod_{\nu=1}^{N-1}\prod_{k\neq l}f(t_{l}^{\nu},t_{k}^{\nu})}{\prod_{\nu=1}^{N-2}f(\bar{t}^{\nu+1},\bar{t}^{\nu})}\det G,\label{eq:norm}
\end{equation}
where $G$ is the Gaudin matrix given by
\begin{equation}
G_{j,k}^{(\mu,\nu)}=-c\frac{\partial\log\Phi_{j}^{(\mu)}}{\partial t_{k}^{\nu}},
\end{equation}
where we defined the expressions
\begin{equation}
\Phi_{k}^{(\mu)}=\alpha_{\mu}(t_{k}^{\mu})\frac{f(\bar{t}_{k}^{\mu},t_{k}^{\mu})}{f(t_{k}^{\mu},\bar{t}_{k}^{\mu})}\frac{f(t_{k}^{\mu},\bar{t}^{\mu-1})}{f(\bar{t}^{\mu+1},t_{k}^{\mu})}.
\end{equation}

We can also define an other monodromy matrix which satisfies the same
$RTT$-algebra \cite{Liashyk:2018egk}. This transfer matrix can be
obtained from one of the quantum minors as
\begin{align}
\widehat{T}_{N+1-j,N+1-i}(u) & =(-1)^{i+j}t_{1,\dots,\hat{i},\dots,N}^{1,\dots,\hat{j},\dots,N}(u-c)\mathrm{qdet}(T(u))^{-1},\\
t_{b_{1},b_{2},\dots,b_{m}}^{a_{1},a_{2},\dots,a_{m}}(u) & =\sum_{p}\mathrm{sgn}(p)T_{a,b_{p(1)}}(u)T_{a,b_{p(2)}}(u-c)\dots T_{a,b_{p(m)}}(u-(m-1)c),\label{eq:Qminors}\\
\mathrm{qdet}(T(u)) & =t_{1,2,\dots,N}^{1,2,\dots,N}(u).
\end{align}
Here $\hat{i}$ an\^{d} $\hat{j}$ mean that the corresponding indices
are omitted. We call $\widehat{T}$ as twisted monodromy matrix. The
twisted monodromy matrix $\widehat{T}$ is also a highest weight representation
of $Y(N)$ with
\begin{equation}
\begin{split}\widehat{T}_{i,i}\left|0\right\rangle  & =\hat{\lambda}_{i}(u)\left|0\right\rangle ,\qquad\text{for }i=1,\dots,N,\\
\widehat{T}_{j,i}\left|0\right\rangle  & =0,\qquad\qquad\text{for }1\leq i<j\leq N,
\end{split}
\end{equation}
where
\begin{equation}
\hat{\lambda}_{i}(u)=\frac{1}{\lambda_{N-i+1}(u-(N-i)c)}\prod_{k=1}^{N-i}\frac{\lambda_{k}(u-kc)}{\lambda_{k}(u-(k-1)c)},
\end{equation}
therefore the ratios of the vacuum eigenvalues have the following
form
\begin{equation}
\hat{\alpha}_{i}(u)=\frac{\hat{\lambda}_{i}(u)}{\hat{\lambda}_{i+1}(u)}=\alpha_{N-i}(u-(N-i)c).\label{eq:idAlph}
\end{equation}
Let $\hat{\mathbb{B}}(\bar{t})$ be the off-shell Bethe vector generated
from $\widehat{T}_{i,j}$. In \cite{Liashyk:2018egk} the connection
between the Bethe vectors $\mathbb{B}(\bar{t})$ and $\hat{\mathbb{B}}(\bar{t})$
was determined
\begin{equation}
\hat{\mathbb{B}}(\bar{t})=(-1)^{\#\bar{t}}\left(\prod_{s=1}^{N-2}f(\bar{t}^{s+1},\bar{t}^{s})\right)^{-1}\mathbb{B}(\mu(\bar{t})),\label{eq:connBhatB-1}
\end{equation}
where
\begin{equation}
\mu(\bar{t})=\{\bar{t}^{N-1}-c,\bar{t}^{N-2}-2c,\dots,\bar{t}^{1}-(N-1)c\}.
\end{equation}
From this identity we can obtain the eigenvalue of the twisted transfer
matrix 
\begin{align}
\widehat{\mathcal{T}}(u) & =\mathrm{tr}\widehat{T}(u)=\sum_{i=1}^{N}\widehat{T}_{i,i}(u),\\
\widehat{\mathcal{T}}(u)\mathbb{B}(\bar{t}) & =\hat{\tau}(u|\bar{t})\mathbb{B}(\bar{t}).
\end{align}
Using (\ref{eq:connBhatB-1}), (\ref{eq:eig}) and the fact that $\widehat{T}$
satisfies the same $RTT$-relation we can obtain that
\begin{equation}
\hat{\tau}(u|\bar{t})=\sum_{i=1}^{N}\hat{\lambda}_{i}(u)f(\bar{t}^{N-i}+(N-i)c,u)f(u,\bar{t}^{N-i+1}+(N-i+1)c).\label{eq:tweig}
\end{equation}

The twisted monodromy matrix $\widehat{T}$ is similar to the inverse
of the original monodromy matrix $T$:
\begin{equation}
V\widehat{T}^{t}(u)VT(u)=1,\label{eq:invemon}
\end{equation}
where $V$ is an off-diagonal $N\times N$ matrix of the auxiliary
space with the components $V_{i,j}=\delta_{i,N+1-j}$ and the superscript
$t$ is the transposition in the auxiliary space, i.e. $\left[\widehat{T}^{t}(u)\right]_{i,j}=\widehat{T}_{j,i}(u)$.
Applying this equation to the $RTT$-relation we obtain the $R\widehat{T}T$-relation
\begin{equation}
\bar{R}_{1,2}(u-v)\widehat{T}_{1}(u)T_{2}(v)=T_{2}(v)\widehat{T}_{1}(u)\bar{R}_{1,2}(u-v),
\end{equation}
where we used the crossed $R$-matrix
\begin{equation}
\bar{R}_{1,2}(u)=V_{2}R_{1,2}^{t_{2}}(-u)V_{2}.
\end{equation}
For a concrete form of the monodromy matrix (\ref{eq:genmod}) we
can obtain the explicit form of twisted monodromy matrix using (\ref{eq:invemon})
\begin{equation}
\widehat{T}_{0}(u)=\widehat{L}_{0,J}^{\Lambda^{(J)}}(u-\xi_{J})\dots\widehat{L}_{0,1}^{\Lambda^{(1)}}(u-\xi_{1}),
\end{equation}
where
\begin{equation}
\widehat{L}_{0,1}^{\Lambda}(u)=V_{0}\left(\left(\widehat{L}_{0,1}^{\Lambda}(u)\right)^{-1}\right)^{t_{0}}V_{0}.\label{eq:crossedLax}
\end{equation}
We can see that twisted monodromy matrix has a same co-product property
i.e.
\begin{equation}
T(u)=T^{(2)}(u)T^{(1)}(u)\Longrightarrow\widehat{T}(u)=\widehat{T}^{(2)}(u)\widehat{T}^{(1)}(u).
\end{equation}

\section{Integrable crosscap states\label{sec:Integrable-crosscap-states}}

In this section we generalize the algebraic framework of the boundary
states \cite{Gombor:2021hmj} for crosscap states.

\subsection{Untwisted case}

At first we divide the quantum space as $\mathcal{H}=\mathcal{H}^{(1)}\otimes\mathcal{H}^{(2)}$.
Let us define the following KT-relation

\begin{equation}
K(u)\langle\mathcal{C}|T^{(1)}(u)=\langle\mathcal{C}|T^{(2)}(-u)K(u),\label{eq:utwKT}
\end{equation}
where $\langle\mathcal{C}|\in\mathcal{H}^{*}$ is the crosscap state
and $K(u)\in GL(N)$ is the $K$-matrix. We can derive a consistency
condition for the $K$-matrix from the following to equations
\begin{multline}
\langle\mathcal{C}|T_{a}^{(1)}(u)T_{b}^{(1)}(v)=R_{a,b}^{-1}(u-v)\langle\mathcal{C}|T_{b}^{(1)}(v)T_{a}^{(1)}(u)R_{a,b}(u-v)=\\
=R_{a,b}^{-1}(u-v)K_{b}^{-1}(v)\langle\mathcal{C}|T_{b}^{(2)}(-v)T_{a}^{(1)}(u)K_{b}(v)R_{a,b}(u-v)=\\
=R_{a,b}^{-1}(u-v)K_{b}^{-1}(v)\langle\mathcal{C}|T_{a}^{(1)}(u)T_{b}^{(2)}(-v)K_{b}(v)R_{a,b}(u-v)=\\
=R_{a,b}^{-1}(u-v)K_{b}^{-1}(v)K_{a}^{-1}(u)\langle\mathcal{C}|T_{a}^{(2)}(-u)T_{b}^{(2)}(-v)K_{a}(u)K_{b}(v)R_{a,b}(u-v),
\end{multline}
and
\begin{multline}
\langle\mathcal{C}|T_{a}^{(1)}(u)T_{b}^{(1)}(v)=K_{a}^{-1}(u)\langle\mathcal{C}|T_{a}^{(2)}(-u)T_{b}^{(1)}(v)K_{a}(u)=\\
=K_{a}^{-1}(u)\langle\mathcal{C}|T_{b}^{(1)}(v)T_{a}^{(2)}(-u)K_{a}(u)=\\
=K_{a}^{-1}(u)K_{b}^{-1}(v)\langle\mathcal{C}|T_{b}^{(2)}(-v)T_{a}^{(2)}(-u)K_{b}(v)K_{a}(u)=\\
=K_{a}^{-1}(u)K_{b}^{-1}(v)R_{a,b}^{-1}(u-v)\langle\mathcal{C}|T_{a}^{(2)}(-u)T_{b}^{(2)}(-v)R_{a,b}(u-v)K_{b}(v)K_{a}(u).
\end{multline}
We can see that the consistency condition is
\begin{equation}
K_{a}(u)K_{b}(v)R_{a,b}(u-v)=R_{a,b}(u-v)K_{a}(u)K_{b}(v).
\end{equation}
Substituting the explicit form of the $R$-matrix we obtain that
\begin{equation}
K_{a}(u)K_{b}(v)=K_{a}(v)K_{b}(u),
\end{equation}
which has the following solution (up to a spectral parameter dependent
normalization)
\begin{equation}
K(u)=K,
\end{equation}
where $K$ is a spectral parameter independent invertible $N\times N$
matrix.

The crosscap states also have co-product property. Let us assume that
crosscap states exist on the spaces $\mathcal{H}^{(a)}=\mathcal{H}^{(1,a)}\otimes\mathcal{H}^{(2,a)}$
and $\mathcal{H}^{(b)}=\mathcal{H}^{(1,b)}\otimes\mathcal{H}^{(2,b)}$
with the same $K$-matrix i.e. 
\begin{equation}
K\langle\mathcal{C}^{(a)}|T^{(1,a)}(u)=\langle\mathcal{C}^{(a)}|T^{(2,a)}(-u)K,\qquad K\langle\mathcal{C}^{(b)}|T^{(1,b)}(u)=\langle\mathcal{C}^{(b)}|T^{(2,b)}(-u)K.
\end{equation}
On the product spaces $\mathcal{H}^{(1)}=\mathcal{H}^{(1,a)}\otimes\mathcal{H}^{(1,b)}$
and $\mathcal{H}^{(2)}=\mathcal{H}^{(2,a)}\otimes\mathcal{H}^{(2,b)}$
the monodromy matrices are
\begin{equation}
T^{(1)}(u)=T^{(1,b)}(u)T^{(1,a)}(u),\qquad T^{(2)}(u)=T^{(2,b)}(u)T^{(2,a)}(u).
\end{equation}
Let us consider the following expression
\begin{multline}
K\langle\mathcal{C}^{(a)}|\langle\mathcal{C}^{(b)}|T^{(1)}(u)=K\langle\mathcal{C}^{(b)}|T^{(1,b)}(u)\langle\mathcal{C}^{(a)}|T^{(1,a)}(u)=\langle\mathcal{C}^{(b)}|T^{(2,b)}(-u)K\langle\mathcal{C}^{(a)}|T^{(1,a)}(u)=\\
=\langle\mathcal{C}^{(b)}|T^{(2,b)}(-u)\langle\mathcal{C}^{(a)}|T^{(2,a)}(-u)K=\langle\mathcal{C}^{(a)}|\langle\mathcal{C}^{(b)}|T^{(2)}(-u)K,
\end{multline}
therefore on the product space $\mathcal{H}=\mathcal{H}^{(1)}\otimes\mathcal{H}^{(2)}=\mathcal{H}^{(1,a)}\otimes\mathcal{H}^{(1,b)}\otimes\mathcal{H}^{(2,a)}\otimes\mathcal{H}^{(2,b)}$
the state $\langle\mathcal{C}^{(a)}|\langle\mathcal{C}^{(b)}|$ is
a crosscap state with $K$-matrix $K$. 

Using this co-product property we can define crosscap states for the
general monodromy matrices (\ref{eq:genmod}) from the solution for
the elementary representations $L^{\Lambda}(u-\xi)$. Let us substitute
$T^{(1)}(u)=L^{\Lambda^{(1)}}(u-\xi_{1})$ and $T^{(2)}(u)=L^{\Lambda^{(2)}}(u-\xi_{2})$
to the KT-relation (\ref{eq:utwKT})
\begin{equation}
K_{0}\langle c|L_{0,1}^{\Lambda^{(1)}}(u-\xi_{1})=\langle c|L_{0,2}^{\Lambda^{(2)}}(-u-\xi_{2})K_{0}.
\end{equation}
Let us use the following parametrization
\begin{equation}
K=\sum_{i,j}K_{i,j}E_{i,j},\qquad\langle c|=\sum_{a,b}\Psi_{b,a}e_{a}^{\Lambda^{(1)}}\otimes e_{b}^{\Lambda^{(2)}},\qquad L^{\Lambda}(u)=\sum_{i,j,a,b}L^{\Lambda}(u)_{i,a}^{j,b}E_{i,j}\otimes E_{a,b}^{\Lambda},
\end{equation}
where $e_{i}^{\Lambda}$ are the canonical basis of the dual space
of $\mathcal{V}^{\Lambda}$. After the substitution we obtain that
\begin{equation}
K_{i,j}\Psi_{a,b}L^{\Lambda^{(1)}}(u-\xi_{1})_{j,c}^{k,b}=L^{\Lambda^{(2)}}(-u-\xi_{2})_{i,a}^{j,b}\Psi_{b,c}K_{j,k}.
\end{equation}
This can be rewritten as
\begin{equation}
K_{1}\Psi_{2}L_{1,2}^{\Lambda^{(1)}}(u-\xi_{1})^{t_{2}}=L_{1,2}^{\Lambda^{(2)}}(-u-\xi_{2})\Psi_{2}K_{1}.
\end{equation}
Using the explicit form of the Lax-operators
\begin{equation}
\frac{c}{u-\xi_{1}}\sum_{i,j}KE_{i,j}\otimes\Psi\left(E_{j,i}^{\Lambda^{(1)}}\right)^{t}=-\frac{c}{u+\xi_{2}}\sum_{i,j}E_{i,j}K\otimes E_{j,i}^{\Lambda^{(2)}}\Psi,
\end{equation}
or equivalently
\begin{equation}
\sum_{i,j}E_{i,j}\otimes E_{j,i}^{\Lambda^{(2)}}=\frac{u+\xi_{2}}{u-\xi_{1}}\sum_{i,j}KE_{i,j}K^{-1}\otimes\Psi\left(-E_{j,i}^{\Lambda^{(1)}}\right)^{t}\Psi^{-1}.
\end{equation}
This equation requires that $\xi_{2}=-\xi_{1}$ and the representation
$\Lambda^{(2)}$ is similar to the contra-gradient representation
of $\Lambda^{(1)}=\Lambda$, i.e. $\Lambda^{(2)}=\bar{\Lambda}$.
Substituting back
\begin{equation}
E_{i,j}^{\Lambda^{(1)}}=E_{i,j}^{\Lambda},\qquad E_{i,j}^{\Lambda^{(2)}}=E_{i,j}^{\bar{\Lambda}}=-V^{\Lambda}E_{j,i}^{\Lambda}V^{\Lambda}=-E_{N+1-j,N+1-i}^{\Lambda},
\end{equation}
therefore we just obtained that
\begin{equation}
\sum_{i,j}E_{i,j}\otimes E_{j,i}^{\Lambda}=\sum_{i,j}K^{-1}E_{i,j}K\otimes\Psi^{-1}E_{N+1-j,N+1-i}^{\Lambda}\Psi=\sum_{i,j}(VK)^{-1}E_{i,j}VK\otimes\Psi^{-1}E_{j,i}^{\Lambda}\Psi.
\end{equation}
It means that the matrix $\Psi=\psi^{\Lambda}$ is the image of the
matrix $\psi=VK$ in the representation $\Lambda$. 

Using this solution, we just obtained a general solution of the $KT$-equation
with the monodromy matrices
\begin{align}
T_{0}^{(1)}(u) & =L_{0,J/2}^{\Lambda^{(J/2)}}(u-\xi_{J/2})\dots L_{0,1}^{\Lambda^{(1)}}(u-\xi_{1}),\label{eq:utwcond1}\\
T_{0}^{(2)}(u) & =L_{0,J}^{\bar{\Lambda}^{(J)}}(u+\xi_{J})\dots L_{0,J/2+1}^{\bar{\Lambda}^{(1)}}(u+\xi_{1}),\label{eq:utwcond2}
\end{align}
and crosscap states 
\begin{equation}
\langle\mathcal{C}|=\prod_{j=1}^{J/2}\langle c|_{j},\qquad\langle c|_{j}=\sum_{a,b}\psi_{b,a}^{\Lambda^{(j)}}\langle a|_{j}\langle b|_{j+\frac{J}{2}}.\label{eq:untwcrosscap}
\end{equation}

Now let us turn on the consequences for the transfer matrices. At
first let us calculate that
\begin{multline}
\langle\mathcal{C}|T_{a}^{(2)}(u)T_{b}^{(1)}(u)=K_{a}\langle\mathcal{C}|T_{a}^{(1)}(-u)T_{b}^{(1)}(u)K_{a}^{-1}=K_{a}R_{ab}(2u)\langle\mathcal{C}|T_{b}^{(1)}(u)T_{a}^{(1)}(-u)R_{ab}^{-1}(2u)K_{a}^{-1}=\\
K_{a}R_{ab}(2u)K_{b}^{-1}\langle\mathcal{C}|T_{b}^{(2)}(-u)T_{a}^{(1)}(-u)K_{b}R_{ab}^{-1}(2u)K_{a}^{-1}.
\end{multline}
Since
\begin{equation}
\mathcal{T}(u)=\mathrm{tr_{a,b}}\left[T_{a}^{(2)}(u)T_{b}^{(1)}(u)P_{a,b}\right],
\end{equation}
we can obtain that
\begin{align}
\langle\mathcal{C}|\mathcal{T}(u)= & \mathrm{tr_{a,b}}\left[K_{a}R_{ab}(2u)K_{b}^{-1}\langle\mathcal{C}|T_{b}^{(2)}(-u)T_{a}^{(1)}(-u)K_{b}R_{ab}^{-1}(2u)K_{a}^{-1}P_{a,b}\right]=\nonumber \\
 & \mathrm{tr_{a,b}}\left[\langle\mathcal{C}|T_{b}^{(2)}(-u)T_{a}^{(1)}(-u)K_{b}R_{ab}^{-1}(2u)K_{a}^{-1}K_{b}R_{ab}(2u)K_{a}^{-1}P_{a,b}\right]=\nonumber \\
 & \mathrm{tr_{a,b}}\left[\langle\mathcal{C}|T_{b}^{(2)}(-u)T_{a}^{(1)}(-u)K_{a}^{-1}R_{ab}^{-1}(2u)K_{b}^{2}R_{ab}(2u)K_{a}^{-1}P_{a,b}\right],
\end{align}
where we used the $GL(N)$ symmetry of the R-matrix
\begin{equation}
K_{a}K_{b}R_{ab}(u)=R_{ab}(u)K_{a}K_{b}.
\end{equation}
Assuming that
\begin{equation}
K^{2}=1,\label{eq:K2utw}
\end{equation}
we obtain the untwisted integrability condition
\begin{equation}
\langle\mathcal{C}|\mathcal{T}(u)=\langle\mathcal{C}|\mathcal{T}(-u).\label{eq:intcondUtw}
\end{equation}
The condition (\ref{eq:K2utw}) means that the $K$-matrix has eigenvalues
$\pm1$ and let $\#(+1)=M$ and $\#(-1)=N-M$. For such $K$-matrices
the residual symmetry is $\mathfrak{gl}(M)\otimes\mathfrak{gl}(N-M)$. 

In \cite{Gombor:2021hmj} it was shown that the integrability condition
(\ref{eq:intcondUtw}) leads to achiral pair structure of Bethe roots
of the on-shell Bethe states with non-vanishing overlaps i.e. $\bar{t}^{\nu}=-\bar{t}^{N-\nu}$
for $\nu<\frac{N}{2}$ and $\bar{t}^{\frac{N}{2}}=\bar{t}^{+,\frac{N}{2}}\cup\bar{t}^{-,\frac{N}{2}}$where
$\bar{t}^{+,\frac{N}{2}}=-\bar{t}^{-,\frac{N}{2}}$ for even $N$.

\subsection{Twisted case\label{subsec:Twisted-case-1}}

Let us define the following twisted KT-relations

\begin{equation}
K(u)\langle\mathcal{C}|T^{(1)}(u)=\lambda_{0}(u)\langle\mathcal{C}|\widehat{T}^{(2)}(-u)K(u),
\end{equation}
where $\langle\mathcal{C}|$ is the twisted crosscap state and $\lambda_{0}(u)$
is spectral parameter dependent function for a proper normalization.
Using (\ref{eq:invemon}) we have an equivalent form
\begin{equation}
\bar{K}(u)\langle\mathcal{C}|T^{(2)}(u)=\lambda_{0}(-u)\langle\mathcal{C}|\widehat{T}^{(1)}(-u)\bar{K}(u),
\end{equation}
where $\bar{K}(u)=VK^{t}(-u)V$.

We can repeat the previous analysis for this twisted equation. For
the consistency condition we obtain that
\begin{equation}
K(u)=K,
\end{equation}
where $K$ is a spectral parameter independent invertible $N\times N$
matrix.

The twisted crosscap states also has a co-product property. Let us
assume that twisted crosscap states exist on the spaces $\mathcal{H}^{(a)}=\mathcal{H}^{(1,a)}\otimes\mathcal{H}^{(2,a)}$
and $\mathcal{H}^{(b)}=\mathcal{H}^{(1,b)}\otimes\mathcal{H}^{(2,b)}$
with the same $K$-matrix i.e. 
\begin{equation}
K\langle\mathcal{C}^{(a)}|T^{(1,a)}(u)=\lambda_{0}^{a}(u)\langle\mathcal{C}^{(a)}|\widehat{T}^{(2,a)}(-u)K,\qquad K\langle\mathcal{C}^{(b)}|T^{(1,b)}(u)=\lambda_{0}^{b}(u)\langle\mathcal{C}^{(b)}|\widehat{T}^{(2,b)}(-u)K.
\end{equation}
One can show that, on the product space $\mathcal{H}=\mathcal{H}^{(1)}\otimes\mathcal{H}^{(2)}=\mathcal{H}^{(1,a)}\otimes\mathcal{H}^{(1,b)}\otimes\mathcal{H}^{(2,a)}\otimes\mathcal{H}^{(2,b)}$,
the state $\langle\mathcal{C}^{(a)}|\langle\mathcal{C}^{(b)}|$ is
a crosscap state with $K$-matrix $K$. 

Using this co-product property we can define crosscap states for the
general monodromy matrices (\ref{eq:genmod}) from the solutions for
the elementary representations $L^{\Lambda}(u-\xi)$. Let us substitute
to the KT-relation
\begin{equation}
K_{0}\langle c|L_{0,1}^{\Lambda^{(1)}}(u-\xi_{1})=\lambda_{0}(u)\langle c|\widehat{L}_{0,2}^{\Lambda^{(2)}}(-u-\xi_{2})K_{0}.
\end{equation}
We know that the crossed Lax operator has the form (\ref{eq:crossedLax})
\begin{equation}
\widehat{L}_{0,1}^{\Lambda}(u)=V_{0}\left(\left(L_{0,1}^{\Lambda}(u)\right)^{-1}\right)^{t_{0}}V_{0}.
\end{equation}
Let us concentrate on the representations with rectangular Young tableaux
i.e let us assume that $\Lambda_{i}^{(2)}=s$ for $i\leq a$ and $\Lambda_{i}=0$
for $i>a$. For these representations, the Lax operators have inversion
property
\begin{equation}
\left(L_{0,1}^{\Lambda^{(2)}}(u)\right)^{-1}=\frac{u}{u+sc}\frac{u+sc-ac}{u-ac}L_{0,1}^{\Lambda^{(2)}}(-u-sc+ac).
\end{equation}
After the substitution we just obtain that
\begin{equation}
V_{0}K_{0}\langle c|L_{0,1}^{\Lambda^{(1)}}(u-\xi_{1})=\langle c|L_{0,1}^{\Lambda^{(2)}}(u+\xi_{2}-sc+ac)^{t_{0}}V_{0}K_{0},
\end{equation}
where we fixed the normalization as
\begin{equation}
\frac{1}{\lambda_{0}(u)}=\frac{-u-\xi_{2}}{-u-\xi_{2}+sc}\frac{-u-\xi_{2}+sc-ac}{-u-\xi_{2}-ac}.
\end{equation}
 Let us use the following parametrization
\begin{equation}
VK=\sum_{i,j}\psi_{i,j}^{-1}E_{i,j},\qquad\langle c|=\sum_{a,b}\Psi_{a,b}e_{a}^{\Lambda^{(1)}}\otimes e_{b}^{\Lambda^{(2)}},\qquad L^{\Lambda}(u)=\sum_{i,j,a,b}L(u)_{i,a}^{j,b}E_{i,j}\otimes E_{a,b}^{\Lambda},
\end{equation}
where $e_{i}^{\Lambda}$ are the canonical basis of the dual space
of $\mathcal{V}^{\Lambda}$. After the substitution we obtain that
\begin{equation}
\psi_{i,j}^{-1}\Psi_{b,c}L^{\Lambda^{(1)}}(u-\xi_{1})_{j,a}^{k,b}=L^{\Lambda^{(2)}}(u+\xi_{2}-sc+ac)_{j,c}^{i,b}\Psi_{a,b}\psi_{j,k}^{-1}=L^{\Lambda^{(2)}}(u+\xi_{2}-sc+ac)_{i,b}^{j,c}\Psi_{a,b}\psi_{j,k}^{-1}.
\end{equation}
This can be rewritten as
\begin{equation}
\psi_{1}^{-1}L_{1,2}^{\Lambda^{(1)}}(u-\xi_{1})\Psi_{2}=\Psi_{2}L_{1,2}^{\Lambda^{(2)}}(u+\xi_{2}-sc+ac)\psi_{1}^{-1}.
\end{equation}
This equation requires that $\xi_{2}=-\xi_{1}+sc-ac$ and the representation
$\Lambda^{(2)}$ is similar to the representation $\Lambda^{(1)}$
i.e. $\Lambda^{(1)}=\Lambda^{(2)}=\Lambda$. The matrix $\Psi=\psi^{\Lambda}$
is the image of the matrix $\psi=(VK)^{-1}$ in the representation
$\Lambda$. 

Using this solution, we just obtained the general solutions of the
twisted $KT$-equation with the monodromy matrices
\begin{align}
T_{0}^{(1)}(u) & =L_{0,J/2}^{\Lambda^{(J/2)}}(u-\xi_{J/2})\dots L_{0,1}^{\Lambda^{(1)}}(u-\xi_{1}),\label{eq:twcond1}\\
T_{0}^{(2)}(u) & =L_{0,J}^{\Lambda^{(J/2)}}(u+\xi_{J}-sc+ac)\dots L_{0,J/2+1}^{\Lambda^{(1)}}(u+\xi_{1}-sc+ac),\label{eq:twcond2}
\end{align}
and crosscap states 
\begin{equation}
\langle\mathcal{C}|=\prod_{j=1}^{J/2}\langle c|_{j},\qquad\langle c|_{j}=\sum_{a,b}\psi_{a,b}^{\Lambda^{(j)}}\langle a|_{j}\langle b|_{j+\frac{J}{2}},\label{eq:twcrosscap}
\end{equation}
where $\Lambda_{i}^{(j)}=s_{j}$ for $i\leq a_{j}$ and $\Lambda_{i}^{(j)}=0$
for $i>a_{j}$.

Now let us turn on the consequences for the transfer matrices. At
first let us calculate
\begin{multline}
\langle\mathcal{C}|T_{a}^{(2)}(u)T_{b}^{(1)}(u)=\lambda_{0}(-u)\bar{K}_{a}^{-1}\langle\mathcal{C}|\widehat{T}_{a}^{(1)}(-u)T_{b}^{(1)}(u)\bar{K}_{a}=\\
=\lambda_{0}(-u)\bar{K}_{a}^{-1}\bar{R}_{ab}^{-1}(-2u)\langle\mathcal{C}|T_{b}^{(1)}(u)\widehat{T}_{a}^{(1)}(-u)\bar{R}_{ab}(-2u)\bar{K}_{a}=\\
\lambda_{0}(u)\lambda_{0}(-u)\bar{K}_{a}^{-1}\bar{R}_{ab}^{-1}(-2u)K_{b}^{-1}\langle\mathcal{C}|\widehat{T}_{b}^{(2)}(-u)\widehat{T}_{a}^{(1)}(-u)K_{b}\bar{R}_{ab}(-2u)\bar{K}_{a}.
\end{multline}
Since
\begin{equation}
\mathcal{T}(u)=\mathrm{tr_{a,b}}\left[T_{a}^{(2)}(u)T_{b}^{(1)}(u)P_{a,b}\right],
\end{equation}
we can obtain that
\begin{align}
\langle\mathcal{C}|\mathcal{T}(u)=\lambda_{0}(u)\lambda_{0}(-u) & \mathrm{tr_{a,b}}\left[\bar{K}_{a}^{-1}\bar{R}_{ab}^{-1}(-2u)K_{b}^{-1}\langle\mathcal{C}|\widehat{T}_{b}^{(2)}(-u)\widehat{T}_{a}^{(1)}(-u)K_{b}\bar{R}_{ab}(-2u)\bar{K}_{a}P_{a,b}\right]=\nonumber \\
\lambda_{0}(u)\lambda_{0}(-u) & \mathrm{tr_{a,b}}\left[\langle\mathcal{C}|T_{b}^{(2)}(-u)T_{a}^{(1)}(-u)K_{b}\bar{R}_{ab}(-2u)\bar{K}_{a}\bar{K}_{b}^{-1}\bar{R}_{ab}^{-1}(-2u)K_{a}^{-1}P_{a,b}\right]\nonumber \\
\lambda_{0}(u)\lambda_{0}(-u) & \mathrm{tr_{a,b}}\left[\langle\mathcal{C}|T_{b}^{(2)}(-u)T_{a}^{(1)}(-u)\bar{K}_{a}\bar{R}_{ab}(-2u)K_{b}\bar{K}_{b}^{-1}\bar{R}_{ab}^{-1}(-2u)K_{a}^{-1}P_{a,b}\right],
\end{align}
where we used the $GL(N)$ symmetry of the crossed R-matrix
\begin{equation}
K_{a}\bar{R}_{ab}(u)V_{b}K_{b}^{t}V_{b}=V_{b}K_{b}^{t}V_{b}\bar{R}_{ab}(u)K_{a},
\end{equation}
and the fact that $\bar{K}=VK^{t}V$. Assuming that
\begin{equation}
K=\pm\bar{K},\label{eq:condKtw}
\end{equation}
we obtain the twisted integrability condition
\begin{equation}
\langle\mathcal{C}|\mathcal{T}(u)=\lambda_{0}(u)\lambda_{0}(-u)\langle\mathcal{C}|\widehat{\mathcal{T}}(-u).\label{eq:intconTw}
\end{equation}
The condition (\ref{eq:condKtw}) is equivalent to
\begin{equation}
VK^{t}V=\pm K,
\end{equation}
therefore we have two classes. For the positive and negative signs
the residual symmetries are $\mathfrak{so}(N)$ and $\mathfrak{sp}(N)$,
respectively. 

In \cite{Gombor:2021hmj} it was shown that the integrability condition
(\ref{eq:intconTw}) leads to chiral pair structure of Bethe roots
of the on-shell Bethe states with non-vanishing overlaps i.e. $\bar{t}^{\nu}=\bar{t}^{+,\nu}\cup\bar{t}^{-,\nu}$where
$\bar{t}^{+,\nu}=-\bar{t}^{-,\nu}-\nu c$.

\section{Application to AdS/CFT\label{sec:Application-to-AdS/CFT}}

In this section we collect the integrable crosscap states which can
be relevant for the AdS/CFT correspondence. We would like to point
out that this paper is not intended to provide an exact holographic
description where these crosscap states appears. We only collect those
crosscap states of the homogeneous $SO(6)$ and the alternating $SU(4)$
spin chains that preserve integrability. These states could be relevant
for the $\mathcal{N}=4$ SYM and ABJM theories since these spin chains
describe the scalar sectors of these field theories at weak coupling.

\subsection{Scalar sector of the $\mathcal{N}=4$ SYM}

For the $SO(6)$ spin chain we have to use the result of the $\mathfrak{gl}(4)$
case. In this situation there are 4 classes of crosscap states.

\subsubsection{Twisted case}

We saw that there are two classes in the twisted case: the $\mathfrak{so}(4)=\mathfrak{so}(3)\oplus\mathfrak{so}(3)$
and the $\mathfrak{sp}(4)=\mathfrak{so}(5)$ case.

The $\mathfrak{so}(3)\oplus\mathfrak{so}(3)$ symmetric crosscap state
in the real basis is built from the vector
\begin{equation}
\begin{split}\langle c|_{j}= & +\langle1|_{j}\langle1|_{j+\frac{J}{2}}+\langle2|_{j}\langle2|_{j+\frac{J}{2}}+\langle3|_{j}\langle3|_{j+\frac{J}{2}}\\
 & -\langle4|_{j}\langle4|_{j+\frac{J}{2}}-\langle5|_{j}\langle5|_{j+\frac{J}{2}}-\langle6|_{j}\langle6|_{j+\frac{J}{2}}.
\end{split}
\label{eq:so3so3}
\end{equation}

The $\mathfrak{so}(5)$ symmetric crosscap state in the real basis
is built from the vector
\begin{equation}
\begin{split}\langle c|_{j}= & +\langle1|_{j}\langle1|_{j+\frac{J}{2}}-\langle2|_{j}\langle2|_{j+\frac{J}{2}}-\langle3|_{j}\langle3|_{j+\frac{J}{2}}\\
 & -\langle4|_{j}\langle4|_{j+\frac{J}{2}}-\langle5|_{j}\langle5|_{j+\frac{J}{2}}-\langle6|_{j}\langle6|_{j+\frac{J}{2}}.
\end{split}
\end{equation}
The global rotations are also integrable crosscap states. \emph{For
these integrable states the pair structure is chiral.} 

\subsubsection{Untwisted case}

We saw that there are two classes in the untwisted case: the $\mathfrak{gl}(4)$,
$\mathfrak{gl}(2)\oplus\mathfrak{gl}(2)$ and the $\mathfrak{gl}(3)$
cases. These symmetries in the $SO(6)$ language are $\mathfrak{so}(6)$,
$\mathfrak{so}(2)\oplus\mathfrak{so}(4)$ and the $\mathfrak{gl}(3)$,
respectively.

The $\mathfrak{so}(6)$ symmetric crosscap state in the real basis
is built from the vector

\begin{equation}
\begin{split}\langle c|_{j}= & +\langle1|_{j}\langle1|_{j+\frac{J}{2}}+\langle2|_{j}\langle2|_{j+\frac{J}{2}}+\langle3|_{j}\langle3|_{j+\frac{J}{2}}\\
 & +\langle4|_{j}\langle4|_{j+\frac{J}{2}}+\langle5|_{j}\langle5|_{j+\frac{J}{2}}+\langle6|_{j}\langle6|_{j+\frac{J}{2}}.
\end{split}
\end{equation}

The $\mathfrak{so}(2)\oplus\mathfrak{so}(4)$ symmetric crosscap state
in the real basis is built from the vector
\begin{equation}
\begin{split}\langle c|_{j}= & +\langle1|_{j}\langle1|_{j+\frac{J}{2}}+\langle2|_{j}\langle2|_{j+\frac{J}{2}}-\langle3|_{j}\langle3|_{j+\frac{J}{2}}\\
 & -\langle4|_{j}\langle4|_{j+\frac{J}{2}}-\langle5|_{j}\langle5|_{j+\frac{J}{2}}-\langle6|_{j}\langle6|_{j+\frac{J}{2}}.
\end{split}
\label{eq:so2so4}
\end{equation}

The $\mathfrak{gl}(3)$ symmetric crosscap state in the real basis
is built from the vector
\begin{equation}
\begin{split}\langle c|_{j}= & +\langle1|_{j}\langle4|_{j+\frac{J}{2}}+\langle2|_{j}\langle5|_{j+\frac{J}{2}}+\langle3|_{j}\langle6|_{j+\frac{J}{2}}\\
 & -\langle4|_{j}\langle1|_{j+\frac{J}{2}}-\langle5|_{j}\langle2|_{j+\frac{J}{2}}-\langle6|_{j}\langle3|_{j+\frac{J}{2}}.
\end{split}
\label{eq:gl3}
\end{equation}
The global rotations are also integrable crosscap states. \emph{For
these integrable states the pair structure is achiral.} 

\subsection{Scalar sector of the ABJM}

For the ABJM we have to use the result of the $\mathfrak{gl}(4)$
case when representations are alternating: $\Lambda^{(2i-1)}=(1,0,0,0)$
and $\Lambda^{(2i)}=(0,0,0,-1)$. In this convention the monodromy
matrix has the form
\begin{equation}
T_{0}(u)=\bar{L}_{0,J}(u+c)L_{0,J-1}(u-c)\dots\bar{L}_{0,2}(u+c)L_{0,1}(u-c),\label{eq:conv1}
\end{equation}
where we also used the conventions $\xi_{2j-1}=c$, $\xi_{2j}=-c$
for the inhomogeneities and $L(u)=L^{(1,0,0,0)}(u)$, $\bar{L}(u)=L^{(0,0,0,-1)}(u)$
see the definition (\ref{eq:Lax}). The vacuum eigenvalues are 
\begin{equation}
\lambda_{1}(u)=\left(\frac{u}{u-c}\right)^{\frac{J}{2}},\quad\lambda_{2}(u)=\lambda_{3}(u)=1,\quad\lambda_{4}(u)=\left(\frac{u}{u+c}\right)^{\frac{J}{2}},
\end{equation}
therefore the Bethe equations are
\begin{align}
\left(\frac{t_{j}^{1}}{t_{j}^{1}-c}\right)^{L} & =-\prod_{k=1}^{r_{1}}\frac{t_{j}^{1}-t_{k}^{1}+c}{t_{j}^{1}-t_{k}^{1}-c}\prod_{k=1}^{r_{2}}\frac{t_{j}^{1}-t_{k}^{2}-c}{t_{j}^{1}-t_{k}^{2}},\\
1 & =-\prod_{k=1}^{r_{2}}\frac{t_{j}^{2}-t_{k}^{2}+c}{t_{j}^{2}-t_{k}^{2}-c}\prod_{k=1}^{r_{1}}\frac{t_{j}^{2}-t_{k}^{1}}{t_{j}^{2}-t_{k}^{1}+c}\prod_{k=1}^{r_{3}}\frac{t_{j}^{2}-t_{k}^{3}-c}{t_{j}^{2}-t_{k}^{3}},\\
\left(\frac{t_{j}^{3}+c}{t_{j}^{3}}\right)^{L} & =-\prod_{k=1}^{r_{3}}\frac{t_{j}^{3}-t_{k}^{3}+c}{t_{j}^{3}-t_{k}^{3}-c}\prod_{k=1}^{r_{2}}\frac{t_{j}^{3}-t_{k}^{2}}{t_{j}^{3}-t_{k}^{2}+c}.
\end{align}
Re-defining the Bethe roots as $t_{j}^{1}=u_{j}+c/2$, $t_{j}^{2}=w_{j}$,
$t_{j}^{3}=v_{j}-c/2$ ($r_{1}=K_{u}$,$r_{2}=K_{w}$,$r_{3}=K_{v}$)
and using the convention $c=i$, the Bethe equations read as
\begin{align}
\left(\frac{u_{j}+i/2}{u_{j}-i/2}\right)^{L} & =-\prod_{k=1}^{K_{u}}\frac{u_{j}-u_{k}+i}{u_{j}-u_{k}-i}\prod_{k=1}^{K_{w}}\frac{u_{j}-w_{k}-i/2}{u_{j}-w_{k}+i/2},\\
1 & =-\prod_{k=1}^{K_{w}}\frac{w_{j}-w_{k}+i}{w_{j}-w_{k}-i}\prod_{k=1}^{K_{u}}\frac{w_{j}-u_{k}-i/2}{w_{j}-u_{k}+i/2}\prod_{k=1}^{K_{v}}\frac{w_{j}-v_{k}-i/2}{w_{j}-v_{k}+i/2},\\
\left(\frac{v_{j}+i/2}{v_{j}-i/2}\right)^{L} & =-\prod_{k=1}^{K_{v}}\frac{v_{j}-v_{k}+i}{v_{j}-v_{k}-i}\prod_{k=1}^{K_{w}}\frac{v_{j}-w_{k}-i/2}{v_{j}-w_{k}+i/2},
\end{align}
which are the Bethe equations for the scalar sector of the ABJM theory
\cite{Yang:2021hrl}. 

We will see that the convention for the monodromy matrix (\ref{eq:conv1})
is compatible with the untwisted crosscap states but there is an alternative
convention for the monodromy matrix (which will be compatible for
the twisted case) since the $(0,0,0,-1)$ and $(1,1,1,0)$ are equivalent
representations. Now let us define the monodromy matrix in the alternative
way
\begin{equation}
\tilde{T}_{0}(u)=\tilde{L}_{0,J}(u+c)L_{0,J-1}(u)\dots\tilde{L}_{0,2}(u+c)L_{0,1}(u),\label{eq:conv2}
\end{equation}
where $\tilde{L}(u)=L^{(1,1,1,0)}(u)$. Using the definition (\ref{eq:Lax})
we can easily show that
\begin{multline}
\tilde{L}(u)=\mathbf{1}+\frac{c}{u}\sum_{i,j=1}^{4}E_{i,j}\otimes E_{j,i}^{(1,1,1,0)}=\mathbf{1}+\frac{c}{u}\sum_{i,j=1}^{4}E_{i,j}\otimes\left(E_{j,i}^{(0,0,0,-1)}+\delta_{i,j}\right)\\
=\left(\frac{u+c}{u}\right)\mathbf{1}+\frac{c}{u}\sum_{i,j=1}^{4}E_{i,j}\otimes E_{j,i}^{(0,0,0,-1)}=\left(\frac{u+c}{u}\right)\bar{L}(u+c),
\end{multline}
therefore the two conventions for the monodromy matrix are really
equivalent as
\begin{equation}
\tilde{T}_{0}(u)=\left(\frac{u+2c}{u+c}\right)^{\frac{J}{2}}\bar{L}_{0,J}(u+2c)L_{0,J-1}(u)\dots\bar{L}_{0,2}(u+2c)L_{0,1}(u)=\left(\frac{u+2c}{u+c}\right)^{\frac{J}{2}}T_{0}(u+c).
\end{equation}

\subsubsection{Twisted case}

In the twisted case we used the following convention for the representations
$\Lambda_{i}^{(j)}=s_{j}$ for $i\leq a_{j}$ and $\Lambda_{i}^{(j)}=0$
for $i>a_{j}$ (see subsection (\ref{subsec:Twisted-case-1})) therefore
we have to use the second convention (\ref{eq:conv2}) for the monodromy
matrix. We saw that in the integrability requires that $\Lambda^{(j)}=\Lambda^{(j+\frac{J}{2})}$
therefore $J/2$ has to be even. We also saw that $\xi_{j+\frac{J}{2}}=-\xi_{j}+s_{j}c-a_{j}c$
(see (\ref{eq:twcond1}),(\ref{eq:twcond2})). Since $s_{j}=1$,$a_{j}=1$
and $\xi_{j}=0$ for the odd sites and $s_{j}=1,a_{j}=3$ and $\xi_{j}=-2c$
for the even sites the monodromy matrix (\ref{eq:conv2}) is compatible
with the twisted crosscap states for even $J/2$. We saw that there
are two classes in the twisted case: the $\mathfrak{so}(4)$ and the
$\mathfrak{sp}(4)$ case.

The $\mathfrak{so}(4)$ symmetric crosscap state is built from the
vector
\begin{equation}
\begin{split}\langle c|_{2j-1} & =\langle Y_{1}|_{2j-1}\langle Y_{1}|_{2j-1+\frac{J}{2}}+\langle Y_{2}|_{2j-1}\langle Y_{2}|_{2j-1+\frac{J}{2}}+\langle Y_{3}|_{2j-1}\langle Y_{3}|_{2j-1+\frac{J}{2}}+\langle Y_{4}|_{2j-1}\langle Y_{4}|_{2j-1+\frac{J}{2}},\\
\langle c|_{2j} & =\langle Y_{1}^{\dagger}|_{2j}\langle Y_{1}^{\dagger}|_{2j+\frac{J}{2}}+\langle Y_{2}^{\dagger}|_{2j}\langle Y_{2}^{\dagger}|_{2j+\frac{J}{2}}+\langle Y_{3}^{\dagger}|_{2j}\langle Y_{3}^{\dagger}|_{2j+\frac{J}{2}}+\langle Y_{4}^{\dagger}|_{2j}\langle Y_{4}^{\dagger}|_{2j+\frac{J}{2}}.
\end{split}
\label{eq:so4}
\end{equation}

The $\mathfrak{sp}(4)$ symmetric crosscap state is built from the
vector
\begin{equation}
\begin{split}\langle c|_{2j-1} & =\langle Y_{1}|_{2j-1}\langle Y_{3}|_{2j-1+\frac{J}{2}}+\langle Y_{2}|_{2j-1}\langle Y_{4}|_{2j-1+\frac{J}{2}}-\langle Y_{3}|_{2j-1}\langle Y_{1}|_{2j-1+\frac{J}{2}}-\langle Y_{4}|_{2j-1}\langle Y_{2}|_{2j-1+\frac{J}{2}},\\
\langle c|_{2j} & =\langle Y_{1}^{\dagger}|_{2j}\langle Y_{3}^{\dagger}|_{2j+\frac{J}{2}}+\langle Y_{2}^{\dagger}|_{2j}\langle Y_{4}^{\dagger}|_{2j+\frac{J}{2}}-\langle Y_{3}^{\dagger}|_{2j}\langle Y_{1}^{\dagger}|_{2j+\frac{J}{2}}-\langle Y_{4}^{\dagger}|_{2j}\langle Y_{2}^{\dagger}|_{2j+\frac{J}{2}}.
\end{split}
\end{equation}
The global rotations are also integrable crosscap states. \emph{For
these integrable states the pair structure is chiral.}

\subsubsection{Untwisted case}

For the untwisted case we saw that $\Lambda^{(j)}=\bar{\Lambda}^{(j+\frac{J}{2})}$
(see (\ref{eq:utwcond1}),(\ref{eq:utwcond2})). This requires that
$J/2$ has to be odd. We also saw that $\xi_{j}=-\xi_{j+\frac{J}{2}}$
(see (\ref{eq:utwcond1}),(\ref{eq:utwcond2})) and it is compatible
with the first convention for the transfer matrix (\ref{eq:conv1})
where $\xi_{2j-1}=c$ and $\xi_{2j}=-c$. We saw that there are three
classes in the untwisted case: the $\mathfrak{gl}(4)$, $\mathfrak{gl}(3)$
and the $\mathfrak{gl}(2)\oplus\mathfrak{gl}(2)$ symmetric states. 

The $\mathfrak{gl}(4)$ symmetric crosscap state is built from the
vector

\begin{equation}
\begin{split}\langle c|_{2j-1} & =\langle Y_{1}|_{2j-1}\langle Y_{1}^{\dagger}|_{2j-1+\frac{J}{2}}+\langle Y_{2}|_{2j-1}\langle Y_{2}^{\dagger}|_{2j-1+\frac{J}{2}}+\langle Y_{3}|_{2j-1}\langle Y_{3}^{\dagger}|_{2j-1+\frac{J}{2}}+\langle Y_{4}|_{2j-1}\langle Y_{4}^{\dagger}|_{2j-1+\frac{J}{2}},\\
\langle c|_{2j} & =\langle Y_{1}^{\dagger}|_{2j}\langle Y_{1}|_{2j+\frac{J}{2}}+\langle Y_{2}^{\dagger}|_{2j}\langle Y_{2}|_{2j+\frac{J}{2}}+\langle Y_{3}^{\dagger}|_{2j}\langle Y_{3}|_{2j+\frac{J}{2}}+\langle Y_{4}^{\dagger}|_{2j}\langle Y_{4}|_{2j+\frac{J}{2}}.
\end{split}
\end{equation}

The $\mathfrak{gl}(3)$ symmetric crosscap state is built from the
vector

\begin{equation}
\begin{split}\langle c|_{2j-1} & =\langle Y_{1}|_{2j-1}\langle Y_{1}^{\dagger}|_{2j-1+\frac{J}{2}}+\langle Y_{2}|_{2j-1}\langle Y_{2}^{\dagger}|_{2j-1+\frac{J}{2}}+\langle Y_{3}|_{2j-1}\langle Y_{3}^{\dagger}|_{2j-1+\frac{J}{2}}-\langle Y_{4}|_{2j-1}\langle Y_{4}^{\dagger}|_{2j-1+\frac{J}{2}},\\
\langle c|_{2j} & =\langle Y_{1}^{\dagger}|_{2j}\langle Y_{1}|_{2j+\frac{J}{2}}+\langle Y_{2}^{\dagger}|_{2j}\langle Y_{2}|_{2j+\frac{J}{2}}+\langle Y_{3}^{\dagger}|_{2j}\langle Y_{3}|_{2j+\frac{J}{2}}-\langle Y_{4}^{\dagger}|_{2j}\langle Y_{4}|_{2j+\frac{J}{2}}.
\end{split}
\end{equation}

The $\mathfrak{gl}(2)\otimes\mathfrak{gl}(2)$ symmetric crosscap
state is built from the vector
\begin{equation}
\begin{split}\langle c|_{2j-1} & =\langle Y_{1}|_{2j-1}\langle Y_{4}^{\dagger}|_{2j-1+\frac{J}{2}}+\langle Y_{2}|_{2j-1}\langle Y_{3}^{\dagger}|_{2j-1+\frac{J}{2}}+\langle Y_{3}|_{2j-1}\langle Y_{2}^{\dagger}|_{2j-1+\frac{J}{2}}+\langle Y_{4}|_{2j-1}\langle Y_{1}^{\dagger}|_{2j-1+\frac{J}{2}},\\
\langle c|_{2j} & =\langle Y_{4}^{\dagger}|_{2j}\langle Y_{1}|_{2j+\frac{J}{2}}+\langle Y_{3}^{\dagger}|_{2j}\langle Y_{2}|_{2j+\frac{J}{2}}+\langle Y_{2}^{\dagger}|_{2j}\langle Y_{3}|_{2j+\frac{J}{2}}+\langle Y_{1}^{\dagger}|_{2j}\langle Y_{4}|_{2j+\frac{J}{2}}.
\end{split}
\label{eq:gl2gl2}
\end{equation}
The global rotations are also integrable crosscap states. \emph{For
these integrable states the pair structure is achiral}.

\section{Sum formulas for the off-shell overlaps\label{sec:Sum-formulas-for}}

In this section we derive a sum formula for the off-shell overlap
\begin{equation}
\langle\mathcal{C}|\mathbb{B}(\bar{w}).
\end{equation}
The method uses the results of \cite{Hutsalyuk:2017tcx,Liashyk:2018egk}
which are reviewed in the appendices \ref{sec:Off-shell-Bethe-vectors}
and \ref{sec:Sum-formula-for}.

\subsection{Untwisted case}

In this section we focus on the crosscap states which correspond to
the $K$-matrix $K=V$. Using this K-matrix the $KT$-relation reads
as
\begin{equation}
\langle\mathcal{C}|T_{i,j}^{(1)}(u)=\langle\mathcal{C}|T_{N+1-i,N+1-j}^{(2)}(-u).
\end{equation}
From this relation we can obtain that
\begin{equation}
\lambda_{i}^{(1)}(u)=\lambda_{N+1-i}^{(2)}(-u),\Longrightarrow\alpha_{i}^{(1)}(u)=\frac{1}{\alpha_{N-i}^{(2)}(-u)}.
\end{equation}

Using the KT-relation and the recurrence relation for the off-shell
Bethe vectors (\ref{eq:rec1-1}) we can obtain that
\begin{equation}
\langle\mathcal{C}|\left(\mathbb{\tilde{\mathbb{B}}}^{(1)}(\bar{s})\otimes\tilde{\mathbb{B}}^{(2)}(\bar{t})\right)=\tilde{\mathbb{\mathbb{C}}}^{(2)}(\pi^{a}(\bar{s}))\tilde{\mathbb{B}}^{(2)}(\bar{t})=\tilde{S}^{(2)}(\pi^{a}(\bar{s})|\bar{t}),
\end{equation}
where
\begin{equation}
\pi^{a}(\bar{s})=(-\bar{s}^{N-1},-\bar{s}^{N-2},\dots,-\bar{s}^{1}).
\end{equation}
The renormalized Bethe vector $\mathbb{\tilde{\mathbb{B}}}(\bar{w})$
is defined in the appendix \ref{sec:Off-shell-Bethe-vectors} The
derivation can be found in the Appendix \ref{subsec:Untwisted-case}.
Using the co-product formula (\ref{eq:coproduct}) of the Bethe state
we can obtain that
\begin{multline}
\langle\mathcal{C}|\tilde{\mathbb{B}}(\bar{w})=\sum_{\mathrm{part}(\bar{w})}\frac{\prod_{\nu=1}^{N-1}\lambda_{\nu}^{(2)}(\bar{s}^{\nu})\lambda_{\nu+1}^{(1)}(\bar{t}^{\nu})f(\bar{t}^{\nu},\bar{s}^{\nu})}{\prod_{\nu=1}^{N-2}f(\bar{t}^{\nu+1},\bar{s}^{\nu})}\langle\mathcal{C}|\mathbb{\tilde{\mathbb{B}}}^{(1)}(\bar{s})\otimes\tilde{\mathbb{B}}^{(2)}(\bar{t})=\\
=\sum_{\mathrm{part}(\bar{w})}\frac{\prod_{\nu=1}^{N-1}\lambda_{\nu}^{(2)}(\bar{s}^{\nu})\lambda_{\nu+1}^{(1)}(\bar{t}^{\nu})f(\bar{t}^{\nu},\bar{s}^{\nu})}{\prod_{\nu=1}^{N-2}f(\bar{t}^{\nu+1},\bar{s}^{\nu})}\tilde{S}^{(2)}(\pi^{a}(\bar{s})|\bar{t}).
\end{multline}
The sum goes over all the partitions of $\bar{w}^{\nu}=\bar{s}^{\nu}\cup\bar{t}^{\nu}$,
$\bar{t}^{\nu}=\bar{t}_{\textsc{i}}^{\nu}\cup\bar{t}_{\textsc{ii}}^{\nu}$
and $\bar{s}^{\nu}=\bar{s}_{\textsc{i}}^{\nu}\cup\bar{s}_{\textsc{ii}}^{\nu}$
where $\#\bar{t}_{\textsc{i}}^{\nu}=\#\bar{s}_{\textsc{i}}^{N-\nu}$
and $\#\bar{t}_{\textsc{ii}}^{\nu}=\#\bar{s}_{\textsc{ii}}^{N-\nu}$.
Using the sum rule of the off-shell scalar product (\ref{eq:sumrule})
we obtain that
\begin{multline}
\langle\mathcal{C}|\tilde{\mathbb{B}}(\bar{w})=\sum_{\mathrm{part}(\bar{w})}\frac{\prod_{\nu=1}^{N-1}\lambda_{\nu}^{(2)}(\bar{s}^{\nu})\lambda_{\nu+1}^{(1)}(\bar{t}^{\nu})f(\bar{t}^{\nu},\bar{s}^{\nu})}{\prod_{\nu=1}^{N-2}f(\bar{t}^{\nu+1},\bar{s}^{\nu})}\times\\
W(\pi^{a}(\bar{s}_{\textsc{i}}),\pi^{a}(\bar{s}_{\textsc{ii}})|\bar{t}_{\textsc{i}},\bar{t}_{\textsc{ii}})\prod_{k=1}^{N-1}\lambda_{k}^{(2)}(-\bar{s}_{\textsc{i}}^{N-k}))\lambda_{k+1}^{(2)}(-\bar{s}_{\textsc{ii}}^{N-k}))\lambda_{k+1}^{(2)}(\bar{t}_{\textsc{i}}^{k})\lambda_{k}^{(2)}(\bar{t}_{\textsc{ii}}^{k}).
\end{multline}
After some simplification we obtain that
\begin{multline}
\langle\mathcal{C}|\tilde{\mathbb{B}}(\bar{w})=\sum_{\mathrm{part}(\bar{w})}\frac{\prod_{\nu=1}^{N-1}\lambda_{\nu}^{(2)}(\bar{s}^{\nu})\lambda_{\nu+1}^{(1)}(\bar{t}^{\nu})f(\bar{t}^{\nu},\bar{s}^{\nu})}{\prod_{\nu=1}^{N-2}f(\bar{t}^{\nu+1},\bar{s}^{\nu})}\times\\
W(\pi^{a}(\bar{s}_{\textsc{i}}),\pi^{a}(\bar{s}_{\textsc{ii}})|\bar{t}_{\textsc{i}},\bar{t}_{\textsc{ii}})\prod_{k=1}^{N-1}\lambda_{k+1}^{(1)}(\bar{s}_{\textsc{i}}^{k})\lambda_{k}^{(1)}(\bar{s}_{\textsc{ii}}^{k})\lambda_{k+1}^{(2)}(\bar{t}_{\textsc{i}}^{k})\lambda_{k}^{(2)}(\bar{t}_{\textsc{ii}}^{k}).
\end{multline}
After a renormalization we have a sum formula for the off-shell overlap
\begin{multline}
\langle\mathcal{C}|\mathbb{B}(\bar{w})=\sum_{\mathrm{part}(\bar{w})}\frac{\prod_{k=1}^{N-1}f(\bar{t}^{k},\bar{s}^{k})f(\bar{s}_{\textsc{i}}^{k},\bar{s}_{\textsc{ii}}^{k})f(\bar{t}_{\textsc{i}}^{k},\bar{t}_{\textsc{ii}}^{k})}{\prod_{k=1}^{N-2}f(\bar{t}^{k+1},\bar{s}^{k})f(\bar{s}_{\textsc{i}}^{k+1},\bar{s}_{\textsc{ii}}^{k})f(\bar{t}_{\textsc{i}}^{k+1},\bar{t}_{\textsc{ii}}^{k})}\times\\
Z(\pi^{a}(\bar{s}_{\textsc{i}})|\bar{t}_{\textsc{i}})Z(\bar{t}_{\textsc{ii}}|\pi^{a}(\bar{s}_{\textsc{ii}}))\prod_{k=1}^{N-1}\alpha_{k}^{(2)}(\bar{s}_{\textsc{i}}^{k})\alpha_{k}(\bar{s}_{\textsc{ii}}^{k})\alpha_{k}^{(2)}(\bar{t}_{\textsc{ii}}^{k}),\label{eq:utwSum}
\end{multline}
where we used the identity (\ref{eq:WZZ}).

\subsection{Twisted case}

In this section we focus on the crosscap states which correspond to
the $K$-matrix $K=V$. Using this K-matrix the twisted $KT$-relations
read as
\begin{equation}
\langle\mathcal{C}|T_{i,j}^{(1)}(u)=\lambda_{0}(u)\langle\mathcal{C}|\widehat{T}_{N+1-i,N+1-j}^{(2)}(-u),\qquad\langle\mathcal{C}|T_{i,j}^{(2)}(u)=\lambda_{0}(-u)\langle\mathcal{C}|\widehat{T}_{N+1-i,N+1-j}^{(1)}(-u).
\end{equation}
From these relations we can obtain that
\begin{equation}
\lambda_{i}^{(1)}(u)=\lambda_{0}(u)\hat{\lambda}_{N+1-i}^{(2)}(-u),\qquad\lambda_{i}^{(2)}(u)=\lambda_{0}(-u)\hat{\lambda}_{N+1-i}^{(1)}(-u),\label{eq:idLam}
\end{equation}
therefore
\[
\alpha_{i}^{(1)}(u)=\frac{1}{\alpha_{i}^{(2)}(-u-ic)}.
\]

Using the recursion formula of the off-shell Bethe state (\ref{eq:rec1-1})
and the KT-relation we can obtain that
\begin{equation}
\langle\mathcal{C}|\left(\tilde{\hat{\mathbb{B}}}^{(1)}(\bar{s})\otimes\tilde{\mathbb{B}}^{(2)}(\bar{t})\right)=\frac{1}{\prod_{\nu=1}^{N-1}\lambda_{0}(\bar{s}^{\nu})}\tilde{\mathbb{\mathbb{C}}}^{(2)}(\pi^{a}(\bar{s}))\tilde{\mathbb{B}}^{(2)}(\bar{t})=\frac{1}{\prod_{\nu=1}^{N-1}\lambda_{0}(\bar{s}^{\nu})}\tilde{S}^{(2)}(\pi^{a}(\bar{s})|\bar{t}).
\end{equation}
The derivation can be found in the Appendix \ref{subsec:Twisted-case}.
Using the co-product formula of the off-shell Bethe state (\ref{eq:coproduct})
and the identity (\ref{eq:connBhatB-1}) we can obtain that
\begin{multline}
\langle\mathcal{C}|\tilde{\mathbb{B}}(\bar{w})=\sum_{\mathrm{part}(\bar{w})}\frac{\prod_{\nu=1}^{N-1}\lambda_{\nu}^{(2)}(\bar{s}^{\nu})\lambda_{\nu+1}^{(1)}(\bar{t}^{\nu})f(\bar{t}^{\nu},\bar{s}^{\nu})}{\prod_{\nu=1}^{N-2}f(\bar{t}^{\nu+1},\bar{s}^{\nu})}\langle\mathcal{C}|\mathbb{\tilde{\mathbb{B}}}^{(1)}(\bar{s})\otimes\tilde{\mathbb{B}}^{(2)}(\bar{t})=\\
=\sum_{\mathrm{part}(\bar{w})}\frac{(-1)^{\#\bar{s}}\prod_{\nu=1}^{N-1}\lambda_{\nu+1}^{(1)}(\bar{s}^{\nu})}{\prod_{\nu=1}^{N-1}\hat{\lambda}_{\nu+1}^{(1)}(\bar{s}^{N-\nu}+(N-\nu)c)\lambda_{0}(\bar{s}^{N-\nu}+(N-\nu)c)}\times\\
\frac{\prod_{\nu=1}^{N-1}\lambda_{\nu}^{(2)}(\bar{s}^{\nu})\lambda_{\nu+1}^{(1)}(\bar{t}^{\nu})f(\bar{t}^{\nu},\bar{s}^{\nu})\prod_{\nu=1}^{N-2}f(\bar{s}^{\nu},\bar{s}^{\nu+1}+c)}{\prod_{\nu=1}^{N-2}f(\bar{t}^{\nu+1},\bar{s}^{\nu})}\tilde{S}^{(2)}(\pi^{c}(\bar{s})|\bar{t}),
\end{multline}
where
\begin{equation}
\pi^{c}(\bar{s})=\pi^{a}(\mu^{-1}(\bar{s}))=\{-\bar{s}^{1}-c,-\bar{s}^{2}-2c,\dots,-\bar{s}^{N-1}-(N-1)c\}.
\end{equation}
The sum goes over all the partitions of $\bar{w}^{\nu}=\bar{s}^{\nu}\cup\bar{t}^{\nu}$,
$\bar{t}^{\nu}=\bar{t}_{\textsc{i}}^{\nu}\cup\bar{t}_{\textsc{ii}}^{\nu}$
and $\bar{s}^{\nu}=\bar{s}_{\textsc{i}}^{\nu}\cup\bar{s}_{\textsc{ii}}^{\nu}$
where $\#\bar{t}_{\textsc{i}}^{\nu}=\#\bar{s}_{\textsc{i}}^{\nu}$
and $\#\bar{t}_{\textsc{ii}}^{\nu}=\#\bar{s}_{\textsc{ii}}^{\nu}$.
Using the sum rule of the scalar product (\ref{eq:sumrule}) we obtain
that
\begin{multline}
\langle\mathcal{C}|\tilde{\mathbb{B}}(\bar{w})=\sum_{\mathrm{part}(\bar{w})}\frac{(-1)^{\#\bar{s}}\prod_{\nu=1}^{N-1}\lambda_{\nu+1}^{(1)}(\bar{s}^{\nu})}{\prod_{\nu=1}^{N-1}\hat{\lambda}_{N-\nu+1}^{(1)}(\bar{s}^{\nu}+\nu c)\lambda_{0}(\bar{s}^{\nu}+\nu c)}\times\\
\frac{\prod_{\nu=1}^{N-1}\lambda_{\nu}^{(2)}(\bar{s}^{\nu})\lambda_{\nu+1}^{(1)}(\bar{t}^{\nu})f(\bar{t}^{\nu},\bar{s}^{\nu})\prod_{\nu=1}^{N-2}f(\bar{s}^{\nu},\bar{s}^{\nu+1}+c)}{\prod_{\nu=1}^{N-2}f(\bar{t}^{\nu+1},\bar{s}^{\nu})}\times\\
W(\pi^{c}(\bar{s}_{\textsc{i}}),\pi^{c}(\bar{s}_{\textsc{ii}})|\bar{t}_{\textsc{i}},\bar{t}_{\textsc{ii}})\prod_{k=1}^{N-1}\lambda_{k}^{(2)}(-\bar{s}_{\textsc{i}}^{k}-kc))\lambda_{k+1}^{(2)}(-\bar{s}_{\textsc{ii}}^{k}-kc))\lambda_{k+1}^{(2)}(\bar{t}_{\textsc{i}}^{k})\lambda_{k}^{(2)}(\bar{t}_{\textsc{ii}}^{k}).
\end{multline}
Using the identities (\ref{eq:idLam}) and (\ref{eq:idAlph}) we obtain
that
\begin{multline}
\langle\mathcal{C}|\tilde{\mathbb{B}}(\bar{w})=\sum_{\mathrm{part}(\bar{w})}\prod_{\nu=1}^{N-1}\lambda_{\nu+1}^{(1)}(\bar{s}^{\nu})(-1)^{\#\bar{s}}\frac{\prod_{\nu=1}^{N-1}\lambda_{\nu}^{(2)}(\bar{s}^{\nu})\lambda_{\nu+1}^{(1)}(\bar{t}^{\nu})f(\bar{t}^{\nu},\bar{s}^{\nu})\prod_{\nu=1}^{N-2}f(\bar{s}^{\nu},\bar{s}^{\nu+1}+c)}{\prod_{\nu=1}^{N-2}f(\bar{t}^{\nu+1},\bar{s}^{\nu})}\times\\
W(\pi^{c}(\bar{s}_{\textsc{i}}),\pi^{c}(\bar{s}_{\textsc{ii}})|\bar{t}_{\textsc{i}},\bar{t}_{\textsc{ii}})\prod_{k=1}^{N-1}\alpha_{k}^{(1)}(\bar{s}_{\textsc{ii}}^{k}))\lambda_{k+1}^{(2)}(\bar{t}_{\textsc{i}}^{k})\lambda_{k}^{(2)}(\bar{t}_{\textsc{ii}}^{k}).
\end{multline}
After a renormalization we have a sum formula for the off-shell overlap
\begin{multline}
\langle\mathcal{C}|\mathbb{B}(\bar{w})=\sum_{\mathrm{part}(\bar{w})}(-1)^{\#\bar{s}}\frac{\prod_{\nu=1}^{N-1}f(\bar{t}^{\nu},\bar{s}^{\nu})f(\bar{s}_{\textsc{i}}^{\nu},\bar{s}_{\textsc{ii}}^{\nu})f(\bar{t}_{\textsc{i}}^{\nu},\bar{t}_{\textsc{ii}}^{\nu})}{\prod_{\nu=1}^{N-2}f(\bar{t}^{\nu+1},\bar{s}^{\nu})}\frac{\prod_{\nu=1}^{N-2}f(\bar{s}^{\nu},\bar{s}^{\nu+1}+c)}{\prod_{k=1}^{N-2}f(\bar{s}_{\textsc{i}}^{k},\bar{s}_{\textsc{ii}}^{k+1}+c)f(\bar{t}_{\textsc{i}}^{k+1},\bar{t}_{\textsc{ii}}^{k})}\times\\
Z(\pi^{c}(\bar{s}_{\textsc{i}})|\bar{t}_{\textsc{i}})Z(\bar{t}_{\textsc{ii}}|\pi^{c}(\bar{s}_{\textsc{ii}}))\prod_{k=1}^{N-1}\alpha_{k}^{(1)}(\bar{s}_{\textsc{ii}}^{k})\alpha_{k}^{(2)}(\bar{s}^{k})\alpha_{k}^{(2)}(\bar{t}_{\textsc{ii}}^{k}),\label{eq:twSum}
\end{multline}
where we used the identity (\ref{eq:WZZ}).

\section{On-shell limit\label{sec:On-shell-limit}}

Let us continue with the on-shell limit of the overlaps. The calculation
of this section is based on the derivation of the norm of the Bethe
states \cite{Hutsalyuk:2017way} and the overlaps of the boundary
states \cite{Gombor:2021hmj}. 

We saw that the non-vanishing overlaps require pair structures $\bar{w}=\bar{w}^{+}\cup\bar{w}^{-}$
These new sets are defined for the untwisted case as
\begin{equation}
\begin{split}w_{k}^{+,\nu} & =w_{k}^{\nu},\quad w_{k}^{-,\nu}=w_{k}^{N-\nu},\quad\nu<N/2,\\
w_{k}^{+,N/2} & =w_{k}^{N/2},\quad w_{k}^{-,N/2}=w_{k+\frac{r_{N/2}}{2}}^{N/2},\quad k=1,\dots,\frac{r_{N/2}}{2},
\end{split}
\end{equation}
and 
\begin{equation}
w_{k}^{+,\nu}=w_{k}^{\nu},\quad w_{k}^{-,\nu}=w_{k+\frac{r_{\nu}}{2}}^{\nu},\quad k=1,\dots,\frac{r_{\nu}}{2},
\end{equation}
for the twisted case. The pair structure limit is $\bar{w}^{-}\to-\bar{w}^{+}$
for untwisted case and $\bar{w}^{-,\nu}\to-\bar{w}^{+,\nu}-\nu c$
for the twisted case. Let us introduce a common notation for the pair
structure limit $\bar{w}^{-}\to\hat{\pi}\left(\bar{w}^{+}\right)$
where $\hat{\pi}\left(\bar{w}^{+}\right)=-\bar{w}^{+}$ or $\hat{\pi}\left(\bar{w}^{+,\nu}\right)=-\bar{w}^{+,\nu}-\nu c$
for the achiral or the chiral pair structures.

\subsection{Gaudin-like determinants\label{subsec:Gaudin-like-determinants}}

We show that the on-shell overlaps are proportional to the Gaudin-like
determinant $\det G^{+}$ where the Gaudin-like matrix is defined
as
\begin{equation}
G_{j,k}^{+,(\mu,\nu)}=-c\left(\frac{\partial}{\partial t_{k}^{+,\nu}}+\frac{\partial}{\partial t_{k}^{-,\nu}}\right)\log\Phi_{j}^{+,(\mu)}\Biggr|_{\bar{w}^{-}=\hat{\pi}\left(\bar{w}^{+}\right)},\label{eq:Gdef}
\end{equation}
where 
\begin{equation}
\Phi_{k}^{+,(\mu)}=\alpha_{\mu}(w_{k}^{+,\mu})\frac{f(\bar{w}_{k}^{\mu},w_{k}^{+,\mu})}{f(w_{k}^{+,\mu},\bar{w}_{k}^{\mu})}\frac{f(w_{k}^{+,\mu},\bar{w}^{\mu-1})}{f(\bar{w}^{\mu+1},w_{k}^{+,\mu})}.
\end{equation}
It is crucial that we first take the derivative in (\ref{eq:Gdef})
and only after take the pair structure limit. The diagonal elements
of this matrix contains derivatives of $\log\alpha$-s for which we
use the following notation
\begin{equation}
X_{j}^{+,\mu}=-c\frac{d}{du}\log\alpha_{\mu}(u)\Biggr|_{u=w_{j}^{+,\mu}}.\label{eq:Xvar}
\end{equation}

We also use the derivatives of $\log\alpha^{(1)}$ and $\log\alpha^{(2)}$
\[
X_{j}^{(1),\mu}=-c\frac{d}{du}\log\alpha_{\mu}^{(1)}(u)\Biggr|_{u=w_{j}^{+,\mu}},\qquad X_{j}^{(2),\mu}=-c\frac{d}{du}\log\alpha_{\mu}^{(2)}(u)\Biggr|_{u=w_{j}^{+,\mu}}.
\]
 We can see that for a specific model the variables $X_{j}^{+,\mu}$
are functions of the Bethe roots. Let us define a more general case
where the variables $X_{j}^{(1),\mu},X_{j}^{(2),\mu}$ and $w_{j}^{+,\mu}$
are independent which gives us a more general version of the Gaudin
determinant where we do not impose (\ref{eq:Xvar})
\begin{equation}
\mathbf{F}^{(\mathbf{r}^{+})}(\bar{X}^{(1)},\bar{X}^{(2)},\bar{w}^{+})=\det G^{+}.
\end{equation}
This function depends on three sets of variables $\bar{X}^{(1,2)}=\cup_{\nu}\bar{X}^{(1,2),\nu}$
and $\bar{w}^{+}=\cup_{\nu}\bar{w}^{+,\nu}$. The superscript denotes
the number of Bethe roots as $\mathbf{r}^{+}=\#\bar{X}^{(1)}=\#\bar{X}^{(2)}=\#\bar{w}^{+}$.

The function $\mathbf{F}^{(\mathbf{r}^{+})}$ obeys the following
Korepin criteria.

\paragraph{Korepin criteria. }
\begin{enumerate}[label=(\roman*)]
\item \label{enum:prop1}The function $\mathbf{F}^{(\mathbf{r}^{+})}(\bar{X}^{(1)},\bar{X}^{(2)},\bar{w}^{+})$
is symmetric over the replacement of the pairs $(X_{j}^{(1),\mu},X_{j}^{(2),\mu},w_{j}^{+,\mu})\leftrightarrow(X_{k}^{(1),\mu},X_{k}^{(2),\mu},w_{k}^{+,\mu})$.
\item \label{enum:prop2}It is linear function of each $X_{j}^{(1),\mu}$
and $X_{j}^{(2),\mu}$.
\item \label{enum:prop3}$\mathbf{F}^{(1)}(X_{1}^{(1),\nu},X_{1}^{(2),\nu},w_{1}^{+,\nu})=X_{1}^{(1),\nu}+X_{1}^{(2),\nu}$.
\item \label{enum:prop4}The coefficient of $X_{j}^{(i),\mu}$ is given
by the function $\mathbf{F}^{(\mathbf{r}^{+}-1)}$ with modified parameters
$X_{k}^{(i),\nu}$
\begin{equation}
\frac{\partial\mathbf{F}^{(\mathbf{r}^{+})}(\bar{X}^{(1)},\bar{X}^{(2)},\bar{w}^{+})}{\partial X_{j}^{(a),\mu}}=\mathbf{F}^{(\mathbf{r}^{+}-1)}(\bar{X}^{(1),\mu,m_{a}}\backslash X_{j}^{(1),\mu,m_{a}},\bar{X}^{(2),\mu,m_{a}}\backslash X_{j}^{(2),\mu,m_{a}},\bar{w}^{+}\backslash w_{j}^{+,\mu}),\label{eq:derivFX}
\end{equation}
where the original variables $X_{k}^{(b),\nu}$ should be replaced
by the modified expression $X_{k}^{(b),\nu,m_{a}}$ which is defined
in the Appendix \ref{sec:Definitions-of-the}. 
\item \label{enum:prop5}$\mathbf{F}^{(\mathbf{r}^{+})}(\bar{X}^{+},\bar{w}^{+})=0$,
if all $X_{j}^{(i),\mu}=0$.
\end{enumerate}
It is easy to show that the functions $\mathbf{F}^{(\mathbf{r}^{+})}(\bar{X}^{(1)},\bar{X}^{(2)},\bar{w}^{+})$
satisfy the Korepin criteria. The reverse statement can be also proven
easily and this proof is the same as the proof of Proposition 4.1.
in \cite{Hutsalyuk:2017way}. 

\subsection{On-shell formulas}

It turns out that the results of the previous subsection are enough
to derive the closed form of the on-shell overlaps. 

Let us introduce a normalized overlap for the untwisted case
\begin{equation}
\mathbb{N}(\bar{w})=\frac{\prod_{\nu=1}^{\frac{N}{2}-1}f(\bar{w}^{+,\nu+1},\bar{w}^{+,\nu})\left[f(\bar{w}^{-,\frac{N}{2}},\bar{w}^{+,\frac{N}{2}-1})\right]^{2}}{\prod_{\nu=1}^{\frac{N}{2}}\prod_{i\neq j}f(w_{i}^{+,\nu},w_{j}^{+,\nu})f(\bar{w}^{-,\frac{N}{2}},\bar{w}^{+,\frac{N}{2}})}\frac{1}{\prod_{\nu=1}^{\frac{N}{2}}\alpha_{\nu}^{(2)}(\bar{w}^{+,\nu})}\langle\mathcal{C}|\mathbb{B}(\bar{w}),\label{eq:normUTweven}
\end{equation}
for even $N$ and

\begin{equation}
\mathbb{N}(\bar{w})=\frac{\prod_{\nu=1}^{\frac{N-1}{2}}f(\bar{w}^{+,\nu+1},\bar{w}^{+,\nu})}{\prod_{\nu=1}^{\frac{N-1}{2}}\prod_{i\neq j}f(w_{i}^{+,\nu},w_{j}^{+,\nu})}\frac{1}{\prod_{\nu=1}^{\frac{N-1}{2}}\alpha_{\nu}^{(2)}(\bar{w}^{+,\nu})}\langle\mathcal{C}|\mathbb{B}(\bar{w}),\label{eq:normUTwodd}
\end{equation}
for odd $N$. For the twisted case the normalized overlap is
\begin{equation}
\mathbb{N}(\bar{w})=\frac{(-1)^{\frac{\#\bar{w}}{2}}}{\prod_{\nu=1}^{N-1}\prod_{i\neq j}f(w_{i}^{+,\nu},w_{j}^{+,\nu})f(\bar{w}^{-,\nu},\bar{w}^{+,\nu})f(\bar{w}^{+,\nu+1},\bar{w}^{-,\nu})\alpha_{\nu}^{(2)}(\bar{w}^{+,\nu})}\langle\mathcal{C}|\mathbb{B}(\bar{w}).\label{eq:normTw}
\end{equation}
In appendix \ref{sec:Normalized-overlaps-and} we show that a re-normalized
version of the overlap in the on-shell limit satisfies the Korepin
criteria therefore it is equal to the Gaudin-like determinant.
\begin{equation}
\mathbb{N}(\bar{w})=\det G^{+}.
\end{equation}
Therefore the untwisted on-shell overlaps are
\begin{equation}
\langle\mathcal{C}|\mathbb{B}(\bar{w})=\frac{\prod_{\nu=1}^{\frac{N}{2}}\prod_{i\neq j}f(w_{i}^{+,\nu},w_{j}^{+,\nu})f(\bar{w}^{-,\frac{N}{2}},\bar{w}^{+,\frac{N}{2}})}{\prod_{\nu=1}^{\frac{N}{2}-1}f(\bar{w}^{+,\nu+1},\bar{w}^{+,\nu})\left[f(\bar{w}^{-,\frac{N}{2}},\bar{w}^{+,\frac{N}{2}-1})\right]^{2}}\prod_{\nu=1}^{\frac{N}{2}}\alpha_{\nu}^{(2)}(\bar{w}^{+,\nu})\det G^{+},
\end{equation}
for even $N$ and

\begin{equation}
\langle\mathcal{C}|\mathbb{B}(\bar{w})=\frac{\prod_{\nu=1}^{\frac{N-1}{2}}\prod_{i\neq j}f(w_{i}^{+,\nu},w_{j}^{+,\nu})}{\prod_{\nu=1}^{\frac{N-1}{2}}f(\bar{w}^{+,\nu+1},\bar{w}^{+,\nu})}\prod_{\nu=1}^{\frac{N-1}{2}}\alpha_{\nu}^{(2)}(\bar{w}^{+,\nu})\det G^{+},
\end{equation}
for odd $N$. For the twisted case the on-shell overlap is
\begin{equation}
\langle\mathcal{C}|\mathbb{B}(\bar{w})=(-1)^{\frac{\#\bar{w}}{2}}\prod_{\nu=1}^{N-1}\prod_{i\neq j}f(w_{i}^{+,\nu},w_{j}^{+,\nu})f(\bar{w}^{-,\nu},\bar{w}^{+,\nu})f(\bar{w}^{+,\nu+1},\bar{w}^{-,\nu})\alpha_{\nu}^{(2)}(\bar{w}^{+,\nu})\det G^{+}.
\end{equation}
The scalar product for the achiral pair structure reads as
\begin{equation}
\mathbb{C}(\bar{w})\mathbb{B}(\bar{w})=\frac{\left[\prod_{\nu=1}^{\frac{N}{2}}\prod_{k\neq l}f(w_{l}^{+,\nu},w_{k}^{+,\nu})\right]^{2}\left[f(\bar{w}^{-,\frac{N}{2}},\bar{w}^{+,\frac{N}{2}})f(\bar{w}^{+,\frac{N}{2}},\bar{w}_{k}^{-,\frac{N}{2}})\right]}{\left[\prod_{\nu=1}^{\frac{N}{2}-1}f(\bar{w}^{+,\nu+1},\bar{w}^{+,\nu})f(\bar{w}^{-,\frac{N}{2}},\bar{w}^{+,\frac{N}{2}-1})\right]^{2}}\det G^{+}\det G^{-},
\end{equation}
for even $N$ and

\begin{equation}
\mathbb{C}(\bar{w})\mathbb{B}(\bar{w})=\frac{\left[\prod_{\nu=1}^{\frac{N-1}{2}}\prod_{k\neq l}f(w_{l}^{+,\nu},w_{k}^{+,\nu})\right]^{2}}{\left[\prod_{\nu=1}^{\frac{N-3}{2}}f(\bar{w}^{+,\nu+1},\bar{w}^{+,\nu})\right]^{2}f(\bar{w}^{\frac{N+1}{2}},\bar{w}^{\frac{N-1}{2}})}\det G^{+}\det G^{-},
\end{equation}
for odd $N$. For the chiral pair structure we have
\begin{equation}
\mathbb{C}(\bar{w})\mathbb{B}(\bar{w})=\prod_{\nu=1}^{N-1}\left[\prod_{k\neq l}f(w_{l}^{+,\nu},w_{k}^{+,\nu})\right]^{2}f(\bar{w}^{+,\nu},\bar{w}^{-,\nu})f(\bar{w}^{-,\nu},\bar{w}^{+,\nu})\det G^{+}\det G^{-}.
\end{equation}
We can also obtain the overlaps for the dual vectors (we only need
to replace $\alpha_{\nu}^{(1)}\leftrightarrow\alpha_{\nu}^{(2)}$).
Therefore, in the untwisted case, we have
\begin{equation}
\mathbb{C}(\bar{w})|\mathcal{C}\rangle=\frac{\prod_{\nu=1}^{\frac{N}{2}}\prod_{i\neq j}f(w_{i}^{+,\nu},w_{j}^{+,\nu})f(\bar{w}^{-,\frac{N}{2}},\bar{w}^{+,\frac{N}{2}})}{\prod_{\nu=1}^{\frac{N}{2}-1}f(\bar{w}^{+,\nu+1},\bar{w}^{+,\nu})\left[f(\bar{w}^{-,\frac{N}{2}},\bar{w}^{+,\frac{N}{2}-1})\right]^{2}}\prod_{\nu=1}^{\frac{N}{2}}\alpha_{\nu}^{(1)}(\bar{w}^{+,\nu})\det G^{+},
\end{equation}
for even $N$ and

\begin{equation}
\mathbb{C}(\bar{w})|\mathcal{C}\rangle=\frac{\prod_{\nu=1}^{\frac{N-1}{2}}\prod_{i\neq j}f(w_{i}^{+,\nu},w_{j}^{+,\nu})}{\prod_{\nu=1}^{\frac{N-1}{2}}f(\bar{w}^{+,\nu+1},\bar{w}^{+,\nu})}\prod_{\nu=1}^{\frac{N-1}{2}}\alpha_{\nu}^{(1)}(\bar{w}^{+,\nu})\det G^{+},
\end{equation}
for odd $N$. For the twisted case we have
\begin{equation}
\mathbb{C}(\bar{w})|\mathcal{C}\rangle=(-1)^{\frac{\#\bar{w}}{2}}\prod_{\nu=1}^{N-1}\prod_{i\neq j}f(w_{i}^{+,\nu},w_{j}^{+,\nu})f(\bar{w}^{-,\nu},\bar{w}^{+,\nu})f(\bar{w}^{+,\nu+1},\bar{w}^{-,\nu})\alpha_{\nu}^{(1)}(\bar{w}^{+,\nu})\det G^{+}.
\end{equation}
For the normalized overlaps, in the untwisted case, we obtain that
\begin{equation}
\frac{\mathbb{C}(\bar{w})|\mathcal{C}\rangle\langle\mathcal{C}|\mathbb{B}(\bar{w})}{\mathbb{C}(\bar{w})\mathbb{B}(\bar{w})}=\left[\frac{1}{f(\bar{w}^{-,\frac{N}{2}},\bar{w}^{\frac{N}{2}-1})}\right]^{2}\frac{f(\bar{w}^{-,\frac{N}{2}},\bar{w}^{+,\frac{N}{2}})}{f(\bar{w}^{+,\frac{N}{2}},\bar{w}^{-,\frac{N}{2}})}\prod_{\nu=1}^{\frac{N}{2}}\alpha_{\nu}(\bar{w}^{+,\nu})\frac{\det G^{+}}{\det G^{-}},\label{eq:nromovTw2}
\end{equation}
for even $N$ in and

\begin{equation}
\frac{\mathbb{C}(\bar{w})|\mathcal{C}\rangle\langle\mathcal{C}|\mathbb{B}(\bar{w})}{\mathbb{C}(\bar{w})\mathbb{B}(\bar{w})}=\frac{1}{f(\bar{w}^{\frac{N-1}{2}},\bar{w}^{\frac{N+1}{2}})}\prod_{\nu=1}^{\frac{N-1}{2}}\alpha_{\nu}(\bar{w}^{+,\nu})\frac{\det G^{+}}{\det G^{-}},\label{eq:normovTw}
\end{equation}
for odd $N$. In the twisted case we have 
\begin{equation}
\frac{\mathbb{C}(\bar{w})|\mathcal{C}\rangle\langle\mathcal{C}|\mathbb{B}(\bar{w})}{\mathbb{C}(\bar{w})\mathbb{B}(\bar{w})}=\prod_{\nu=1}^{N-1}\frac{f(\bar{w}^{-,\nu},\bar{w}^{+,\nu})}{f(\bar{w}^{+,\nu},\bar{w}^{-,\nu})}\left[f(\bar{w}^{+,\nu+1},\bar{w}^{-,\nu})\right]^{2}\alpha_{\nu}(\bar{w}^{+,\nu})\frac{\det G^{+}}{\det G^{-}}.
\end{equation}
Using the identity
\begin{equation}
f(\bar{w}^{+,\nu+1},\bar{w}^{-,\nu})=f(-\bar{w}^{-,\nu+1}-c,-\bar{w}^{+,\nu})=f(\bar{w}^{+,\nu},\bar{w}^{-,\nu+1}+c)=\frac{1}{f(\bar{w}^{-,\nu+1},\bar{w}^{+,\nu})},
\end{equation}
we can obtain that
\begin{equation}
\frac{\mathbb{C}(\bar{w})|\mathcal{C}\rangle\langle\mathcal{C}|\mathbb{B}(\bar{w})}{\mathbb{C}(\bar{w})\mathbb{B}(\bar{w})}=\prod_{\nu=1}^{N-1}\frac{f(\bar{w}^{-,\nu},\bar{w}^{+,\nu})}{f(\bar{w}^{+,\nu},\bar{w}^{-,\nu})}\frac{f(\bar{w}^{+,\nu},\bar{w}^{-,\nu-1})}{f(\bar{w}^{-,\nu+1},\bar{w}^{+,\nu})}\alpha_{\nu}(\bar{w}^{+,\nu})\frac{\det G^{+}}{\det G^{-}}.\label{eq:normovUTw}
\end{equation}

Using the Bethe Ansatz equations, the on-shell formulas (\ref{eq:nromovTw2}),
(\ref{eq:normovTw}) and (\ref{eq:normovUTw}) can be written in the
following universal form
\begin{equation}
\boxed{\frac{\mathbb{C}(\bar{w})|\mathcal{C}\rangle\langle\mathcal{C}|\mathbb{B}(\bar{w})}{\mathbb{C}(\bar{w})\mathbb{B}(\bar{w})}=\frac{\det G^{+}}{\det G^{-}}.}\label{eq:final}
\end{equation}

We just derived the overlap formulas which can be applied for a wide
class of integrable crosscap states of the $\mathfrak{gl}(N)$ symmetric
spin chains. For the AdS/CFT point of view it can be applied for the
crosscap states (\ref{eq:so3so3}) (chiral pair structure) and (\ref{eq:so2so4})
(achiral pair structure) in the $\mathcal{N}=4$ SYM. In the ABJM
the formula (\ref{eq:final}) can be applied for the crosscap states
(\ref{eq:so4}) (chiral pair structure) and for the crosscap state
(\ref{eq:gl2gl2}) (achiral pair structure).

\section{Conclusion}

In this paper we generalized the algebraic method of \cite{Gombor:2021hmj}
to the crosscap states of the $\mathfrak{gl}(N)$ symmetric spin chains.
The main advantage of this approach is that it is independent from
the quantum space. We classified the crosscap states and collected
the states which can be relevant in the AdS/CFT correspondence. We
also investigated the overlaps between the crosscap and Bethe states.
Our main result is the equation (\ref{eq:final}) which give the exact
overlaps for the $\mathfrak{so}(N)$ and $\mathfrak{gl}(\left\lfloor \frac{N}{2}\right\rfloor )\otimes\mathfrak{gl}(\left\lceil \frac{N}{2}\right\rceil )$
symmetric crosscap states. 

Finally we collect some possible future directions for research.
\begin{itemize}
\item Our proof is not complete even for the $\mathfrak{gl}(N)$ symmetric
spin chains since it does not work for the $\mathfrak{sp}(N)$ and
the $\mathfrak{gl}(M)\otimes\mathfrak{gl}(N-M)$ symmetric two-site
states where $M<\left\lfloor \frac{N}{2}\right\rfloor $. It would
be nice to find an extension of the method of this paper which contains
these remaining cases.
\item Our method is based on the $KT$-relation and the recurrence relations
of the off-shell Bethe states. For the $U_{q}(\mathfrak{gl}(N))$
and $\mathfrak{so}(2N+1)$-invariant spin chains the recurrence relations
are available \cite{Liashyk:2019owy,Liashyk:2020xsv,Liashyk:2021tth}.
Using these results it would be interesting to generalize our method
to the $q$-deformed and orthogonal spin chains. 
\item For the applications in the AdS/CFT duality, it would be interesting
to generalize our results to $\mathfrak{gl}(N|M)$ symmetric spin
chains. For the context of the $AdS_{5}/CFT_{4}$ duality the relevant
case is the $\mathfrak{gl}(4|4)$. 
\item It would be also interesting to combine the crosscap states and the
SoV framework, like it was done for the boundary states in \cite{Gombor:2021uxz}.
This method would give an alternative formula for the crosscap overlap
which would contain Vandermonde determinants instead of Gaudin determinants.
\item An other possible direction is to generalize the integrable crosscap
states for long range spin chains which are relevant in $\mathcal{N}=4$
SYM at higher loops. Combining this to the method of \cite{Gombor:2022lco}
we could investigate the wrapping corrections of the overlaps between
crosscap and Bethe states.
\end{itemize}

\section*{Acknowledgments}

I thank Shota Komatsu, Balázs Pozsgay, and Zoltán Bajnok for the useful
discussions and the NKFIH grant K134946 for support.

\appendix

\section{Off-shell Bethe vectors\label{sec:Off-shell-Bethe-vectors}}

In this section we review the recurrence and action formulas of the
off-shell Bethe vectors which are used in our derivations of the overlaps.
The formulas of this section can be found in \cite{Hutsalyuk:2017tcx,Hutsalyuk:2017way,Liashyk:2018egk,Hutsalyuk:2020dlw}.

The off-shell Bethe vectors can be calculated from the following sum
formula \cite{Hutsalyuk:2017tcx}

\begin{equation}
\mathbb{B}(\{z,\bar{t}^{1}\},\left\{ \bar{t}^{k}\right\} _{k=2}^{N-1})=\sum_{j=2}^{N}\frac{T_{1,j}(z)}{\lambda_{2}(z)}\sum_{\mathrm{part}(\bar{t})}\mathbb{B}(\bar{t}^{1},\left\{ \bar{t}_{\textsc{ii}}^{k}\right\} _{k=2}^{j-1},\left\{ \bar{t}^{k}\right\} _{k=j}^{N-1})\frac{\prod_{\nu=2}^{j-1}\alpha_{\nu}(\bar{t}_{\textsc{i}}^{\nu})g(\bar{t}_{\textsc{i}}^{\nu},\bar{t}_{\textsc{i}}^{\nu-1})f(\bar{t}_{\textsc{ii}}^{\nu},\bar{t}_{\textsc{i}}^{\nu})}{\prod_{\nu=1}^{j-1}f(\bar{t}^{\nu+1},\bar{t}_{\textsc{i}}^{\nu})},\label{eq:rec1}
\end{equation}
where the sum goes over all the possible partitions $\bar{t}^{\nu}=\bar{t}_{\textsc{i}}^{\nu}\cup\bar{t}_{\textsc{ii}}^{\nu}$
for $\nu=2,\dots,j-1$ where $\bar{t}_{\textsc{i}}^{\nu},\bar{t}_{\textsc{ii}}^{\nu}$
are disjoint subsets and $\#\bar{t}_{\textsc{i}}^{\nu}=1$. We set
by definition $\bar{t}_{\textsc{i}}^{1}=\{z\}$ and $\bar{t}^{N}=\emptyset$. 

There is another sum formula
\begin{equation}
\mathbb{B}(\left\{ \bar{t}^{k}\right\} _{k=1}^{N-2},\{z,\bar{t}^{N-1}\})=\sum_{j=1}^{N-1}\frac{T_{j,N}(z)}{\lambda_{N}(z)}\sum_{\mathrm{part}(\bar{t})}\mathbb{B}(\left\{ \bar{t}^{k}\right\} _{k=1}^{j-1},\left\{ \bar{t}_{\textsc{ii}}^{k}\right\} _{k=j}^{N-2},\bar{t}^{N-1})\frac{\prod_{\nu=j}^{N-2}g(\bar{t}_{\textsc{i}}^{\nu+1},\bar{t}_{\textsc{i}}^{\nu})f(\bar{t}_{\textsc{i}}^{\nu},\bar{t}_{\textsc{ii}}^{\nu})}{\prod_{\nu=j}^{N-1}f(\bar{t}_{\textsc{i}}^{\nu},\bar{t}^{\nu-1})},\label{eq:rec2}
\end{equation}
where the sum goes over all the possible partitions $\bar{t}^{\nu}=\bar{t}_{\textsc{i}}^{\nu}\cup\bar{t}_{\textsc{ii}}^{\nu}$
for $\nu=j,\dots,N-2$ where $\bar{t}_{\textsc{i}}^{\nu},\bar{t}_{\textsc{ii}}^{\nu}$
are disjoint subsets and $\#\bar{t}_{\textsc{i}}^{\nu}=1$. We set
by definition $\bar{t}_{\textsc{i}}^{N-1}=\{z\}$ and $\bar{t}^{0}=\emptyset$.

We also use the following action formula for the off-shell Bethe vectors
\cite{Hutsalyuk:2020dlw}

\begin{multline}
T_{i,j}(z)\mathbb{B}(\bar{t})=\lambda_{N}(z)\sum_{\mathrm{part}(\bar{w})}\mathbb{B}(\bar{w}_{\textsc{ii}})\frac{\prod_{s=j}^{i-1}f(\bar{w}_{\textsc{i}}^{s},\bar{w}_{\textsc{iii}}^{s})}{\prod_{s=j}^{i-2}f(\bar{w}_{\textsc{i}}^{s+1},\bar{w}_{\textsc{iii}}^{s})}\times\\
\prod_{s=1}^{i-1}\frac{f(\bar{w}_{\textsc{i}}^{s},\bar{w}_{\textsc{ii}}^{s})}{h(\bar{w}_{\textsc{i}}^{s},\bar{w}_{\textsc{i}}^{s-1})f(\bar{w}_{\textsc{i}}^{s},\bar{w}_{\textsc{ii}}^{s-1})}\prod_{s=j}^{N-1}\frac{\alpha_{s}(\bar{w}_{\textsc{iii}}^{s})f(\bar{w}_{\textsc{ii}}^{s},\bar{w}_{\textsc{iii}}^{s})}{h(\bar{w}_{\textsc{iii}}^{s+1},\bar{w}_{\textsc{iii}}^{s})f(\bar{w}_{\textsc{ii}}^{s+1},\bar{w}_{\textsc{iii}}^{s})},\label{eq:act}
\end{multline}
where $\bar{w}^{\nu}=\{z,\bar{t}^{\nu}\}$. The sum goes over all
the partitions of $\bar{w}^{\nu}=\bar{w}_{\textsc{i}}^{\nu}\cup\bar{w}_{\textsc{ii}}^{\nu}\cup\bar{w}_{\textsc{iii}}^{\nu}$
where $\bar{w}_{\textsc{i}}^{\nu},\bar{w}_{\textsc{ii}}^{\nu},\bar{w}_{\textsc{iii}}^{\nu}$
are disjoint sets for a fixed $\nu$ and $\#\bar{w}_{\textsc{i}}^{\nu}=\Theta(i-1-\nu)$,
$\#\bar{w}_{\textsc{iii}}^{\nu}=\Theta(\nu-j)$. We also set $\bar{w}_{\textsc{i}}^{0}=\bar{w}_{\textsc{iii}}^{N}=\{z\}$
and $\bar{w}_{\textsc{ii}}^{0}=\bar{w}_{\textsc{iii}}^{0}=\bar{w}_{\textsc{i}}^{N}=\bar{w}_{\textsc{ii}}^{N}=\emptyset$.
We also used the unit step function $\Theta(k)$ which is defined
as $\Theta(k)=1$ for $k\geq0$ and $\Theta(k)=0$ for $k<0$.

We can see that the diagonal elements $T_{i,i}(u)$ do not change
the quantum numbers $r_{j}$. The creation operators i.e. $T_{i,j}(u)$
where $i<j$ increase the quantum numbers $r_{i},r_{i+1},\dots,r_{j-1}$
by one. The annihilation operators i.e. $T_{j,i}(u)$ where $i<j$
decrease the quantum numbers $r_{i},r_{i+1},\dots,r_{j-1}$ by one. 

In some parts of the derivations it is more convenient to use the
following re-normalized Bethe states
\begin{equation}
\tilde{\mathbb{B}}(\bar{t})=\prod_{\nu=1}^{N-1}\lambda_{\nu+1}(\bar{t}^{\nu})\mathbb{B}(\bar{t}),\label{eq:renOffShell}
\end{equation}
for which the recurrence and action formulas for can be written as
\begin{equation}
\tilde{\mathbb{B}}(\{z,\bar{t}^{1}\},\left\{ \bar{t}^{k}\right\} _{k=2}^{N-1})=\sum_{j=2}^{N}T_{1,j}(z)\sum_{\mathrm{part}(\bar{t})}\tilde{\mathbb{B}}(\bar{t}^{1},\left\{ \bar{t}_{\textsc{ii}}^{k}\right\} _{k=2}^{j-1},\left\{ \bar{t}^{k}\right\} _{k=j}^{N-1})\frac{\prod_{\nu=2}^{j-1}\lambda_{\nu}(\bar{t}_{\textsc{i}}^{\nu})g(\bar{t}_{\textsc{i}}^{\nu},\bar{t}_{\textsc{i}}^{\nu-1})f(\bar{t}_{\textsc{ii}}^{\nu},\bar{t}_{\textsc{i}}^{\nu})}{\prod_{\nu=1}^{j-1}f(\bar{t}^{\nu+1},\bar{t}_{\textsc{i}}^{\nu})},\label{eq:rec1-1}
\end{equation}
\begin{multline}
\tilde{\mathbb{B}}(\left\{ \bar{t}^{k}\right\} _{k=1}^{N-2},\{z,\bar{t}^{N-1}\})=\sum_{j=1}^{N-1}T_{j,N}(z)\sum_{\mathrm{part}(\bar{t})}\tilde{\mathbb{B}}(\left\{ \bar{t}^{k}\right\} _{k=1}^{j-1},\left\{ \bar{t}_{\textsc{ii}}^{k}\right\} _{k=j}^{N-2},\bar{t}^{N-1})\times\\
\frac{\prod_{\nu=j}^{N-2}\lambda_{\nu+1}(\bar{t}_{\textsc{i}}^{\nu})g(\bar{t}_{\textsc{i}}^{\nu+1},\bar{t}_{\textsc{i}}^{\nu})f(\bar{t}_{\textsc{i}}^{\nu},\bar{t}_{\textsc{ii}}^{\nu})}{\prod_{\nu=j}^{N-1}f(\bar{t}_{\textsc{i}}^{\nu},\bar{t}^{\nu-1})},\label{eq:rec2-1}
\end{multline}
and
\begin{multline}
T_{i,j}(z)\tilde{\mathbb{B}}(\bar{t})=\sum_{\mathrm{part}(\bar{w})}\frac{\prod_{s=1}^{i-1}\lambda_{s+1}(\bar{w}_{\textsc{i}}^{s})\prod_{s=j}^{N-1}\lambda_{s}(\bar{w}_{\textsc{iii}}^{s})}{\prod_{s=1}^{N-2}\lambda_{s+1}(z)}\tilde{\mathbb{B}}(\bar{w}_{\textsc{ii}})\\
\frac{\prod_{s=j}^{i-1}f(\bar{w}_{\textsc{i}}^{s},\bar{w}_{\textsc{iii}}^{s})}{\prod_{s=j}^{i-2}f(\bar{w}_{\textsc{i}}^{s+1},\bar{w}_{\textsc{iii}}^{s})}\prod_{s=1}^{i-1}\frac{f(\bar{w}_{\textsc{i}}^{s},\bar{w}_{\textsc{ii}}^{s})}{h(\bar{w}_{\textsc{i}}^{s},\bar{w}_{\textsc{i}}^{s-1})f(\bar{w}_{\textsc{i}}^{s},\bar{w}_{\textsc{ii}}^{s-1})}\prod_{s=j}^{N-1}\frac{f(\bar{w}_{\textsc{ii}}^{s},\bar{w}_{\textsc{iii}}^{s})}{h(\bar{w}_{\textsc{iii}}^{s+1},\bar{w}_{\textsc{iii}}^{s})f(\bar{w}_{\textsc{ii}}^{s+1},\bar{w}_{\textsc{iii}}^{s})}.
\end{multline}

We will also use the co-product formula of the off-shell Bethe vectors.
Let $\mathcal{H}^{(1)},\mathcal{H}^{(2)}$ be two quantum spaces for
which $\mathcal{H}=\mathcal{H}^{(1)}\otimes\mathcal{H}^{(2)}$ and
the corresponding off-shell states are $\tilde{\mathbb{B}}^{(1)}(\bar{t}),\tilde{\mathbb{B}}^{(2)}(\bar{t})$.
The co-product formula reads as \cite{Hutsalyuk:2016srn}
\begin{equation}
\tilde{\mathbb{B}}(\bar{t})=\sum_{\mathrm{part}(\bar{t})}\frac{\prod_{\nu=1}^{N-1}\lambda_{\nu}^{(2)}(\bar{t}_{\mathrm{i}}^{\nu})\lambda_{\nu+1}^{(1)}(\bar{t}_{\mathrm{ii}}^{\nu})f(\bar{t}_{\mathrm{ii}}^{\nu},\bar{t}_{\mathrm{i}}^{\nu})}{\prod_{\nu=1}^{N-2}f(\bar{t}_{\mathrm{ii}}^{\nu+1},\bar{t}_{\mathrm{i}}^{\nu})}\mathbb{\tilde{\mathbb{B}}}^{(1)}(\bar{t}_{\mathrm{i}})\otimes\tilde{\mathbb{B}}^{(2)}(\bar{t}_{\mathrm{ii}}),\label{eq:coproduct}
\end{equation}
where the sum goes over all the possible partitions $\bar{t}^{\nu}=\bar{t}_{\mathrm{i}}^{\nu}\cup\bar{t}_{\mathrm{ii}}^{\nu}$
where $\bar{t}_{\mathrm{i}}^{\nu},\bar{t}_{\mathrm{ii}}^{\nu}$ are
disjoint subsets.

We will also need the recursion formulas for the left off-shell vectors
\begin{multline}
\tilde{\mathbb{C}}(\left\{ \bar{s}^{k}\right\} _{k=2}^{N-1},\{z,\bar{s}^{N-1}\})=\sum_{j=1}^{N-1}\sum_{\mathrm{part}(\bar{t})}\tilde{\mathbb{C}}(\left\{ \bar{s}^{k}\right\} _{k=1}^{j-1},\left\{ \bar{s}_{\textsc{ii}}^{k}\right\} _{k=j}^{N-2},\bar{s}^{N-1})T_{N,j}(z)\times\\
\frac{\prod_{\nu=j}^{N-2}\lambda_{\nu+1}(\bar{t}_{\textsc{i}}^{\nu})g(\bar{s}_{\textsc{i}}^{\nu+1},\bar{s}_{\textsc{i}}^{\nu})f(\bar{s}_{\textsc{i}}^{\nu},\bar{s}_{\textsc{ii}}^{\nu})}{\prod_{\nu=j}^{N-1}f(\bar{s}_{\textsc{i}}^{\nu},\bar{s}^{\nu-1})}.\label{eq:recC1}
\end{multline}

\section{Sum formula for the scalar product\label{sec:Sum-formula-for}}

In \cite{Hutsalyuk:2017tcx} it was shown that the off-shell scalar
product has the following sum rule

\begin{equation}
S(\bar{s}|\bar{t})=\mathbb{C}(\bar{s})\mathbb{B}(\bar{t})=\sum W(\bar{s}_{\textsc{i}},\bar{s}_{\textsc{ii}}|\bar{t}_{\textsc{i}},\bar{t}_{\textsc{ii}})\prod_{k=1}^{N-1}\alpha_{k}(\bar{s}_{\textsc{i}}^{k})\alpha_{k}(\bar{t}_{\textsc{ii}}^{k}).
\end{equation}
The sum goes over all the partitions of $\bar{t}^{\nu}=\bar{t}_{\textsc{i}}^{\nu}\cup\bar{t}_{\textsc{ii}}^{\nu}$
and $\bar{s}^{\nu}=\bar{s}_{\textsc{i}}^{\nu}\cup\bar{s}_{\textsc{ii}}^{\nu}$
where $\#\bar{t}_{\textsc{i}}^{\nu}=\#\bar{s}_{\textsc{i}}^{\nu}$
and $\#\bar{t}_{\textsc{ii}}^{\nu}=\#\bar{s}_{\textsc{ii}}^{\nu}$.
For the renormalized Bethe vector we obtain that
\begin{equation}
\tilde{S}(\bar{s}|\bar{t})=\tilde{\mathbb{C}}(\bar{s})\tilde{\mathbb{B}}(\bar{t})=\sum W(\bar{s}_{\textsc{i}},\bar{s}_{\textsc{ii}}|\bar{t}_{\textsc{i}},\bar{t}_{\textsc{ii}})\prod_{k=1}^{N-1}\lambda_{k}(\bar{s}_{\textsc{i}}^{k})\lambda_{k+1}(\bar{s}_{\textsc{ii}}^{k})\lambda_{k+1}(\bar{t}_{\textsc{i}}^{k})\lambda_{k}(\bar{t}_{\textsc{ii}}^{k}).\label{eq:sumrule}
\end{equation}
One can introduce the highest coefficients (HC)
\begin{equation}
W(\bar{s},\emptyset|\bar{t},\emptyset)=Z(\bar{s}|\bar{t}),
\end{equation}
\begin{equation}
W(\emptyset,\bar{s}|\emptyset,\bar{t})=\bar{Z}(\bar{s}|\bar{t}),
\end{equation}
for which we have the identity 
\begin{equation}
\bar{Z}(\bar{s}|\bar{t})=Z(\bar{t}|\bar{s}).
\end{equation}
The coefficients $W$ can be expressed with the HC-s as
\begin{equation}
W(\bar{s}_{\textsc{i}},\bar{s}_{\textsc{ii}}|\bar{t}_{\textsc{i}},\bar{t}_{\textsc{ii}})=Z(\bar{s}_{\textsc{i}}|\bar{t}_{\textsc{i}})Z(\bar{t}_{\textsc{ii}}|\bar{s}_{\textsc{ii}})\frac{\prod_{k=1}^{N-1}f(\bar{s}_{\textsc{ii}}^{k},\bar{s}_{\textsc{i}}^{k})f(\bar{t}_{\textsc{i}}^{k},\bar{t}_{\textsc{ii}}^{k})}{\prod_{k=1}^{N-2}f(\bar{s}_{\textsc{ii}}^{k+1},\bar{s}_{\textsc{i}}^{k})f(\bar{t}_{\textsc{i}}^{k+1},\bar{t}_{\textsc{ii}}^{k})}.\label{eq:WZZ}
\end{equation}
A recurrence relation for the HC-s can be found in \cite{Hutsalyuk:2017tcx}
but in this paper we only need the pole structure of the HC-s \cite{Hutsalyuk:2017way}
\begin{equation}
Z(\bar{s}|\bar{t})\Biggr|_{s_{j}^{\mu}\to t_{j}^{\mu}}=g(t_{j}^{\mu},s_{j}^{\mu})\frac{f(\bar{t}_{j}^{\mu},t_{j}^{\mu})f(s_{j}^{\mu},\bar{s}_{j}^{\mu})}{f(\bar{t}_{j}^{\mu+1},t_{j}^{\mu})f(s_{j}^{\mu},\bar{s}_{j}^{\mu-1})}Z(\bar{s}\backslash\{s_{j}^{\mu}\}|\bar{t}\backslash\{t_{j}^{\mu}\})+reg.\label{eq:poleZ}
\end{equation}
where the $"reg"$ contains the regular terms in the limit $s_{j}^{\mu}\to t_{j}^{\mu}$.

\section{Relation between crosscap overlaps and scalar products}

In this section we prove the identity
\begin{equation}
\langle\mathcal{C}|\left(\mathbb{\tilde{\mathbb{B}}}^{(1)}(\bar{s})\otimes\tilde{\mathbb{B}}^{(2)}(\bar{t})\right)=\tilde{\mathbb{\mathbb{C}}}^{(2)}(\pi^{a}(\bar{s}))\tilde{\mathbb{B}}^{(2)}(\bar{t}),\label{eq:utw}
\end{equation}
for the untwisted and the identity
\begin{equation}
\langle\mathcal{C}|\left(\tilde{\hat{\mathbb{B}}}^{(1)}(\bar{s})\otimes\tilde{\mathbb{B}}^{(2)}(\bar{t})\right)=\frac{1}{\prod_{\nu=1}^{N-1}\lambda_{0}(\bar{s}^{\nu})}\tilde{\mathbb{\mathbb{C}}}^{(2)}(\pi^{a}(\bar{s}))\tilde{\mathbb{B}}^{(2)}(\bar{t}),\label{eq:tw}
\end{equation}
for the twisted case. We use induction in the number of roots $r_{1}=\#\bar{t}^{1}=\#\bar{s}^{1}$.
Let assume that (\ref{eq:utw}) and (\ref{eq:tw}) is true up to $r_{1}$
Bethe roots. Let us increase the number of roots and using the recursion
formula (\ref{eq:recC1}) we obtain a recursion for the scalar product
\begin{multline}
\tilde{\mathbb{C}}^{(2)}(\left\{ \bar{s}^{k}\right\} _{k=2}^{N-1},\{z,\bar{s}^{N-1}\})\tilde{\mathbb{B}}^{(2)}(\bar{t})=\\
\sum_{j=1}^{N-1}\sum_{\mathrm{part}(\bar{t})}\tilde{\mathbb{C}}^{(2)}(\left\{ \bar{s}^{k}\right\} _{k=1}^{j-1},\left\{ \bar{s}_{\textsc{ii}}^{k}\right\} _{k=j}^{N-2},\bar{s}^{N-1})T_{N,j}^{(2)}(z)\tilde{\mathbb{B}}^{(2)}(\bar{t})\frac{\prod_{\nu=j}^{N-2}\lambda_{\nu+1}^{(2)}(\bar{s}_{\textsc{i}}^{\nu})g(\bar{s}_{\textsc{i}}^{\nu+1},\bar{s}_{\textsc{i}}^{\nu})f(\bar{s}_{\textsc{i}}^{\nu},\bar{s}_{\textsc{ii}}^{\nu})}{\prod_{\nu=j}^{N-1}f(\bar{s}_{\textsc{i}}^{\nu},\bar{s}^{\nu-1})},\label{eq:recscal}
\end{multline}
where $\#\bar{s}^{1}=r_{1}$ and $\#\bar{t}^{1}=r_{1}+1$.

\subsection{Untwisted case\label{subsec:Untwisted-case}}

Now we derive a recursion for the crosscap overlap

\begin{equation}
\langle\mathcal{C}|\left(\tilde{\mathbb{B}}^{(1)}(\{z,\bar{s}^{1}\},\left\{ \bar{s}^{k}\right\} _{k=2}^{N-1})\otimes\tilde{\mathbb{B}}^{(2)}(\bar{t})\right).
\end{equation}
Using the recursion formula (\ref{eq:rec1-1}) we can obtain that

\begin{multline}
\langle\mathcal{C}|\left(\tilde{\mathbb{B}}^{(1)}(\{z,\bar{s}^{1}\},\left\{ \bar{s}^{k}\right\} _{k=2}^{N-1})\otimes\tilde{\mathbb{B}}^{(2)}(\bar{t})\right)=\\
\sum_{j=2}^{N}\langle\mathcal{C}|T_{1,j}^{(1)}(z)\sum_{\mathrm{part}(\bar{t})}\tilde{\mathbb{B}}^{(1)}(\bar{s}^{1},\left\{ \bar{s}_{\textsc{ii}}^{k}\right\} _{k=2}^{j-1},\left\{ \bar{s}^{k}\right\} _{k=j}^{N-1})\otimes\tilde{\mathbb{B}}^{(2)}(\bar{t})\frac{\prod_{\nu=2}^{j-1}\lambda_{\nu}^{(1)}(\bar{s}_{\textsc{i}}^{\nu})g(\bar{s}_{\textsc{i}}^{\nu},\bar{s}_{\textsc{i}}^{\nu-1})f(\bar{s}_{\textsc{ii}}^{\nu},\bar{s}_{\textsc{i}}^{\nu})}{\prod_{\nu=1}^{j-1}f(\bar{s}^{\nu+1},\bar{s}_{\textsc{i}}^{\nu})}.
\end{multline}
Using the KT-relation we obtain that
\begin{multline}
\langle\mathcal{C}|\left(\tilde{\mathbb{B}}^{(1)}(\{z,\bar{s}^{1}\},\left\{ \bar{s}^{k}\right\} _{k=2}^{N-1})\otimes\tilde{\mathbb{B}}^{(2)}(\bar{t})\right)=\\
\sum_{j=2}^{N}\sum_{\mathrm{part}(\bar{t})}\langle\mathcal{C}|\tilde{\mathbb{B}}^{(1)}(\bar{s}^{1},\left\{ \bar{s}_{\textsc{ii}}^{k}\right\} _{k=2}^{j-1},\left\{ \bar{s}^{k}\right\} _{k=j}^{N-1})\otimes T_{N,N+1-j}^{(2)}(-z)\tilde{\mathbb{B}}^{(2)}(\bar{t})\frac{\prod_{\nu=2}^{j-1}\lambda_{\nu}^{(1)}(\bar{s}_{\textsc{i}}^{\nu})g(\bar{s}_{\textsc{i}}^{\nu},\bar{s}_{\textsc{i}}^{\nu-1})f(\bar{s}_{\textsc{ii}}^{\nu},\bar{s}_{\textsc{i}}^{\nu})}{\prod_{\nu=1}^{j-1}f(\bar{s}^{\nu+1},\bar{s}_{\textsc{i}}^{\nu})}.
\end{multline}
After some rearrangements we obtain that
\begin{multline}
\langle\mathcal{C}|\left(\tilde{\mathbb{B}}^{(1)}(\{z,\bar{s}^{1}\},\left\{ \bar{s}^{k}\right\} _{k=2}^{N-1})\otimes\tilde{\mathbb{B}}^{(2)}(\bar{t})\right)=\\
\sum_{j=1}^{N-1}\sum_{\mathrm{part}(\bar{t})}\langle\mathcal{C}|\tilde{\mathbb{B}}^{(1)}(\bar{s}^{1},\left\{ \bar{s}_{\textsc{ii}}^{k}\right\} _{k=2}^{N-j},\left\{ \bar{s}^{k}\right\} _{k=N+1-j}^{N-1})\otimes T_{N,j}^{(2)}(-z)\tilde{\mathbb{B}}^{(2)}(\bar{t})\times\\
\frac{\prod_{\nu=j}^{N-2}\lambda_{\nu+1}^{(2)}(-\bar{s}_{\textsc{i}}^{N-\nu})g(-\bar{s}_{\textsc{i}}^{N-\nu-1},-\bar{s}_{\textsc{i}}^{N-\nu})f(-\bar{s}_{\textsc{i}}^{N-\nu},-\bar{s}_{\textsc{ii}}^{N-\nu})}{\prod_{\nu=j}^{N-1}f(-\bar{s}_{\textsc{i}}^{N-\nu},-\bar{s}^{N-\nu+1})}.
\end{multline}
We can see that this recursion is the same what we obtained for the
scalar product (\ref{eq:recscal}) therefore we just derived (\ref{eq:utw}).

\subsection{Twisted case\label{subsec:Twisted-case}}

Now we derive a recursion for the crosscap overlap

\begin{equation}
\langle\mathcal{C}|\left(\tilde{\hat{\mathbb{B}}}^{(1)}(\{z,\bar{s}^{1}\},\left\{ \bar{s}^{k}\right\} _{k=2}^{N-1})\otimes\tilde{\mathbb{B}}^{(2)}(\bar{t})\right).
\end{equation}
Using the recursion formula (\ref{eq:rec1-1}) we can obtain that

\begin{multline}
\langle\mathcal{C}|\left(\tilde{\hat{\mathbb{B}}}^{(1)}(\{z,\bar{s}^{1}\},\left\{ \bar{s}^{k}\right\} _{k=2}^{N-1})\otimes\tilde{\mathbb{B}}^{(2)}(\bar{t})\right)=\sum_{j=2}^{N}\langle\mathcal{C}|\widehat{T}_{1,j}^{(1)}(z)\sum_{\mathrm{part}(\bar{t})}\tilde{\hat{\mathbb{B}}}^{(1)}(\bar{s}^{1},\left\{ \bar{s}_{\textsc{ii}}^{k}\right\} _{k=2}^{j-1},\left\{ \bar{s}^{k}\right\} _{k=j}^{N-1})\otimes\tilde{\mathbb{B}}^{(2)}(\bar{t})\times\\
\frac{\prod_{\nu=2}^{j-1}\hat{\lambda}_{\nu}^{(1)}(\bar{s}_{\textsc{i}}^{\nu})g(\bar{s}_{\textsc{i}}^{\nu},\bar{s}_{\textsc{i}}^{\nu-1})f(\bar{s}_{\textsc{ii}}^{\nu},\bar{s}_{\textsc{i}}^{\nu})}{\prod_{\nu=1}^{j-1}f(\bar{s}^{\nu+1},\bar{s}_{\textsc{i}}^{\nu})}.
\end{multline}
Using the twisted KT-relation we obtain that
\begin{multline}
\langle\mathcal{C}|\left(\tilde{\hat{\mathbb{B}}}^{(1)}(\{z,\bar{s}^{1}\},\left\{ \bar{s}^{k}\right\} _{k=2}^{N-1})\otimes\tilde{\mathbb{B}}^{(2)}(\bar{t})\right)=\\
\sum_{j=2}^{N}\sum_{\mathrm{part}(\bar{t})}\langle\mathcal{C}|\tilde{\hat{\mathbb{B}}}^{(1)}(\bar{s}^{1},\left\{ \bar{s}_{\textsc{ii}}^{k}\right\} _{k=2}^{j-1},\left\{ \bar{s}^{k}\right\} _{k=j}^{N-1})\otimes T_{N,N+1-j}^{(2)}(-z)\tilde{\mathbb{B}}^{(2)}(\bar{t})\times\\
\frac{1}{\lambda_{0}(z)}\frac{\prod_{\nu=2}^{j-1}\hat{\lambda}_{\nu}^{(1)}(\bar{s}_{\textsc{i}}^{\nu})g(\bar{s}_{\textsc{i}}^{\nu},\bar{s}_{\textsc{i}}^{\nu-1})f(\bar{s}_{\textsc{ii}}^{\nu},\bar{s}_{\textsc{i}}^{\nu})}{\prod_{\nu=1}^{j-1}f(\bar{s}^{\nu+1},\bar{s}_{\textsc{i}}^{\nu})}.
\end{multline}
After some rearrangements we obtain that
\begin{multline}
\langle\mathcal{C}|\left(\tilde{\hat{\mathbb{B}}}^{(1)}(\{z,\bar{s}^{1}\},\left\{ \bar{s}^{k}\right\} _{k=2}^{N-1})\otimes\tilde{\mathbb{B}}^{(2)}(\bar{t})\right)=\\
\sum_{j=1}^{N-1}\sum_{\mathrm{part}(\bar{t})}\langle\mathcal{C}|\tilde{\hat{\mathbb{B}}}^{(1)}(\bar{s}^{1},\left\{ \bar{s}_{\textsc{ii}}^{k}\right\} _{k=2}^{N-j},\left\{ \bar{s}^{k}\right\} _{k=N+1-j}^{N-1})\otimes T_{N,j}^{(2)}(-z)\tilde{\mathbb{B}}^{(2)}(\bar{t})\times\\
\frac{1}{\lambda_{0}(z)\prod_{\nu=2}^{N-j}\lambda_{0}(\bar{s}_{\textsc{i}}^{\nu})}\frac{\prod_{\nu=j}^{N-2}\lambda_{\nu+1}^{(2)}(-\bar{s}_{\textsc{i}}^{N-\nu})g(-\bar{s}_{\textsc{i}}^{N-\nu-1},-\bar{s}_{\textsc{i}}^{N-\nu})f(-\bar{s}_{\textsc{i}}^{N-\nu},-\bar{s}_{\textsc{ii}}^{N-\nu})}{\prod_{\nu=j}^{N-1}f(-\bar{s}_{\textsc{i}}^{N-\nu},-\bar{s}^{N-\nu+1})}.
\end{multline}
We can see that this recursion is the same what we obtained for the
scalar product (\ref{eq:recscal}) therefore we just derived (\ref{eq:tw}).

\section{Definitions of the modified $X$-s\label{sec:Definitions-of-the}}

In the equation (\ref{eq:derivFX}) the modification is trivial for
$|\nu-\mu|>1$ i.e. $X_{k}^{(b),\nu,m_{a}}=X_{k}^{(b),\nu}$ for $|\nu-\mu|>1$.
For the other cases we define the modified $X_{k}^{(b),\nu}$-s, separately.
Let us define the following expressions
\begin{align}
F^{\pm,\mu} & =-c\frac{d}{du}\log\frac{f(w_{j}^{\pm,\mu},u)}{f(u,w_{j}^{\pm,\mu})}\Biggr|_{u=w_{k}^{+,\mu}},\\
G^{\pm,\mu+1} & =-c\frac{d}{du}\log f(u,w_{j}^{\pm,\mu})\Biggr|_{u=w_{k}^{+,\mu+1}},\\
H^{\pm,\mu-1} & =+c\frac{d}{du}\log f(w_{j}^{\pm,\mu},u)\Biggr|_{u=w_{k}^{+,\mu-1}}.
\end{align}

For the twisted case the modified $X_{k}^{(b),\nu}$-s in the equation
(\ref{eq:derivFX}) are defined as
\begin{align}
X_{k}^{(1),\mu,m_{1}} & =X_{k}^{(1),\mu}+F^{+,\mu}, & X_{k}^{(2),\mu,m_{1}} & =X_{k}^{(2),\mu}+F^{-,\mu},\nonumber \\
X_{k}^{(1),\mu+1,m_{1}} & =X_{k}^{(1),\mu+1}+G^{+,\mu+1}, & X_{k}^{(2),\mu+1,m_{1}} & =X_{k}^{(2),\mu+1}+G^{-,\mu+1},\nonumber \\
X_{k}^{(1),\mu-1,m_{1}} & =X_{k}^{(1),\mu-1}+H^{+,\mu-1}, & X_{k}^{(2),\mu-1,m_{1}} & =X_{k}^{(2),\mu-1}+H^{-,\mu-1},\nonumber \\
X_{k}^{(2),\mu,m_{2}} & =X_{k}^{(2),\mu}+F^{+,\mu}, & X_{k}^{(1),\mu,m_{2}} & =X_{k}^{(1),\mu}+F^{-,\mu},\\
X_{k}^{(2),\mu+1,m_{2}} & =X_{k}^{(2),\mu+1}+G^{+,\mu+1}, & X_{k}^{(1),\mu+1,m_{2}} & =X_{k}^{(1),\mu+1}+G^{-,\mu+1},\nonumber \\
X_{k}^{(2),\mu-1,m_{2}} & =X_{k}^{(2),\mu-1}+H^{+,\mu-1}. & X_{k}^{(1),\mu-1,m_{2}} & =X_{k}^{(1),\mu-1}+H^{-,\mu-1}.\nonumber 
\end{align}
For the untwisted case the modified $X_{k}^{(b),\mu}$-s in the equation
(\ref{eq:derivFX}) are defined as
\begin{align}
X_{k}^{(1),\mu,m_{1}} & =X_{k}^{(1),\mu}+F^{+,\mu}, & X_{k}^{(2),\mu,m_{2}} & =X_{k}^{(2),\mu}+F^{+,\mu},\nonumber \\
X_{k}^{(1),\mu+1,m_{1}} & =X_{k}^{(1),\mu+1}+G^{+,\mu+1}, & X_{k}^{(2),\mu+1,m_{2}} & =X_{k}^{(2),\mu+1}+G^{+,\mu+1},\\
X_{k}^{(1),\mu-1,m_{1}} & =X_{k}^{(1),\mu-1}+H^{+,\mu-1}, & X_{k}^{(2),\mu-1,m_{2}} & =X_{k}^{(2),\mu-1}+H^{+,\mu-1},\nonumber 
\end{align}
and $X_{k}^{(1),\nu,m_{2}}=X_{k}^{(1),\nu}$, $X_{k}^{(2),\nu,mod_{1}}=X_{k}^{(2),\nu}$
for $\mu<\frac{N}{2}-1$ and
\begin{align}
X_{k}^{(1),\frac{N}{2}-1,m_{1}} & =X_{k}^{(1),\frac{N}{2}-1}+F^{+,\frac{N}{2}}, & X_{k}^{(2),\frac{N}{2}-1,m_{1}} & =X_{k}^{(2),\frac{N}{2}-1},\nonumber \\
X_{k}^{(1),\frac{N}{2},m_{1}} & =X_{k}^{(1),\frac{N}{2}}+G^{+,\frac{N}{2}}, & X_{k}^{(2),\frac{N}{2},m_{1}} & =X_{k}^{(2),\frac{N}{2}}+H^{-,\frac{N}{2}},\nonumber \\
X_{k}^{(1),\frac{N}{2}-2,m_{1}} & =X_{k}^{(1),\frac{N}{2}-2}+H^{+,\frac{N}{2}-2}, & X_{k}^{(2),\frac{N}{2}-2,m_{1}} & =X_{k}^{(2),\frac{N}{2}-2},\\
X_{k}^{(2),\frac{N}{2}-1,m_{2}} & =X_{k}^{(2),\frac{N}{2}-1}+F^{+,\frac{N}{2}}, & X_{k}^{(1),\frac{N}{2}-1,m_{2}} & =X_{k}^{(1),\frac{N}{2}-1},\nonumber \\
X_{k}^{(2),\frac{N}{2},m_{2}} & =X_{k}^{(2),\frac{N}{2}}+G^{+,\frac{N}{2}}, & X_{k}^{(1),\frac{N}{2},m_{2}} & =X_{k}^{(1),\frac{N}{2}}+H^{-,\frac{N}{2}},\nonumber \\
X_{k}^{(2),\frac{N}{2}-2,m_{2}} & =X_{k}^{(2),\frac{N}{2}-2}+H^{+,\frac{N}{2}-2}, & X_{k}^{(1),\frac{N}{2}-2,m_{2}} & =X_{k}^{(1),\frac{N}{2}-2},\nonumber 
\end{align}
for $\mu=\frac{N}{2}-1$ and
\begin{align}
X_{k}^{(1),\frac{N}{2},m_{1}} & =X_{k}^{(1),\frac{N}{2}}+F^{+,\frac{N}{2}}, & X_{k}^{(2),\frac{N}{2},m_{1}} & =X_{k}^{(2),\frac{N}{2}}+F^{-,\frac{N}{2}},\nonumber \\
X_{k}^{(1),\frac{N}{2}-1,m_{1}} & =X_{k}^{(1),\frac{N}{2}-1}+H^{+,\frac{N}{2}-1}, & X_{k}^{(2),\frac{N}{2}-1,m_{1}} & =X_{k}^{(2),\frac{N}{2}-1}+H^{-,\frac{N}{2}-1},\nonumber \\
X_{k}^{(2),\frac{N}{2},m_{2}} & =X_{k}^{(2),\frac{N}{2}}+F^{+,\frac{N}{2}}, & X_{k}^{(1),\frac{N}{2},m_{2}} & =X_{k}^{(1),\frac{N}{2}}+F^{-,\frac{N}{2}},\\
X_{k}^{(2),\frac{N}{2}-1,m_{2}} & =X_{k}^{(2),\frac{N}{2}-1}+H^{+,\frac{N}{2}-1}, & X_{k}^{(1),\frac{N}{2}-1,m_{2}} & =X_{k}^{(1),\frac{N}{2}-1}+H^{-,\frac{N}{2}-1},\nonumber 
\end{align}
for $\mu=\frac{N}{2}$ and
\begin{align}
X_{k}^{(1),\frac{N-1}{2},m_{1}} & =X_{k}^{(1),\frac{N-1}{2}}+F^{+,\frac{N-1}{2}}, & X_{k}^{(2),\frac{N-1}{2},m_{1}} & =X_{k}^{(2),\frac{N-1}{2}}+H^{-,\frac{N-1}{2}},\nonumber \\
X_{k}^{(1),\frac{N-3}{2},m_{1}} & =X_{k}^{(1),\frac{N-3}{2}}+H^{+,\frac{N-3}{2}}, & X_{k}^{(2),\frac{N-3}{2},m_{1}} & =X_{k}^{(2),\frac{N-3}{2}},\nonumber \\
X_{k}^{(2),\frac{N-1}{2},m_{2}} & =X_{k}^{(2),\frac{N-1}{2}}+F^{+,\frac{N-1}{2}}, & X_{k}^{(1),\frac{N-1}{2},m_{2}} & =X_{k}^{(1),\frac{N-1}{2}}+H^{-,\frac{N-1}{2}},\\
X_{k}^{(2),\frac{N-3}{2},m_{2}} & =X_{k}^{(2),\frac{N-3}{2}}+H^{+,\frac{N-3}{2}}, & X_{k}^{(1),\frac{N-3}{2},m_{2}} & =X_{k}^{(1),\frac{N-3}{2}},\nonumber 
\end{align}
for $\mu=\frac{N-1}{2}$.

\section{Pair structure limit\label{sec:Pair-offshell}}

In this section we calculate the pair structure limits of the overlaps.
In the untwisted case we have achiral pair structure for which we
have to take the limit $w_{k}^{-,\nu}\to\pi^{a}(w_{k}^{+,\nu})=-w_{k}^{+,\nu}$.
In the twisted case we have chiral pair structure for which we have
to take the limit $w_{k}^{-,\nu}\to\pi^{c}(w_{k}^{+,\nu})=-w_{k}^{+,\nu}-\nu c$.

\subsection{Twisted case}

Let us start with the twisted case. Using the sum formula (\ref{eq:twSum})
and normalization (\ref{eq:normTw}), the normalized overlap reads
as
\begin{equation}
\mathbb{N}(\bar{w})=F(\bar{w}^{+},\bar{w}^{-})\sum_{\mathrm{part}(\bar{w})}G(\bar{s}_{\textsc{i}},\bar{s}_{\textsc{ii}}|\bar{t}_{\textsc{i}},\bar{t}_{\textsc{ii}})\prod_{\nu=1}^{N-1}\alpha_{\nu}^{(1)}(\bar{s}_{\textsc{ii}}^{\nu})\frac{\alpha_{\nu}^{(2)}(\bar{s}^{\nu})\alpha_{\nu}^{(2)}(\bar{t}_{\textsc{ii}}^{\nu})}{\alpha_{\nu}^{(2)}(\bar{w}^{+,\nu})},\label{eq:sumruleN}
\end{equation}
where
\begin{equation}
F(\bar{w}^{+},\bar{w}^{-})=\frac{1}{\prod_{\nu=1}^{N-1}\prod_{i\neq j}f(w_{i}^{+,\nu},w_{j}^{+,\nu})f(\bar{w}^{-,\nu},\bar{w}^{+,\nu})f(\bar{w}^{+,\nu+1},\bar{w}^{-,\nu})},\label{eq:Ff}
\end{equation}
and
\begin{multline}
G(\bar{s}_{\textsc{i}},\bar{s}_{\textsc{ii}}|\bar{t}_{\textsc{i}},\bar{t}_{\textsc{ii}})=\frac{\prod_{\nu=1}^{N-1}f(\bar{t}^{\nu},\bar{s}^{\nu})f(\bar{s}_{\textsc{i}}^{\nu},\bar{s}_{\textsc{ii}}^{\nu})f(\bar{t}_{\textsc{i}}^{\nu},\bar{t}_{\textsc{ii}}^{\nu})}{\prod_{\nu=1}^{N-2}f(\bar{t}^{\nu+1},\bar{s}^{\nu})}\frac{\prod_{\nu=1}^{N-2}f(\bar{s}^{\nu},\bar{s}^{\nu+1}+c)}{\prod_{\nu=1}^{N-2}f(\bar{s}_{\textsc{i}}^{\nu},\bar{s}_{\textsc{ii}}^{\nu+1}+c)f(\bar{t}_{\textsc{i}}^{\nu+1},\bar{t}_{\textsc{ii}}^{\nu})}\times\\
Z(\pi^{c}(\bar{s}_{\textsc{i}})|\bar{t}_{\textsc{i}})Z(\bar{t}_{\textsc{ii}}|\pi^{c}(\bar{s}_{\textsc{ii}})).
\end{multline}
Now we take the $w_{k}^{+,\nu}+w_{k}^{-,\nu}+\nu c\to0$ limit. Let
us define the sets $\bar{\omega}=\bar{w}\backslash\left\{ w_{k}^{+,\nu},w_{k}^{-,\nu}\right\} $,
$\bar{\omega}^{+}=\bar{w}^{+}\backslash\left\{ w_{k}^{+,\nu}\right\} $
and $\bar{\omega}^{-}=\bar{w}^{-}\backslash\left\{ w_{k}^{-,\nu}\right\} $.
We can see that the formal poles appear only for the partitions (see
equation (\ref{eq:poleZ})) for the partitions where
\begin{itemize}
\item $w_{k}^{-,\nu}\in\bar{t}_{\textsc{i}}^{\nu},w_{k}^{+,\nu}\in\bar{s}_{\textsc{i}}^{\nu}$
or 
\item $w_{k}^{-,\nu}\in\bar{t}_{\textsc{ii}}^{\nu},w_{k}^{+,\nu}\in\bar{s}_{\textsc{ii}}^{\nu}$
or 
\item $w_{k}^{-,\nu}\in\bar{s}_{\textsc{i}}^{\nu},w_{k}^{+,\nu}\in\bar{t}_{\textsc{i}}^{\nu}$
or
\item $w_{k}^{-,\nu}\in\bar{t}_{\textsc{ii}}^{\nu},w_{k}^{+,\nu}\in\bar{s}_{\textsc{ii}}^{\nu}$. 
\end{itemize}
Let us fix a partition as $\bar{\omega}=\bar{\sigma}_{\textsc{i}}\cup\bar{\sigma}_{\textsc{ii}}\cup\bar{\tau}_{\textsc{i}}\cup\bar{\tau}_{\textsc{ii}}$.
Let us get the G-term with $\bar{t}_{\textsc{i}}=\left\{ w_{k}^{-,\nu}\right\} \cup\bar{\tau}_{\textsc{i}},\bar{s}_{\textsc{i}}=\left\{ w_{k}^{+,\nu}\right\} \cup\bar{\sigma}_{\textsc{i}}$:
\begin{align}
G(\bar{s}_{\textsc{i}},\bar{s}_{\textsc{ii}}|\bar{t}_{\textsc{i}},\bar{t}_{\textsc{ii}})\Biggr|_{w_{k}^{-,\nu}\to\pi^{c}(w_{k}^{+,\nu})}\to G_{1} & =g(w_{k}^{-,\nu},-w_{k}^{+,\nu}-\nu c)f(w_{k}^{-,\nu},w_{k}^{+,\nu})\nonumber \\
 & \frac{f(\bar{\tau}^{\nu},w_{k}^{+,\nu})f(\bar{\sigma}_{\textsc{i}}^{\nu},w_{k}^{+,\nu})f(w_{k}^{+,\nu},\bar{\sigma}_{\textsc{ii}}^{\nu})}{f(\bar{\tau}^{\nu+1},w_{k}^{+,\nu})f(\bar{\sigma}_{\textsc{i}}^{\nu+1},w_{k}^{+,\nu})f(w_{k}^{+,\nu},\bar{\sigma}_{\textsc{ii}}^{\nu-1})}\times\\
 & \frac{f(w_{k}^{-,\nu},\bar{\sigma}^{\nu})f(\bar{\tau}_{\textsc{i}}^{\nu},w_{k}^{-,\nu})f(w_{k}^{-,\nu},\bar{\tau}_{\textsc{ii}}^{\nu})}{f(w_{k}^{-,\nu},\bar{\sigma}^{\nu-1})f(\bar{\tau}_{\textsc{i}}^{\nu+1},w_{k}^{-,\nu})f(w_{k}^{-,\nu},\bar{\tau}_{\textsc{ii}}^{\nu-1})}G(\bar{\sigma}_{\textsc{i}},\bar{\sigma}_{\textsc{ii}}|\bar{\tau}_{\textsc{i}},\bar{\tau}_{\textsc{ii}})+reg.\nonumber 
\end{align}
where we used (\ref{eq:poleZ}).  The $\alpha$ dependence for this
partition is
\begin{equation}
\prod_{\mu=1}^{N-1}\alpha_{\mu}^{(1)}(\bar{\sigma}_{\textsc{ii}}^{\mu})\frac{\alpha_{\mu}^{(2)}(\bar{\sigma}^{\mu})\alpha_{\mu}^{(2)}(\bar{\tau}_{\textsc{ii}}^{\mu})}{\alpha_{\mu}^{(2)}(\bar{\omega}^{+,\mu})}.
\end{equation}

For the partition $\bar{t}_{\textsc{ii}}=\left\{ w_{k}^{-,\nu}\right\} \cup\bar{\tau}_{\textsc{ii}},\bar{s}_{\textsc{ii}}=\left\{ w_{k}^{+,\nu}\right\} \cup\bar{\sigma}_{\textsc{ii}}$:
\begin{align}
G(\bar{s}_{\textsc{i}},\bar{s}_{\textsc{ii}}|\bar{t}_{\textsc{i}},\bar{t}_{\textsc{ii}})\Biggr|_{w_{k}^{-,\nu}\to\pi^{c}(w_{k}^{+,\nu})}\to G_{2} & =g(-w_{k}^{+,\nu}-\nu c,w_{k}^{-,\nu})f(w_{k}^{-,\nu},w_{k}^{+,\nu})\times\nonumber \\
 & \frac{f(\bar{\tau}^{\nu},w_{k}^{+,\nu})f(\bar{\sigma}_{\textsc{i}}^{\nu},w_{k}^{+,\nu})f(w_{k}^{+,\nu},\bar{\sigma}_{\textsc{ii}}^{\nu})}{f(\bar{\tau}^{\nu+1},w_{k}^{+,\nu})f(\bar{\sigma}_{\textsc{i}}^{\nu+1},w_{k}^{+,\nu})f(w_{k}^{+,\nu},\bar{\sigma}_{\textsc{ii}}^{\nu-1})}\times\\
 & \frac{f(w_{k}^{-,\nu},\bar{\sigma}^{\nu})f(\bar{\tau}_{\textsc{i}}^{\nu},w_{k}^{-,\nu})f(w_{k}^{-,\nu},\bar{\tau}_{\textsc{ii}}^{\nu})}{f(w_{k}^{-,\nu},\bar{\sigma}^{\nu-1})f(\bar{\tau}_{\textsc{i}}^{\nu+1},w_{k}^{-,\nu})f(w_{k}^{-,\nu},\bar{\tau}_{\textsc{ii}}^{\nu-1})}G(\bar{\sigma}_{\textsc{i}},\bar{\sigma}_{\textsc{ii}}|\bar{\tau}_{\textsc{i}},\bar{\tau}_{\textsc{ii}})+reg.\nonumber 
\end{align}
where we used (\ref{eq:poleZ}). The $\alpha$ dependence for this
partition is
\begin{equation}
\alpha_{\nu}^{(1)}(w_{k}^{+,\nu})\alpha_{\nu}^{(2)}(w_{k}^{-,\nu})\prod_{\mu=1}^{N-1}\alpha_{\mu}^{(1)}(\bar{\sigma}_{\textsc{ii}}^{\mu})\frac{\alpha_{\mu}^{(2)}(\bar{\sigma}^{\mu})\alpha_{\mu}^{(2)}(\bar{\tau}_{\textsc{ii}}^{\mu})}{\alpha_{\mu}^{(2)}(\bar{\omega}^{+,\mu})}.
\end{equation}
Let us take the sum of the two terms corresponding the previous two
partitions
\begin{multline}
G_{1}\prod_{\mu=1}^{N-1}\alpha_{\mu}^{(1)}(\bar{\sigma}_{\textsc{ii}}^{\mu})\frac{\alpha_{\mu}^{(2)}(\bar{\sigma}^{\mu})\alpha_{\mu}^{(2)}(\bar{\tau}_{\textsc{ii}}^{\mu})}{\alpha_{\mu}^{(2)}(\bar{\omega}^{+,\mu})}+G_{2}\alpha_{\nu}^{(1)}(w_{k}^{+,\nu})\alpha_{\nu}^{(2)}(w_{k}^{-,\nu})\prod_{\mu=1}^{N-1}\alpha_{\mu}^{(1)}(\bar{\sigma}_{\textsc{ii}}^{\mu})\frac{\alpha_{\mu}^{(2)}(\bar{\sigma}^{\mu})\alpha_{\mu}^{(2)}(\bar{\tau}_{\textsc{ii}}^{\mu})}{\alpha_{\mu}^{(2)}(\bar{\omega}^{+,\mu})}=\\
\left(G_{1}+G_{2}\alpha_{\nu}^{(1)}(w_{k}^{+,\nu})\alpha_{\nu}^{(2)}(w_{k}^{-,\nu})\right)\prod_{\mu=1}^{N-1}\alpha_{\mu}^{(1)}(\bar{\sigma}_{\textsc{ii}}^{\mu})\frac{\alpha_{\mu}^{(2)}(\bar{\sigma}^{\mu})\alpha_{\mu}^{(2)}(\bar{\tau}_{\textsc{ii}}^{\mu})}{\alpha_{\mu}^{(2)}(\bar{\omega}^{+,\mu})}.
\end{multline}
In this sum, the following expression appears
\begin{equation}
g(w_{k}^{-,\nu},-w_{k}^{+,\nu}-\nu c)(1-\alpha_{\nu}^{(1)}(w_{k}^{+,\nu})\alpha_{\nu}^{(2)}(w_{k}^{-,\nu}))\to-c\frac{d}{du}\log\alpha_{\nu}^{(1)}(u)\Biggr|_{u=w_{k}^{+,\nu}}=X_{k}^{(1),\nu},\label{eq:limX}
\end{equation}
therefore the previous sum simplifies as
\begin{multline}
\lim_{w_{k}^{-,\nu}=\pi^{c}(w_{k}^{+,\nu})}\left(G_{1}+G_{2}\alpha_{\nu}^{(1)}(w_{k}^{+,\nu})\alpha_{\nu}^{(2)}(w_{k}^{-,\nu})\right)\prod_{\mu=1}^{N-1}\alpha_{\mu}^{(1)}(\bar{\sigma}_{\textsc{ii}}^{\mu})\frac{\alpha_{\mu}^{(2)}(\bar{\sigma}^{\mu})\alpha_{\mu}^{(2)}(\bar{\tau}_{\textsc{ii}}^{\mu})}{\alpha_{\mu}^{(2)}(\bar{\omega}^{+,\mu})}=X_{k}^{(1),\nu}\times\\
f(w_{k}^{-,\nu},w_{k}^{+,\nu})\frac{f(\bar{\tau}^{\nu},w_{k}^{+,\nu})f(\bar{\sigma}_{\textsc{i}}^{\nu},w_{k}^{+,\nu})f(w_{k}^{+,\nu},\bar{\sigma}_{\textsc{ii}}^{\nu})}{f(\bar{\tau}^{\nu+1},w_{k}^{+,\nu})f(\bar{\sigma}_{\textsc{i}}^{\nu+1},w_{k}^{+,\nu})f(w_{k}^{+,\nu},\bar{\sigma}_{\textsc{ii}}^{\nu-1})}\frac{f(w_{k}^{-,\nu},\bar{\sigma}^{\nu})f(\bar{\tau}_{\textsc{i}}^{\nu},w_{k}^{-,\nu})f(w_{k}^{-,\nu},\bar{\tau}_{\textsc{ii}}^{\nu})}{f(w_{k}^{-,\nu},\bar{\sigma}^{\nu-1})f(\bar{\tau}_{\textsc{i}}^{\nu+1},w_{k}^{-,\nu})f(w_{k}^{-,\nu},\bar{\tau}_{\textsc{ii}}^{\nu-1})}\times\\
G(\bar{\sigma}_{\textsc{i}},\bar{\sigma}_{\textsc{ii}}|\bar{\tau}_{\textsc{i}},\bar{\tau}_{\textsc{ii}})\prod_{\mu=1}^{N-1}\alpha_{\mu}^{(1)}(\bar{\sigma}_{\textsc{ii}}^{\mu})\frac{\alpha_{\mu}^{(2)}(\bar{\sigma}^{\mu})\alpha_{\mu}^{(2)}(\bar{\tau}_{\textsc{ii}}^{\mu})}{\alpha_{\mu}^{(2)}(\bar{\omega}^{+,\mu})}+reg.
\end{multline}
 For the partition $\bar{t}_{\textsc{i}}=\left\{ w_{k}^{+,\nu}\right\} \cup\bar{\tau}_{\textsc{i}},\bar{s}_{\textsc{i}}=\left\{ w_{k}^{-,\nu}\right\} \cup\bar{\sigma}_{\textsc{i}}$:
\begin{align}
G(\bar{s}_{\textsc{i}},\bar{s}_{\textsc{ii}}|\bar{t}_{\textsc{i}},\bar{t}_{\textsc{ii}})\Biggr|_{w_{k}^{-,\nu}\to\pi^{c}(w_{k}^{+,\nu})}\to G_{3} & =g(w_{k}^{+,\nu},-w_{k}^{-,\nu}-\nu c)f(w_{k}^{+,\nu},w_{k}^{-,\nu})\times\nonumber \\
 & \frac{f(w_{k}^{+,\nu},\bar{\sigma}^{\nu})f(\bar{\tau}_{\textsc{i}}^{\nu},w_{k}^{+,\nu})f(w_{k}^{+,\nu},\bar{\tau}_{\textsc{ii}}^{\nu})}{f(w_{k}^{+,\nu},\bar{\sigma}^{\nu-1})f(\bar{\tau}_{\textsc{i}}^{\nu+1},w_{k}^{+,\nu})f(w_{k}^{+,\nu},\bar{\tau}_{\textsc{ii}}^{\nu-1})}\times\\
 & \frac{f(\bar{\sigma}_{\textsc{i}}^{\nu},w_{k}^{-,\nu})f(w_{k}^{-,\nu},\bar{\sigma}_{\textsc{ii}}^{\nu})f(\bar{\tau}^{\nu},w_{k}^{-,\nu})}{f(\bar{\sigma}_{\textsc{i}}^{\nu+1},w_{k}^{-,\nu})f(w_{k}^{-,\nu},\bar{\sigma}_{\textsc{ii}}^{\nu-1})f(\bar{\tau}^{\nu+1},w_{k}^{-,\nu})}G(\bar{\sigma}_{\textsc{i}},\bar{\sigma}_{\textsc{ii}}|\bar{\tau}_{\textsc{i}},\bar{\tau}_{\textsc{ii}})+reg.\nonumber 
\end{align}
where we used (\ref{eq:poleZ}). The $\alpha$ dependence for this
partition is
\begin{equation}
\frac{\alpha_{\nu}^{(2)}(w_{k}^{-,\nu})}{\alpha_{\nu}^{(2)}(w_{k}^{+,\nu})}\prod_{\mu=1}^{N-1}\alpha_{\mu}^{(1)}(\bar{\sigma}_{\textsc{ii}}^{\mu})\frac{\alpha_{\mu}^{(2)}(\bar{\sigma}^{\mu})\alpha_{\mu}^{(2)}(\bar{\tau}_{\textsc{ii}}^{\mu})}{\alpha_{\mu}^{(2)}(\bar{\omega}^{+,\mu})}.
\end{equation}

For the partition $\bar{t}_{\textsc{ii}}=\left\{ w_{k}^{+,\nu}\right\} \cup\bar{\tau}_{\textsc{ii}},\bar{s}_{\textsc{ii}}=\left\{ w_{k}^{-,\nu}\right\} \cup\bar{\sigma}_{\textsc{ii}}$:
\begin{align}
G(\bar{s}_{\textsc{i}},\bar{s}_{\textsc{ii}}|\bar{t}_{\textsc{i}},\bar{t}_{\textsc{ii}})\Biggr|_{w_{k}^{-,\nu}\to\pi^{c}(w_{k}^{+,\nu})}\to G_{4} & =g(-w_{k}^{-,\nu}-\nu c,w_{k}^{+,\nu})f(w_{k}^{+,\nu},w_{k}^{-,\nu})\times\nonumber \\
 & \frac{f(w_{k}^{+,\nu},\bar{\sigma}^{\nu})f(w_{k}^{+,\nu},\bar{\tau}_{\textsc{ii}}^{\nu})f(\bar{\tau}_{\textsc{i}}^{\nu},w_{k}^{+,\nu})}{f(w_{k}^{+,\nu},\bar{\sigma}^{\nu-1})f(w_{k}^{+,\nu},\bar{\tau}_{\textsc{ii}}^{\nu-1})f(\bar{\tau}_{\textsc{i}}^{\nu+1},w_{k}^{+,\nu})}\times\\
 & \frac{f(w_{k}^{-,\nu},\bar{\sigma}_{\textsc{ii}}^{\nu})f(\bar{\sigma}_{\textsc{i}}^{\nu},w_{k}^{-,\nu})f(\bar{\tau}^{\nu},w_{k}^{-,\nu})}{f(w_{k}^{-,\nu},\bar{\sigma}_{\textsc{ii}}^{\nu-1})f(\bar{\sigma}_{\textsc{i}}^{\nu+1},w_{k}^{-,\nu})f(\bar{\tau}^{\nu+1},w_{k}^{-,\nu})}G(\bar{\sigma}_{\textsc{i}},\bar{\sigma}_{\textsc{ii}}|\bar{\tau}_{\textsc{i}},\bar{\tau}_{\textsc{ii}})+reg.\nonumber 
\end{align}
where we used (\ref{eq:poleZ}). The $\alpha$ dependence for this
partition is
\begin{equation}
\alpha_{\nu}^{(1)}(w_{k}^{-,\nu})\alpha_{\nu}^{(2)}(w_{k}^{-,\nu})\prod_{\mu=1}^{N-1}\alpha_{\mu}^{(1)}(\bar{\sigma}_{\textsc{ii}}^{\mu})\frac{\alpha_{\mu}^{(2)}(\bar{\sigma}^{\mu})\alpha_{\mu}^{(2)}(\bar{\tau}_{\textsc{ii}}^{\mu})}{\alpha_{\mu}^{(2)}(\bar{\omega}^{+,\mu})}.
\end{equation}

An analogous way we can obtain that 
\begin{align}
\lim_{w_{k}^{-,\nu}=\pi^{c}(w_{k}^{+,\nu})}= & \left(G_{3}\frac{\alpha_{\nu}^{(2)}(w_{k}^{-,\nu})}{\alpha_{\nu}^{(2)}(w_{k}^{+,\nu})}+G_{4}\alpha_{\nu}^{(1)}(w_{k}^{-,\nu})\alpha_{\nu}^{(2)}(w_{k}^{-,\nu})\right)\prod_{\mu=1}^{N-1}\alpha_{\mu}^{(1)}(\bar{\sigma}_{\textsc{ii}}^{\mu})\frac{\alpha_{\mu}^{(2)}(\bar{\sigma}^{\mu})\alpha_{\mu}^{(2)}(\bar{\tau}_{\textsc{ii}}^{\mu})}{\alpha_{\mu}^{(2)}(\bar{\omega}^{+,\mu})}\nonumber \\
 & \alpha_{\nu}(w_{k}^{-,\nu})X_{k}^{(2),\nu}f(w_{k}^{+,\nu},w_{k}^{-,\nu})\times\nonumber \\
 & \frac{f(w_{k}^{+,\nu},\bar{\sigma}^{\nu})f(w_{k}^{+,\nu},\bar{\tau}_{\textsc{ii}}^{\nu})f(\bar{\tau}_{\textsc{i}}^{\nu},w_{k}^{+,\nu})}{f(w_{k}^{+,\nu},\bar{\sigma}^{\nu-1})f(w_{k}^{+,\nu},\bar{\tau}_{\textsc{ii}}^{\nu-1})f(\bar{\tau}_{\textsc{i}}^{\nu+1},w_{k}^{+,\nu})}\frac{f(w_{k}^{-,\nu},\bar{\sigma}_{\textsc{ii}}^{\nu})f(\bar{\sigma}_{\textsc{i}}^{\nu},w_{k}^{-,\nu})f(\bar{\tau}^{\nu},w_{k}^{-,\nu})}{f(w_{k}^{-,\nu},\bar{\sigma}_{\textsc{ii}}^{\nu-1})f(\bar{\sigma}_{\textsc{i}}^{\nu+1},w_{k}^{-,\nu})f(\bar{\tau}^{\nu+1},w_{k}^{-,\nu})}\times\\
 & G(\bar{\sigma}_{\textsc{i}},\bar{\sigma}_{\textsc{ii}}|\bar{\tau}_{\textsc{i}},\bar{\tau}_{\textsc{ii}})\prod_{\mu=1}^{N-1}\alpha_{\mu}^{(1)}(\bar{\sigma}_{\textsc{ii}}^{\mu})\frac{\alpha_{\mu}^{(2)}(\bar{\sigma}^{\mu})\alpha_{\mu}^{(2)}(\bar{\tau}_{\textsc{ii}}^{\mu})}{\alpha_{\mu}^{(2)}(\bar{\omega}^{+,\mu})}+reg.\nonumber 
\end{align}
Substituting back we obtain that
\begin{align}
\mathbb{N}(\bar{w}) & \Biggr|_{w_{k}^{-,\nu}=\pi^{c}(w_{k}^{+,\nu})}\to\nonumber \\
 & X_{k}^{(1),\nu}f(w_{k}^{-,\nu},w_{k}^{+,\nu})F(\bar{w}^{+},\bar{w}^{-})\frac{f(\bar{\omega}^{\nu},w_{k}^{+,\nu})f(\bar{\omega}^{\nu},w_{k}^{-,\nu})}{f(\bar{\omega}^{\nu+1},w_{k}^{+,\nu})f(\bar{\omega}^{\nu+1},w_{k}^{-,\nu})}\times\nonumber \\
 & \sum_{\mathrm{part}(\bar{\omega})}\frac{f(w_{k}^{+,\nu},\bar{\sigma}_{\textsc{ii}}^{\nu})f(\bar{\sigma}_{\textsc{ii}}^{\nu+1},w_{k}^{+,\nu})}{f(\bar{\sigma}_{\textsc{ii}}^{\nu},w_{k}^{+,\nu})f(w_{k}^{+,\nu},\bar{\sigma}_{\textsc{ii}}^{\nu-1})}\frac{f(w_{k}^{-,\nu},\bar{\sigma}^{\nu})f(\bar{\sigma}^{\nu+1},w_{k}^{-,\nu})}{f(\bar{\sigma}^{\nu},w_{k}^{-,\nu})f(w_{k}^{-,\nu},\bar{\sigma}^{\nu-1})}\frac{f(w_{k}^{-,\nu},\bar{\tau}_{\textsc{ii}}^{\nu})f(\bar{\tau}_{\textsc{ii}}^{\nu+1},w_{k}^{-,\nu})}{f(\bar{\tau}_{\textsc{ii}}^{\nu},w_{k}^{-,\nu})f(w_{k}^{-,\nu},\bar{\tau}_{\textsc{ii}}^{\nu-1})}\times\nonumber \\
 & \qquad\qquad\qquad\qquad\qquad\qquad\qquad\times G(\bar{\sigma}_{\textsc{i}},\bar{\sigma}_{\textsc{ii}}|\bar{\tau}_{\textsc{i}},\bar{\tau}_{\textsc{ii}})\prod_{\nu=1}^{N-1}\alpha_{\nu}^{(1)}(\bar{\sigma}_{\textsc{ii}}^{\nu})\frac{\alpha_{\nu}^{(2)}(\bar{\sigma}^{\nu})\alpha_{\nu}^{(2)}(\bar{\tau}_{\textsc{ii}}^{\nu})}{\alpha_{\nu}^{(2)}(\bar{\omega}^{+,\nu})}+\\
+ & X_{k}^{(2),\nu}f(w_{k}^{+,\nu},w_{k}^{-,\nu})F(\bar{w}^{+},\bar{w}^{-})\frac{f(\bar{\omega}^{\nu},w_{k}^{+,\nu})f(\bar{\omega}^{\nu},w_{k}^{-,\nu})}{f(\bar{\omega}^{\nu+1},w_{k}^{+,\nu})f(\bar{\omega}^{\nu+1},w_{k}^{-,\nu})}\alpha_{\nu}(w_{k}^{-,\nu})\times\nonumber \\
 & \sum_{\mathrm{part}(\bar{\omega})}\frac{f(w_{k}^{-,\nu},\bar{\sigma}_{\textsc{ii}}^{\nu})f(\bar{\sigma}_{\textsc{ii}}^{\nu+1},w_{k}^{-,\nu})}{f(\bar{\sigma}_{\textsc{ii}}^{\nu},w_{k}^{-,\nu})f(w_{k}^{-,\nu},\bar{\sigma}_{\textsc{ii}}^{\nu-1})}\frac{f(w_{k}^{+,\nu},\bar{\sigma}^{\nu})f(\bar{\sigma}^{\nu+1},w_{k}^{+,\nu})}{f(\bar{\sigma}^{\nu},w_{k}^{+,\nu})f(w_{k}^{+,\nu},\bar{\sigma}^{\nu-1})}\frac{f(w_{k}^{-,\nu},\bar{\tau}_{\textsc{ii}}^{\nu})f(\bar{\tau}_{\textsc{ii}}^{\nu+1},w_{k}^{-,\nu})}{f(\bar{\tau}_{\textsc{ii}}^{\nu},w_{k}^{-,\nu})f(w_{k}^{-,\nu},\bar{\tau}_{\textsc{ii}}^{\nu-1})}\times\nonumber \\
 & \qquad\qquad\qquad\qquad\qquad\qquad\qquad\times G(\bar{\sigma}_{\textsc{i}},\bar{\sigma}_{\textsc{ii}}|\bar{\tau}_{\textsc{i}},\bar{\tau}_{\textsc{ii}})\prod_{\nu=1}^{N-1}\alpha_{\nu}^{(1)}(\bar{\sigma}_{\textsc{ii}}^{\nu})\frac{\alpha_{\nu}^{(2)}(\bar{\sigma}^{\nu})\alpha_{\nu}^{(2)}(\bar{\tau}_{\textsc{ii}}^{\nu})}{\alpha_{\nu}^{(2)}(\bar{\omega}^{+,\nu})}+\tilde{\mathbb{N}},\nonumber 
\end{align}
where $\tilde{\mathbb{N}}$ is do not depend on $X_{k}^{(1),\nu}$
and $X_{k}^{(2),\nu}$. Using the following form of the $F$-term
(\ref{eq:Fterm})
\begin{equation}
F(\bar{w}^{+},\bar{w}^{-})=\frac{1}{f(w_{k}^{-,\nu},w_{k}^{+,\nu})f(\bar{\omega}^{\nu},w_{k}^{+,\nu})f(\bar{\omega}^{\nu},w_{k}^{-,\nu})}\frac{f(\bar{\omega}^{+,\nu},w_{k}^{-,\nu})f(w_{k}^{-,\nu},\bar{\omega}^{+,\nu-1})}{f(w_{k}^{-,\nu},\bar{\omega}^{+,\nu})f(\bar{\omega}^{+,\nu+1},w_{k}^{-,\nu})}F(\bar{\omega}^{+},\bar{\omega}^{-}),\label{eq:Fterm}
\end{equation}
and the Bethe equation
\begin{equation}
\alpha_{\nu}(w_{k}^{-,\nu})=\frac{f(w_{k}^{-,\nu},w_{k}^{+,\nu})}{f(w_{k}^{+,\nu},w_{k}^{-,\nu})}\frac{f(w_{k}^{-,\nu},\bar{\omega}^{\nu})}{f(\bar{\omega}^{\nu},w_{k}^{-,\nu})}\frac{f(\bar{\omega}^{\nu+1},w_{k}^{-,\nu})}{f(w_{k}^{-,\nu},\bar{\omega}^{\nu-1})},
\end{equation}
we obtain that

\begin{align}
\mathbb{N}(\bar{w}) & \Biggr|_{w_{k}^{-,\nu}=\pi^{c}(w_{k}^{+,\nu})}\to\nonumber \\
 & X_{k}^{(1),\nu}F(\bar{\omega}^{+},\bar{\omega}^{-})\frac{f(\bar{\omega}^{+,\nu},w_{k}^{-,\nu})f(w_{k}^{-,\nu},\bar{\omega}^{+,\nu-1})}{f(w_{k}^{-,\nu},\bar{\omega}^{+,\nu})f(\bar{\omega}^{+,\nu+1},w_{k}^{-,\nu})}\times\nonumber \\
 & \sum_{\mathrm{part}(\bar{\omega})}\frac{f(w_{k}^{+,\nu},\bar{\sigma}_{\textsc{ii}}^{\nu})f(\bar{\sigma}_{\textsc{ii}}^{\nu+1},w_{k}^{+,\nu})}{f(\bar{\sigma}_{\textsc{ii}}^{\nu},w_{k}^{+,\nu})f(w_{k}^{+,\nu},\bar{\sigma}_{\textsc{ii}}^{\nu-1})}\frac{f(w_{k}^{-,\nu},\bar{\sigma}^{\nu})f(\bar{\sigma}^{\nu+1},w_{k}^{-,\nu})}{f(\bar{\sigma}^{\nu},w_{k}^{-,\nu})f(w_{k}^{-,\nu},\bar{\sigma}^{\nu-1})}\frac{f(w_{k}^{-,\nu},\bar{\tau}_{\textsc{ii}}^{\nu})f(\bar{\tau}_{\textsc{ii}}^{\nu+1},w_{k}^{-,\nu})}{f(\bar{\tau}_{\textsc{ii}}^{\nu},w_{k}^{-,\nu})f(w_{k}^{-,\nu},\bar{\tau}_{\textsc{ii}}^{\nu-1})}\nonumber \\
 & \qquad\qquad\qquad\qquad\qquad\qquad\qquad\times G(\bar{\sigma}_{\textsc{i}},\bar{\sigma}_{\textsc{ii}}|\bar{\tau}_{\textsc{i}},\bar{\tau}_{\textsc{ii}})\prod_{\nu=1}^{N-1}\alpha_{\nu}^{(1)}(\bar{\sigma}_{\textsc{ii}}^{\nu})\frac{\alpha_{\nu}^{(2)}(\bar{\sigma}^{\nu})\alpha_{\nu}^{(2)}(\bar{\tau}_{\textsc{ii}}^{\nu})}{\alpha_{\nu}^{(2)}(\bar{\omega}^{+,\nu})}+\\
+ & X_{k}^{(2),\nu}F(\bar{\omega}^{+},\bar{\omega}^{-})\frac{f(\bar{\omega}^{+,\nu},w_{k}^{+,\nu})}{f(w_{k}^{+,\nu},\bar{\omega}^{+,\nu})}\frac{f(w_{k}^{+,\nu},\bar{\omega}^{+,\nu-1})}{f(\bar{\omega}^{+,\nu+1},w_{k}^{+,\nu})}\times\nonumber \\
 & \sum_{\mathrm{part}(\bar{\omega})}\frac{f(w_{k}^{-,\nu},\bar{\sigma}_{\textsc{ii}}^{\nu})f(\bar{\sigma}_{\textsc{ii}}^{\nu+1},w_{k}^{-,\nu})}{f(\bar{\sigma}_{\textsc{ii}}^{\nu},w_{k}^{-,\nu})f(w_{k}^{-,\nu},\bar{\sigma}_{\textsc{ii}}^{\nu-1})}\frac{f(w_{k}^{+,\nu},\bar{\sigma}^{\nu})f(\bar{\sigma}^{\nu+1},w_{k}^{+,\nu})}{f(\bar{\sigma}^{\nu},w_{k}^{+,\nu})f(w_{k}^{+,\nu},\bar{\sigma}^{\nu-1})}\frac{f(w_{k}^{-,\nu},\bar{\tau}_{\textsc{ii}}^{\nu})f(\bar{\tau}_{\textsc{ii}}^{\nu+1},w_{k}^{-,\nu})}{f(\bar{\tau}_{\textsc{ii}}^{\nu},w_{k}^{-,\nu})f(w_{k}^{-,\nu},\bar{\tau}_{\textsc{ii}}^{\nu-1})}\nonumber \\
 & \qquad\qquad\qquad\qquad\qquad\qquad\qquad\times G(\bar{\sigma}_{\textsc{i}},\bar{\sigma}_{\textsc{ii}}|\bar{\tau}_{\textsc{i}},\bar{\tau}_{\textsc{ii}})\prod_{\nu=1}^{N-1}\alpha_{\nu}^{(1)}(\bar{\sigma}_{\textsc{ii}}^{\nu})\frac{\alpha_{\nu}^{(2)}(\bar{\sigma}^{\nu})\alpha_{\nu}^{(2)}(\bar{\tau}_{\textsc{ii}}^{\nu})}{\alpha_{\nu}^{(2)}(\bar{\omega}^{+,\nu})}+\tilde{\mathbb{N}},\nonumber 
\end{align}
where we also used the identities
\begin{equation}
f(-u,-v)=f(v,u),\qquad f(u,v+c)=f(u-c,v)=\frac{1}{f(v,u)}.
\end{equation}
This limit can be simplified as
\begin{align}
\mathbb{N}(\bar{w}) & \Biggr|_{w_{k}^{-,\nu}=\pi^{c}(w_{k}^{+,\nu})}\to\nonumber \\
 & +X_{k}^{(1),\nu}F(\bar{\omega}^{+},\bar{\omega}^{-})\times\sum_{\mathrm{part}(\bar{\omega})}G(\bar{\sigma}_{\textsc{i}},\bar{\sigma}_{\textsc{ii}}|\bar{\tau}_{\textsc{i}},\bar{\tau}_{\textsc{ii}})\prod_{\nu=1}^{N-1}\alpha_{\nu}^{(1),mod_{1}}(\bar{\sigma}_{\textsc{ii}}^{\nu})\frac{\alpha_{\nu}^{(2),mod_{1}}(\bar{\sigma}^{\nu})\alpha_{\nu}^{(2),mod_{1}}(\bar{\tau}_{\textsc{ii}}^{\nu})}{\alpha_{\nu}^{(2),mod_{1}}(\bar{\omega}^{+,\nu})}+\nonumber \\
 & +X_{k}^{(2),\nu}F(\bar{\omega}^{+},\bar{\omega}^{-})\times\sum_{\mathrm{part}(\bar{\omega})}G(\bar{\sigma}_{\textsc{i}},\bar{\sigma}_{\textsc{ii}}|\bar{\tau}_{\textsc{i}},\bar{\tau}_{\textsc{ii}})\prod_{\nu=1}^{N-1}\alpha_{\nu}^{(1),mod_{2}}(\bar{\sigma}_{\textsc{ii}}^{\nu})\frac{\alpha_{\nu}^{(2),mod_{2}}(\bar{\sigma}^{\nu})\alpha_{\nu}^{(2),mod_{2}}(\bar{\tau}_{\textsc{ii}}^{\nu})}{\alpha_{\nu}^{(2),mod_{2}}(\bar{\omega}^{+,\nu})}+\tilde{\mathbb{N}},\label{eq:Nlim}
\end{align}
where we introduced the modified $\alpha$ variables as 
\begin{align}
\alpha_{\nu}^{(1),mod_{1}}(u) & =\alpha_{\nu}^{(1)}(u)\frac{f(w_{k}^{+,\nu},u)}{f(u,w_{k}^{+,\nu})}, & \alpha_{\nu}^{(2),mod_{1}}(u) & =\alpha_{\nu}^{(2)}(u)\frac{f(w_{k}^{-,\nu},u)}{f(u,w_{k}^{-,\nu})},\nonumber \\
\alpha_{\nu+1}^{(1),mod_{1}}(u) & =\alpha_{\nu+1}^{(1)}(u)f(u,w_{k}^{+,\nu}), & \alpha_{\nu+1}^{(2),mod_{1}}(u) & =\alpha_{\nu+1}^{(2)}(u)f(u,w_{k}^{-,\nu}),\\
\alpha_{\nu-1}^{(1),mod_{1}}(u) & =\alpha_{\nu-1}^{(1)}(u)\frac{1}{f(w_{k}^{+,\nu},u)}, & \alpha_{\nu-1}^{(2),mod_{1}}(u) & =\alpha_{\nu-1}^{(2)}(u)\frac{1}{f(w_{k}^{-,\nu},u)},\nonumber 
\end{align}
and
\begin{align}
\alpha_{\nu}^{(2),mod_{2}}(u) & =\alpha_{\nu}^{(2)}(u)\frac{f(w_{k}^{+,\nu},u)}{f(u,w_{k}^{+,\nu})}, & \alpha_{\nu}^{(1),mod_{2}}(u) & =\alpha_{\nu}^{(1)}(u)\frac{f(w_{k}^{-,\nu},u)}{f(u,w_{k}^{-,\nu})},\nonumber \\
\alpha_{\nu+1}^{(2),mod_{2}}(u) & =\alpha_{\nu+1}^{(2)}(u)f(u,w_{k}^{+,\nu}), & \alpha_{\nu+1}^{(1),mod_{2}}(u) & =\alpha_{\nu+1}^{(1)}(u)f(u,w_{k}^{-,\nu}),\\
\alpha_{\nu-1}^{(2),mod_{2}}(u) & =\alpha_{\nu-1}^{(2)}(u)\frac{1}{f(w_{k}^{+,\nu},u)}, & \alpha_{\nu-1}^{(1),mod_{2}}(u) & =\alpha_{\nu-1}^{(1)}(u)\frac{1}{f(w_{k}^{-,\nu},u)},\nonumber 
\end{align}
and $\alpha_{\mu}^{(i),mod_{1}}(u)=\alpha_{\mu}^{(i),mod_{2}}(u)=\alpha_{\mu}(u)$
for $|\mu-\nu|>1$ and $i=1,2$. In the rhs of (\ref{eq:Nlim}) we
can recognize the sum rule of the normalized overlap (\ref{eq:sumruleN})
therefore the $X_{k}^{(1),\nu}X_{k}^{(2),\nu}$ dependence simplifies
as
\begin{equation}
\mathbb{N}(\bar{w})\Biggr|_{w_{k}^{-,\nu}=\pi^{c}(w_{k}^{+,\nu})}\to X_{k}^{(1),\nu}\mathbb{N}(\bar{\omega})^{mod_{1}}+X_{k}^{(2),\nu}\mathbb{N}(\bar{\omega})^{mod_{2}}+\tilde{\mathbb{N}}.\label{eq:recN}
\end{equation}

\subsection{Untwisted case}

Let us continue with the untwisted case. Using the sum formula (\ref{eq:twSum})
and normalization (\ref{eq:normTw}), the normalized overlap reads
as
\begin{equation}
\mathbb{N}(\bar{w})=F(\bar{w}^{+},\bar{w}^{-})\sum_{\mathrm{part}(\bar{w})}G(\bar{s}_{\textsc{i}},\bar{s}_{\textsc{ii}}|\bar{t}_{\textsc{i}},\bar{t}_{\textsc{ii}})\prod_{\nu=1}^{N-1}\alpha_{\nu}^{(1)}(\bar{s}_{\textsc{ii}}^{\nu})\frac{\alpha_{\nu}^{(2)}(\bar{s}^{\nu})\alpha_{\nu}^{(2)}(\bar{t}_{\textsc{ii}}^{\nu})}{\alpha_{\nu}^{(2)}(\bar{w}^{+,\nu})},\label{eq:sumruleN-1}
\end{equation}
where
\begin{equation}
F(\bar{w}^{+},\bar{w}^{-})=\begin{cases}
\frac{\prod_{\nu=1}^{\frac{N}{2}-1}f(\bar{w}^{+,\nu+1},\bar{w}^{+,\nu})\left[f(\bar{w}^{-,\frac{N}{2}},\bar{w}^{+,\frac{N}{2}-1})\right]^{2}}{\prod_{\nu=1}^{\frac{N}{2}}\prod_{i\neq j}f(w_{i}^{+,\nu},w_{j}^{+,\nu})f(\bar{w}^{-,\frac{N}{2}},\bar{w}^{+,\frac{N}{2}})}, & \text{for even }N,\\
\frac{\prod_{\nu=1}^{\frac{N-1}{2}}f(\bar{w}^{+,\nu+1},\bar{w}^{+,\nu})}{\prod_{\nu=1}^{\frac{N-1}{2}}\prod_{i\neq j}f(w_{i}^{+,\nu},w_{j}^{+,\nu})}, & \text{for odd }N,
\end{cases}
\end{equation}
and
\begin{equation}
G(\bar{s}_{\textsc{i}},\bar{s}_{\textsc{ii}}|\bar{t}_{\textsc{i}},\bar{t}_{\textsc{ii}})=\frac{\prod_{k=1}^{N-1}f(\bar{t}^{k},\bar{s}^{k})f(\bar{s}_{\textsc{i}}^{k},\bar{s}_{\textsc{ii}}^{k})f(\bar{t}_{\textsc{i}}^{k},\bar{t}_{\textsc{ii}}^{k})}{\prod_{k=1}^{N-2}f(\bar{t}^{k+1},\bar{s}^{k})f(\bar{s}_{\textsc{i}}^{k+1},\bar{s}_{\textsc{ii}}^{k})f(\bar{t}_{\textsc{i}}^{k+1},\bar{t}_{\textsc{ii}}^{k})}Z(\pi^{a}(\bar{s}_{\textsc{i}})|\bar{t}_{\textsc{i}})Z(\bar{t}_{\textsc{ii}}|\pi^{a}(\bar{s}_{\textsc{ii}}).
\end{equation}
We can repeat the calculation of the previous section for limit $w_{k}^{+,\nu}+w_{k}^{-,\nu}\to0$
of the untwisted normalized overlap. The calculation is long but straightforward
(there is nothing new in it, compared to the previous one). At the
end we obtain the following formula
\begin{equation}
\mathbb{N}(\bar{w})\Biggr|_{w_{k}^{-,\nu}\to\pi^{a}(w_{k}^{+,\nu})}\to X_{k}^{(1),\nu}\mathbb{N}(\bar{\omega})^{mod_{1}}+X_{k}^{(2),\nu}\mathbb{N}(\bar{\omega})^{mod_{2}}+\tilde{\mathbb{N}},
\end{equation}
where $\tilde{\mathbb{N}}$ does not depend on $X_{k}^{(1),\nu},X_{k}^{(2),\nu}$
and $\mathbb{N}(\bar{\omega})^{mod_{1}}$ , $\mathbb{N}(\bar{\omega})^{mod_{2}}$
contain the following modified $\alpha$-s:
\begin{align}
\alpha_{\nu}^{(1),mod_{1}}(u) & =\alpha_{\nu}^{(1)}(u)\frac{f(w_{k}^{+,\nu},u)}{f(u,w_{k}^{+,\nu})}, & \alpha_{\nu}^{(2),mod_{2}}(u) & =\alpha_{\nu}^{(2)}(u)\frac{f(w_{k}^{+,\nu},u)}{f(u,w_{k}^{+,\nu})},\nonumber \\
\alpha_{\nu+1}^{(1),mod_{1}}(u) & =\alpha_{\nu+1}^{(1)}(u)f(u,w_{k}^{+,\nu}), & \alpha_{\nu+1}^{(2),mod_{2}}(u) & =\alpha_{\nu+1}^{(2)}(u)f(u,w_{k}^{+,\nu}),\\
\alpha_{\nu-1}^{(1),mod_{1}}(u) & =\alpha_{\nu-1}^{(1)}(u)\frac{1}{f(w_{k}^{+,\nu},u)}, & \alpha_{\nu-1}^{(2),mod_{2}}(u) & =\alpha_{\nu-1}^{(2)}(u)\frac{1}{f(w_{k}^{+,\nu},u)},\nonumber 
\end{align}
and
\begin{align}
\alpha_{N-\nu}^{(2),mod_{1}}(u) & =\alpha_{N-\nu}^{(2)}(u)\frac{f(w_{k}^{-,\nu},u)}{f(u,w_{k}^{-,\nu})}, & \alpha_{N-\nu}^{(1),mod_{2}}(u) & =\alpha_{N-\nu}^{(1)}(u)\frac{f(w_{k}^{-,\nu},u)}{f(u,w_{k}^{-,\nu})},\nonumber \\
\alpha_{N-\nu+1}^{(2),mod_{1}}(u) & =\alpha_{N-\nu+1}^{(2)}(u)f(u,w_{k}^{-,\nu}), & \alpha_{N-\nu+1}^{(1),mod_{2}}(u) & =\alpha_{N-\nu+1}^{(1)}(u)f(u,w_{k}^{-,\nu}),\\
\alpha_{\nu-1}^{(2),mod_{1}}(u) & =\alpha_{N-\nu-1}^{(2)}(u)\frac{1}{f(w_{k}^{-,\nu},u)}, & \alpha_{N-\nu-1}^{(1),mod_{2}}(u) & =\alpha_{N-\nu-1}^{(1)}(u)\frac{1}{f(w_{k}^{-,\nu},u)},\nonumber 
\end{align}
and $\alpha_{\mu}^{(i),mod_{1}}(u)=\alpha_{\mu}^{(i),mod_{2}}(u)=\alpha_{\mu}(u)$
for $|\mu-\nu|>1$ and $|\mu+\nu-N|>1$. 

\section{Normalized overlaps and the Korepin criteria\label{sec:Normalized-overlaps-and}}

In this section we show that the normalized overlaps $\mathbb{N}(\bar{w})$
satisfy the Korepin criteria. Let us start with the twisted case.
The criteria \ref{enum:prop1} is obviously true since the Bethe states
and the normalization factors in (\ref{eq:normTw}) are symmetric
over the replacement $w_{j}^{+,\mu}\leftrightarrow w_{k}^{+,\mu}$.
The properties \ref{enum:prop2},\ref{enum:prop4} follow from (\ref{eq:recN}).
The proof that the normalized crosscap overlaps satisfy the criteria
\ref{enum:prop5} is the same as the proofs for the scalar products
\cite{Hutsalyuk:2017way} and the boundary state overlaps \cite{Gombor:2021hmj}.
The proof is based on the fact that on-shell overlaps are non-zero
only for Bethe states with pair structure, i.e., the on-shell overlaps
vanish for general Bethe states.

Only the criteria \ref{enum:prop3} remains to prove. We can substitute
to the sum rule (\ref{eq:sumruleN}). For the the magnon state $\bar{w}^{\nu}=\left\{ w_{1}^{\nu},w_{2}^{\nu}\right\} $
and $\bar{w}^{\mu}=\emptyset$ for $\mu\neq\nu$, the sum formula
simplifies as 
\begin{align}
\mathbb{N}(\left\{ w_{1}^{\nu},w_{2}^{\nu}\right\} ) & =\frac{1}{f(w_{2}^{\nu},w_{1}^{\nu})}G(\left\{ w_{1}^{\nu}\right\} ,\emptyset|\left\{ w_{2}^{\nu}\right\} ,\emptyset)+\nonumber \\
 & \frac{1}{f(w_{2}^{\nu},w_{1}^{\nu})}G(\emptyset,\left\{ w_{1}^{\nu}\right\} |\emptyset,\left\{ w_{2}^{\nu}\right\} )\alpha_{\nu}^{(1)}(w_{1}^{\nu})\alpha_{\nu}^{(2)}(w_{2}^{\nu})+\nonumber \\
 & \frac{1}{f(w_{2}^{\nu},w_{1}^{\nu})}G(\left\{ w_{2}^{\nu}\right\} ,\emptyset|\left\{ w_{1}^{\nu}\right\} ,\emptyset)\frac{\alpha_{\nu}^{(2)}(w_{2}^{\nu})}{\alpha_{\nu}^{(2)}(w_{1}^{\nu})}+\\
 & \frac{1}{f(w_{2}^{\nu},w_{1}^{\nu})}G(\emptyset,\left\{ w_{2}^{\nu}\right\} |\emptyset,\left\{ w_{1}^{\nu}\right\} )\alpha_{\nu}^{(1)}(w_{2}^{\nu})\alpha_{\nu}^{(2)}(w_{2}^{\nu}),\nonumber 
\end{align}
and
\begin{align}
G(\left\{ w_{1}^{\nu}\right\} ,\emptyset|\left\{ w_{2}^{\nu}\right\} ,\emptyset) & =f(w_{2}^{\nu},w_{1}^{\nu})Z(-w_{1}^{\nu}-\nu c|w_{2}^{\nu})Z(\emptyset|\emptyset)=f(w_{2}^{\nu},w_{1}^{\nu})g(w_{2}^{\nu},-w_{1}^{\nu}-\nu c),\\
G(\emptyset,\left\{ w_{1}^{\nu}\right\} |\emptyset,\left\{ w_{2}^{\nu}\right\} ) & =f(w_{2}^{\nu},w_{1}^{\nu})Z(\emptyset|\emptyset)Z(w_{2}^{\nu}|-w_{1}^{\nu}-\nu c)=f(w_{2}^{\nu},w_{1}^{\nu})g(-w_{1}^{\nu}-\nu c,w_{2}^{\nu}),
\end{align}
where we used that
\begin{equation}
Z(\emptyset|\emptyset)=1,\qquad Z(\left\{ s\right\} ,\left\{ t\right\} )=g(t,s).
\end{equation}
Substituting back we obtain that
\begin{align}
\mathbb{N}(\left\{ w_{1}^{\nu},w_{2}^{\nu}\right\} ) & =g(w_{2}^{\nu},-w_{1}^{\nu}-\nu c)(1-\alpha_{\nu}^{(1)}(w_{1}^{\nu})\alpha_{\nu}^{(2)}(w_{2}^{\nu}))+\nonumber \\
 & \frac{f(w_{1}^{\nu},w_{2}^{\nu})}{f(w_{2}^{\nu},w_{1}^{\nu})}\frac{\alpha_{\nu}^{(2)}(w_{2}^{\nu})}{\alpha_{\nu}^{(2)}(w_{1}^{\nu})}g(w_{1}^{\nu},-w_{2}^{\nu}-\nu c)(1-\alpha_{\nu}^{(2)}(w_{1}^{\nu})\alpha_{\nu}^{(1)}(w_{2}^{\nu})).
\end{align}
Taking the limit $w_{2}^{\nu}\to-w_{1}^{\nu}-\nu c$ we obtain that
\begin{equation}
\mathbb{N}(\left\{ w_{1}^{\nu},w_{2}^{\nu}\right\} )=X_{k}^{(1),\nu}+\frac{f(w_{1}^{\nu},w_{2}^{\nu})}{f(w_{2}^{\nu},w_{1}^{\nu})}\frac{1}{\alpha_{\nu}(w_{1}^{\nu})}X_{k}^{(2),\nu},
\end{equation}
where we used (\ref{eq:limX}). Now we can take the on-shell limit
where the following Bethe equation is satisfied
\begin{equation}
\alpha_{\nu}(w_{1}^{\nu})=\frac{f(w_{1}^{\nu},w_{2}^{\nu})}{f(w_{2}^{\nu},w_{1}^{\nu})}.
\end{equation}
Substituting back we just obtained that
\begin{equation}
\mathbb{N}(\left\{ w_{1}^{\nu},w_{2}^{\nu}\right\} )=X_{k}^{(1),\nu}+X_{k}^{(2),\nu},
\end{equation}
therefore we just proved the last criteria \ref{enum:prop3}. Since
the normalized on-shell overlaps satisfy the Korepin criteria, they
can be written as a Gaudin-like determinant:
\begin{equation}
\mathbb{N}(\bar{w})=\det G_{+}.
\end{equation}

The proof in the untwisted case is completely analogous.

\bibliographystyle{elsarticle-num}
\bibliography{ref}

\end{document}